\documentstyle[epsfig,mnras_cite,times]{mn}
\title{The Oxford-Dartmouth Thirty Degree Survey I: Observations and Calibration of a Wide-Field Multi-Band Survey}
\author[MacDonald et al.]{Emily C.
MacDonald$^{1}$\thanks{ecm@astro.ox.ac.uk}, Paul Allen$^{1,2}$, 
Gavin Dalton$^{1,3}$, Leonidas A. Moustakas$^{1,4}$, \newauthor
Catherine Heymans$^{1,5}$, Edward Edmondson$^{1}$, Chris Blake$^{6}$,  Lee Clewley$^{1}$, \newauthor 
Molly C. Hammell$^{7}$, Ed Olding$^{1}$,  Lance Miller$^{1}$, Steve Rawlings$^{1}$, Jasper Wall$^{1}$,\newauthor Gary Wegner$^{7}$ \& Christian Wolf$^{1}$. \\ 
$^1$ University of Oxford, Astrophysics, Keble Road, Oxford, OX1 3RH, UK. \\
$^2$ Research School of Astronomy and Astrophysics, The Australian National University, Mount Stromlo Observatory, Cotter Rd, Weston, ACT 2611, Australia.\\
$^3$ Space Science and Technology Division, Rutherford Appleton Lab., Didcot, OX11 OQX, UK.\\
$^4$ Space Telescope Science Institute, 3700 San Martin Drive, Baltimore, MD 21218. \\
$^5$ Max-Planck-Institut f\"{u}r Astronomie, K\"{o}nigstuhl, D-69117, Heidelberg, Germany.\\
$^6$ Institute for Astronomy, School of Physics A28, University of New South Wales, New South Wales 2006, Australia.\\
$^7$ Dartmouth College, 6127 Wilder Laboratory, Hanover, NH 03755-3528. \\}

\begin{document}

\maketitle
\pagestyle{myheadings}
\markboth{MacDonald et al.} {The Oxford-Dartmouth Thirty Degree Survey I: Observations and Calibration}

\begin{abstract}
The Oxford Dartmouth Thirty Degree Survey (ODTS) is a deep, wide,
multi-band imaging survey designed to cover a total of 30 square
degrees in $BVRi^{\prime}Z$, with a subset of $U$ and $K$ band data, in four separate fields of 5-10
deg$^{2}$ centred at 00:18:24 +34:52, 09:09:45 +40:50, 13:40:00
+02:30 and 16:39:30 +45:24. Observations have been made using the Wide
Field Camera on the 2.5-m Isaac Newton Telescope in La Palma to average limiting depths 
($5\sigma$ Vega, aperture magnitudes) of $U$=24.8, $B$ = 25.6, $V$ = 25.0, $R$ =
24.6, and $i^{\prime}$ = 23.5, with observations taken in ideal conditions reaching the target depths of $U$=25.3, $B$ = 26.2, $V$ = 25.7, $R$ =25.4, and  $i^{\prime}$ = 24.6. The INT $Z$ band data was found to be severely effected by fringing and, consequently, is now being obtained at the MDM observatory in Arizona. A complementary $K$-band survey has also been carried out at MDM, reaching an average depth of $K_{5\sigma} \approx 18.5$. At present, approximately 23 deg$^{2}$ of the
ODTS have been observed, with 3.5 deg$^{2}$ of the 
K band survey completed. This paper details the survey goals, field
selection, observation strategy and data reduction procedure, focusing on the photometric calibration and catalogue
construction.  Preliminary photometric redshifts have been obtained for a subsample of the objects with $R \leq 23$. These results are presented alongside a brief description of the photometric redshift determination technique used. The median redshift of the survey is estimated to be z $\approx 0.7$ from a combination of the ODTS photometric redshifts and comparison with the redshift distributions of other surveys. Finally, galaxy number counts for the ODTS are
presented which are found to be in excellent agreement with previous
studies.

\end{abstract}

 
\section{Introduction} 

Understanding the origin and evolution of galaxies and large scale
structure within the Universe remains one of the most challenging
areas in modern cosmology. With the completion of 2 Degree Field Galaxy
Redshift Survey (2dFGRS, \pcite{Colless}) and the imminent conclusion of the Sloan
Digital Sky Survey (SDSS, \pcite{stoughton}), we are witnessing the emergence of
an accurate and detailed model of structure in the nearby
Universe. However, to gain insight into the evolution of the Universe
out to higher redshifts requires the advent of deep surveys, with
substantial areal coverage to ensure large number statistics and avoid
cosmic variance. A number of such surveys have been initiated in
recent years, each with individual goals but all hoping to shed light on
the structure and formation of the high redshift Universe. For example,
the ESO Imaging Survey (EIS, \pcite{nonino}) is a 24 deg$^{2}$
moderately deep I band survey ($I_{5\sigma \rm AB} \approx 23.7$) with additional
limited coverage in B, V and the infrared. A sub-area of the EIS (0.25 deg$^{2}$) is covered in UBVRI to $I_{5\sigma \rm AB} \approx 24.7$ forming the EIS-Deep survey \cite{eisdeep}. The Canada-France Deep
Fields Survey (CFDF, \pcite{McCracken}), consists of 4 deep fields
totalling 1 deg$^{2}$, all covered in V and I (I$_{5\sigma \rm AB} \approx 25.5$),
with additional U and B coverage. Also ongoing are the Combo-17 survey \cite{combo}, where fields totalling $\approx$ 1 deg$^{2}$ have been observed 
through 17 medium-band filters to a  limiting magnitude of $R_{5\sigma \rm AB}=26.2$, and the NOAO
Deep Wide-Field Survey (NDWFS, \pcite{ndwfs}) which will cover 24 deg$^{2}$ in
BRIJH and K to  $R_{5\sigma \rm AB}=26$. Here we describe the Oxford-Dartmouth Thirty
Degree Survey (ODTS) which aims to provide multi band observations, to
allow for the determination of photometric redshifts, to
depths comparable with the deepest wide field surveys to date and over
a wider area. The Wide Field Camera (WFC) on the Isaac Newton Telescope (INT)
on la Palma has been used to observe $23$ deg$^{2}$ (out of the total $30$ deg$^{2}$) of $BVR$ ($R_{5\sigma \rm AB} \approx 25.6$ assuming $1^{\prime\prime}$ seeing) and $i^{\prime}$ imaging, with a subset of $U$ band data. The Z band data, although initially planned for
observation at the INT, are now currently being acquired using the
2.4m Hiltner Telescope at the MDM observatory, Kitt Peak (see
section~\ref{sec:datared}). Also, a K-Band survey, designed to run in
parallel with and be complementary to the optical ODTS, is currently
being carried out using the 1.3m McGraw-Hill Telescope at MDM, to a
depth of K$\approx$18.5 \cite{olding}. In addition, deep radio data have been
obtained from the VLA, covering a total of 2 square degrees of the
ODTS to a $5\sigma$ flux density limit of 100 micro-Jy at 1.4GHz with a
resolution of $1.5\arcsec$, and with deeper 1.4 GHz data and lower frequency (e.g.
74MHz and 330MHz) radio data over some fraction of the area. Part of the ODTS data also
overlaps with the Texas-Oxford One Thousand (TOOT) redshift survey of
radio sources \cite{Hill}, allowing us to obtain spectroscopic
redshift measurements for a number of sources in the ODTS.

The ODTS was initially designed with the following goals: 

\noindent{\bf (1) Clustering of Bright Lyman Break Galaxies (LBGs)} (Allen et al, submitted.): Lyman Break Galaxies exhibit a break in their spectra short wards of the Lyman limit, $912 \rm \AA$ in the rest frame. For LBGs at high redshift, z=3 and 4, this results in a $U$ and $B$ band drop out thus permitting the detection of Lyman Break candidates using the ODTS multi band data. Due to the extent of the ODTS, the clustering of LBGs can be studied over a much larger area than previously available, thus minimising the effects of cosmic variance.

\noindent{\bf (2) Clustering Properties of Faint Galaxies} (MacDonald et al, in prep.): Previous surveys have lacked the combination of both depth and width required to explore the large scale clustering and evolution of faint galaxies as a function of magnitude, colour and redshift. The size of the ODTS potentially allows the measurement of the angular correlation function up to degree scales.   

\noindent{\bf (3) Clustering of Extremely Red Objects (EROs)} \cite{olding}: Matching the ODTS $R$ band data with the MDM K-band data will allow for the selection of a large number of EROs, the criteria being an $R - K > 5$,  over a relatively large area. The ODTS can be used to estimate their space density and photometric redshifts will allow for clustering and evolution to be analysed.  

\noindent{\bf (4) Detection of High Redshift (z $ > $ 5) Quasars}: Extremely high redshift QSOs are rare but known to exist, as confirmed by the SDSS \cite{anderson}. Uncertainties still surround issues such as the shape of their luminosity function at the faint end and their evolution at high redshift. Colour selection methods, via the (V-i) vs (i-Z) colour-colour relation, will allow for the selection of faint QSO's at z$ >$ 5.

\noindent{\bf (5) High Redshift Galaxy Clusters}: Combining information about the cluster colour-magnitude relation \cite{gladders} with photometric redshifts and a search for spatial over-densities, clusters with 0.2 $< z <$ 1.2 can be selected from the ODTS. The aim is to use this sample to investigate the cluster colour-magnitude relation as a function of redshift and cluster mass (Hammell et al, in prep.). 

This paper summarises the ODTS and presents a full description and
characterisation of the survey data, structured as follows. Section 2
discusses the survey design, specifically the criteria adopted for the
field selection, the filter set used, and the observation
strategy. Section 3 outlines the basic data reduction process. Section
4 describes the photometry and source extraction, and section 5
outlines the astrometry. Section 6 details the photometric calibration for each
band and section 7 outlines the algorithm used to generate the final
matched catalogue. In section 8 a brief summary of the
photometric redshift determination is given and in section 9 galaxy number counts
are presented and compared with previous studies. Finally, a
summary is given in Section 10.

\begin{figure}
\begin{center}
{\leavevmode \epsfxsize=8.cm \epsfysize=7.cm \epsfbox{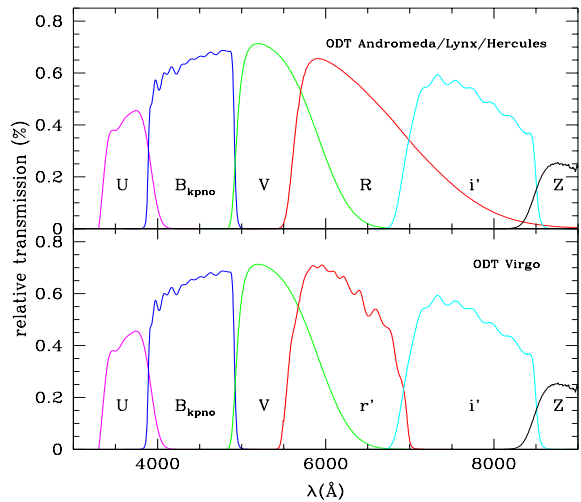}} 
\end{center}
\caption{The system response, determined from the filter transmission curves and the CCD response, for the full set of ODTS filters ($UBVRi^{\prime}Z$).}
\label{fig:filters}
\end{figure}

\section{The Oxford-Dartmouth Thirty Degree Survey Design} 
\label{sec:ODT}

The ODTS was initiated in August 1998, and was designed to make use of
the Wide Field Camera located at the prime focus of the 2.5m Isaac
Newton Telescope on La Palma. It is a deep-wide survey facilitating 6
filters per field; namely RGO $U$, Kitt Peak $B$, Harris $V$, Harris
or Sloan $R$, Sloan $i^{\prime}$ and RGO $Z$, the system response
for which are provided in Figure~\ref{fig:filters}. Six filters were
selected to allow for the determination of photometric redshifts for
objects in the survey. In particular, the Kitt Peak $B$ and Sloan-Gunn
$i^{\prime}$ filters were adopted, as opposed to their Johnson-Cousins
counterparts, because they have higher throughput and simpler, more
box-like transmission curves which are better for photometric redshift
determination and allow for cleaner colour-colour selection of
objects with sensitive spectral features, such as LBGs. In addition
the $i^{\prime}$ filter is less susceptible to fringing caused by sky
emission lines (see section~\ref{sec:datared}). Initially the
Sloan-Gunn $r^{\prime}$ filter was not available at the INT, so
observations were made using the Harris $R$ filter which has a long red
tail, and so suffers from noticeable fringing. However, the
$r{^\prime}$ filter became available at the beginning of the Virgo
field (see table~\ref{tab:fields}) observations and, consequently, was
adopted for this field only.  Initially, the ODTS was designed to
cover $\sim$30 deg$^2$ to estimated depths (2$^{\prime\prime}$ aperture Vega magnitudes) of 
$B$ = 26.2, $V$=25.7, $R$=25.4 and $i^{\prime}$=24.6 and $Z$=22.0, assuming a 5$\sigma$
detection threshold and the median INT seeing of $1\arcsec$, with an
additional sub area of $\approx 5$ deg$^{2}$ in $U$(= 25.3) assuming ideal
(seeing $< 0.7^{\prime\prime}$) conditions. The initial Vega and AB magnitude limits for the 
survey are given in table~\ref{tab:filters}.
\vspace{-0.3cm}

\subsection{Field Selection} 
\label{sec:fieldsel}
Initially the Andromeda, Hercules and Lynx fields were chosen from a
number of potential fields primarily for their observability during
the first few allocated INT runs; their centres are shown in
table~\ref{tab:fields}. The Virgo field was added at a later date to
optimise year round observability and provide a field visible from the
Southern hemisphere. In addition, the selected fields had to have low extinction,
 overlap with existing (multi-wavelength) datasets and have a 
lack of bright
stars, nearby galaxies and bright, large clusters. The mean
extinction, $E(B-V)$, in each field was determined from the DIRBE
corrected IRAS $100 \mu m$ maps of \scite{schlegel} and was found to
be $<$ 0.06 in Andromeda and $<$ 0.02 in the remaining fields. The
large ($4^{\prime}$) pixel scale of the IRAS maps provides little
information about the small scale distribution of the Galactic dust,
and therefore making extinction corrections using $4^{\prime}$ cells,
as opposed to individual galaxy corrections, could imprint a low level
spurious clustering pattern. This potential bias combined with the low
galactic extinction values meant no extinction corrections for
galactic dust were applied.

\begin{table}
\begin{center}
\begin{tabular}{lcccc} \\ \hline
{\bf Field} & {\bf $ \bf \alpha$(J2000)} & {\bf $ \bf \delta$(J2000)} & $\bf l$ &  $\bf b$ \\\hline
Andromeda & 00 18 24 & +34 52 00 & 115 & -27\\
Lynx & 09 09 45 & +40 50 00 & 181 & +42 \\
Hercules & 16 39 30 & +45 24 00 & 70 & +41\\
Virgo & 13 40 00 & +02 30 00 & 330 & +62\\\hline\\ 
\end{tabular}
\caption{Field centres for the ODT survey in equatorial and Galactic coordinates.}
\label{tab:fields}
\end{center}
\end{table}	

\begin{table}
\begin{center}
\vspace{-0.4cm} 
\begin{tabular}{c l c c}            \hline
\bf Filter   &\bf Exposure & \multicolumn{2}{|c|} {\bf{$\bf 5\sigma$ Limiting depths using $\bf 2^{\prime\prime}$ Aperture}}   \\
 & \bf $\:\:\:$ Time & \bf AB Magnitudes & \bf Vega Magnitudes \\ \hline
$U$  & 6 x 1200s          & 26.1  & 25.3   \\
$B$  & 3 x $\,\,\,$900s   & 26.1  & 26.2   \\ 
$V$  & 3 x 1000s          & 25.7  & 25.7   \\
$R$  & 3 x 1200s          & 25.6  & 25.4   \\
$i^{\prime}$ & 3 x 1100s  & 25.0  & 24.6   \\
$Z$  & 1 x $\,\,\,$600s   & 22.4  & 21.9   \\ \hline
\end{tabular}
\caption{The exposure times and expected detection limits ($5\sigma$
  aperture Vega and AB magnitudes) for the various pass bands used for the
  ODTS. With the exception of the $U$ and $Z$ images, each pointing is
  segregated into three separate exposures with 5\arcsec\ offsets of
  the telescope.}
\label{tab:filters}
\end{center}
\end{table}

\subsection{Observations}
\label{sec:obs}

\begin{figure*}
\vspace{0.3cm}\centerline{\epsfig{file=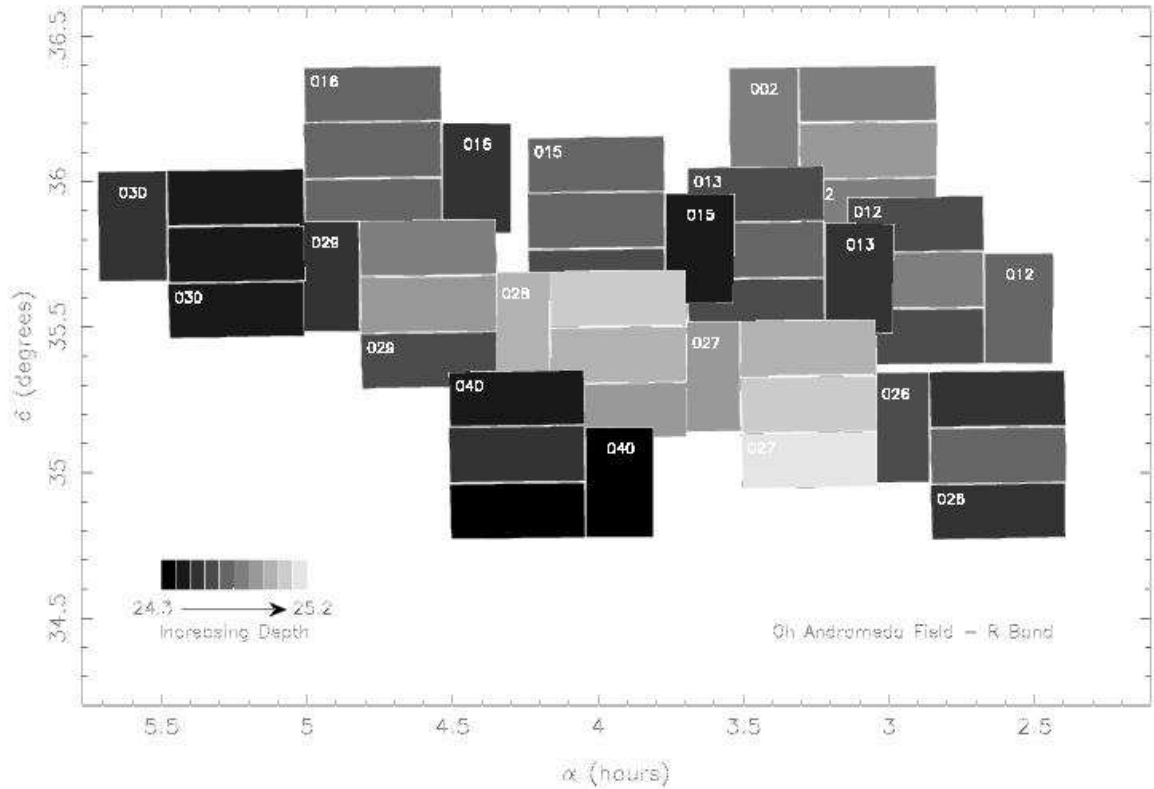, width=11.cm, angle=270}}\vspace{0.5cm}
\caption{A grey-scale image of the reduced $R$ band Andromeda data (totalling 2.78 square degrees), demonstrating how the depths reached vary across the field. Frame depths vary from 24.3 to 25.2 magnitudes where the lighter the shade, the greater the depth. Variable conditions and individual chip response differences lead to slight changes of depth within each pointing as expected.  It should be noted that there are small gaps ($\approx 1^\prime $) present between the WFC chips.}
\label{fig:greyscale}
\end{figure*}

\begin{figure}
\begin{center}
\epsfig{file=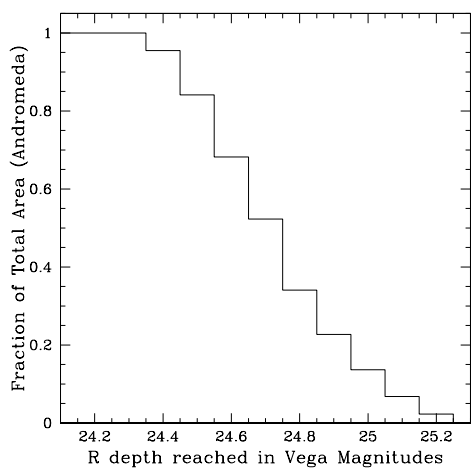, width=6.5cm, angle=0}
\end{center}
\caption{The cumulative distribution of the fraction of area (totalling 2.78 square degrees) versus $R$ depth reached for the reduced Andromeda data. Frame depths vary from 24.3 to 25.2.}
\label{fig:cumulative}
\end{figure}

Observations were made over 63 nights between August 1998 and March
2003, during which time several nights were lost due in part to instrument problems,
but mostly due to bad weather conditions. The data were obtained using the WFC which
is a mosaic of four 4096 x 2048 pixel CCD chips, each chip covering
of $22.8^{\prime} \times 11.4^{\prime}$ ($\approx0.29$ degs$^{2}$ per WFC
pointing) with a pixel scale of $0.33^{\prime\prime} $ per pixel. At
the beginning and end of each night bias, dark and twilight flat-field
frames were acquired. Landolt standard star frames were also observed
several times throughout each night (see section~\ref{sec:photo}). On
the whole, data were taken when conditions were either photometric or
light cirrus was present, with variable seeing across the
fields. Assuming the median INT seeing of 1\arcsec\, the $5\sigma$
depths shown in table~\ref{tab:filters} implied a total of 3.7 hours
per pointing were required to obtain the multi band data
($BVRi^{\prime}Z$), resulting in a survey speed of $\sim 1.2$ deg$^{2}$
per night (not including the $U$ data). The median seeing values actually obtained for each field
in each band are presented in table~\ref{tab:seeing} alongside the
median depth reached in each band in each field. There are no values
for the $Z$ band data at present because the fringing proved too
severe (see section~\ref{sec:datared}). Figure~\ref{fig:greyscale} illustrates how the $R$ band depths vary across the Andromeda field (fully reduced data only) and figure~\ref{fig:cumulative} shows the corresponding cumulative distribution of fraction of total area versus depth in $R$.

\begin{table}
\begin{center}
\begin{tabular}{c c c c }            \hline
 & \bf Andromeda &\bf  Lynx & \bf Hercules \\ 
\bf Filter & \bf Seeing / Depth &\bf Seeing / Depth &\bf Seeing / Depth \\ \hline
$U$  & 1.00 $\: \:$ 24.8 & n/a & n/a  \\ 
$B$  & 1.31 $\: \:$ 25.3 & 1.46 $\: \:$ 25.61 & 1.24 $\: \:$ 25.8  \\
$V$  & 1.24 $\: \:$ 24.8 & 1.23 $\: \:$ 25.2 & 1.66 $\: \:$ 24.9  \\
$R$  & 1.13 $\: \:$ 24.3 & 1.36 $\: \:$ 24.7 & 1.32 $\: \:$ 24.7  \\
$i^{\prime}$ & 1.10 $\: \:$ 23.4 & 1.55 $\: \:$ 23.7 & 1.64 $\: \:$ 23.5  \\ \hline
\end{tabular}
\caption{Median seeing and median depths reached (Vega magnitudes) for the reduced field data, where the depth of each frame has been determined from the turn over in number counts.}
\label{tab:seeing}
\end{center}
\end{table}

\begin{table}
\begin{center}
\vspace{-0.6cm} 
\begin{tabular}{l c c c c c}            \hline
\multicolumn{6}{|c|} {\bf Area in Square Degrees - Observed/Reduced}\\
\bf Field& \bf $U$ & \bf $B$ & \bf $V$ & \bf $R$ & \bf $i^{\prime}$\ \\ \hline
And  & 1.16/1.16 & 6.96/2.31 & 6.96/2.58 & 6.96/2.78 & 6.96/2.81  \\ 
Lynx & n/a & 7.83/1.79 & 7.83/1.77 &7.83/1.78 & 7.83/1.79\\ 
Herc & 0.29/0 & 4.06/1.33 &4.06/1.75 & 4.06/1.30 & 4.06/1.33 \\ 
Virgo & n/a & 3.77/0 & 3.77/0 & 3.77/0 & 3.77/0 \\ \hline
\end{tabular}
\caption{Current areas of observed and reduced multi-band ($UBVRi^{\prime}$) data for the 4 ODTS fields.}
\label{tab:area}
\end{center}
\end{table}

\begin{figure*}
\begin{center}
\hspace{0.7cm}\epsfig{file=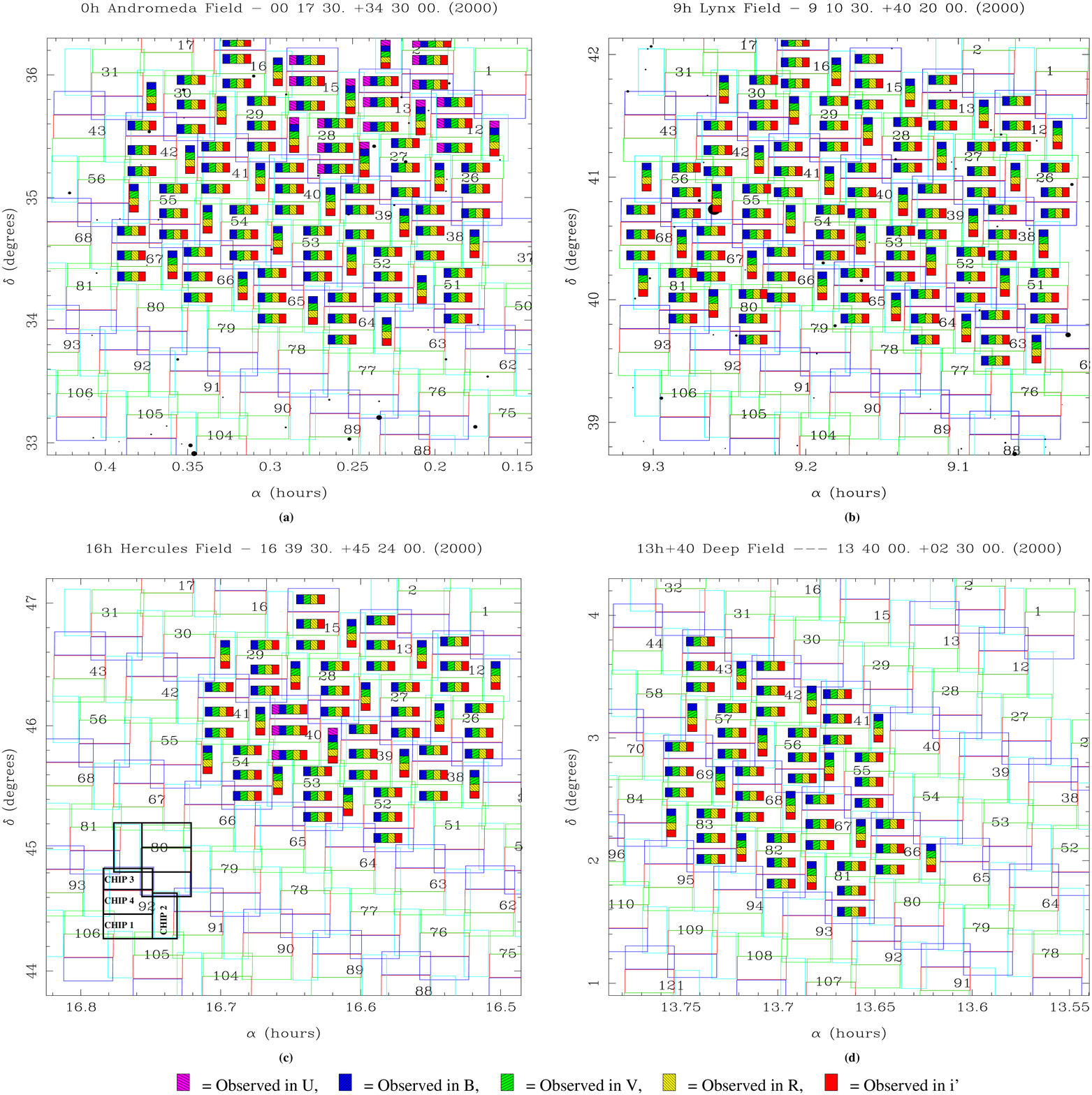, width=17.6cm, angle=0}
\end{center}
\caption{The (a) 0h Andromeda, (b) 9h Lynx, (c) 16h Hercules and (d)
13h Virgo fields divided into diagonal grids, where each numbered grid
element represents one WFC pointing. The 4 individual chips of the WFC
can be seen and in the bottom left corner of diagram (c) two of the
WFC pointings are outlined in black for clarity, and the chip numbers
have been marked. Coloured coded bars illustrate the coverage of each
sub-field to date.}
\label{fig:fieldplots}
\end{figure*}

In order to cover each of the ODTS fields efficiently with the WFC
focal plane geometry, a tiling pattern involving using the camera in two
rotational positions, $180^\circ$ apart, was employed. The fields were
covered by a diagonal grid, with the rotation alternating with each
row. Each grid element, representing one pointing of the INT WFC, was
assigned a unique ODTS identification
number. Figure~\ref{fig:fieldplots} depicts the sub-fields (i.e. the
WFC camera pointings), where the coloured bands within the grid
elements indicate which filters each of the sub-fields have been
observed in thus far.  A couple of WFC pointings have been highlighted
for clarity in figure~\ref{fig:fieldplots} (c) and the WFC chips have
been labelled. In this paper, the terminology {\it pointing} refers to
one entire WFC image and {\it frame} refers to one chip of the WFC.\\

Observations for the optical (INT) portion of the ODTS were completed
in March 2003 with approximately $23$\,deg$^2$ of the survey observed
in $BVRi^{\prime}$, although the data reduction is still ongoing. The
$U$ data, taken in the best observing conditions, currently cover
$\sim$1 deg$^2$ in the Andromeda field. The total coverage of the
fully reduced data is summarised in table~\ref{tab:area}.

\section{Data Reduction}
\subsection{Preprocessing}
\label{sec:datared}

Standard {\tt IRAF}\footnote{{\tt IRAF} is distributed by the NOAO, which is operated by the Association of Universities for Research in Astronomy, Inc. under cooperative
agreement with the National Science Foundation.} data reduction
routines were implemented to remove the instrumental signature. Bias
subtraction was made using the chip over-scan regions and master bias
frames. The dark current was found to be negligible. Small
non-linearities were known to exist in the WFC chips, the corrections for
which were determined by the Cambridge Astronomical Survey
Unit\footnote{The values of the linearity corrections used, along with
other information about the INT WFC, are available on the CASU website
- http://www.ast.cam.ac.uk/$\sim$wfcsur/index.php.} and were
applied appropriate to the time of observation.

Corrections for variations in the pixel to pixel sensitivity and
vignetting caused within the optics were made by dividing frames by
the appropriate master flat field, created by median combining the object
frames, or the twilight flat frames in the case of those bands
subject to fringing. On comparison of the $B$ and $V$ band twilight
flats with the deep sky flats obtained from the $B$ and $V$ object frames, 
it was found that no illumination correction was necessary.

\begin{figure}
\centerline{\psfig{file=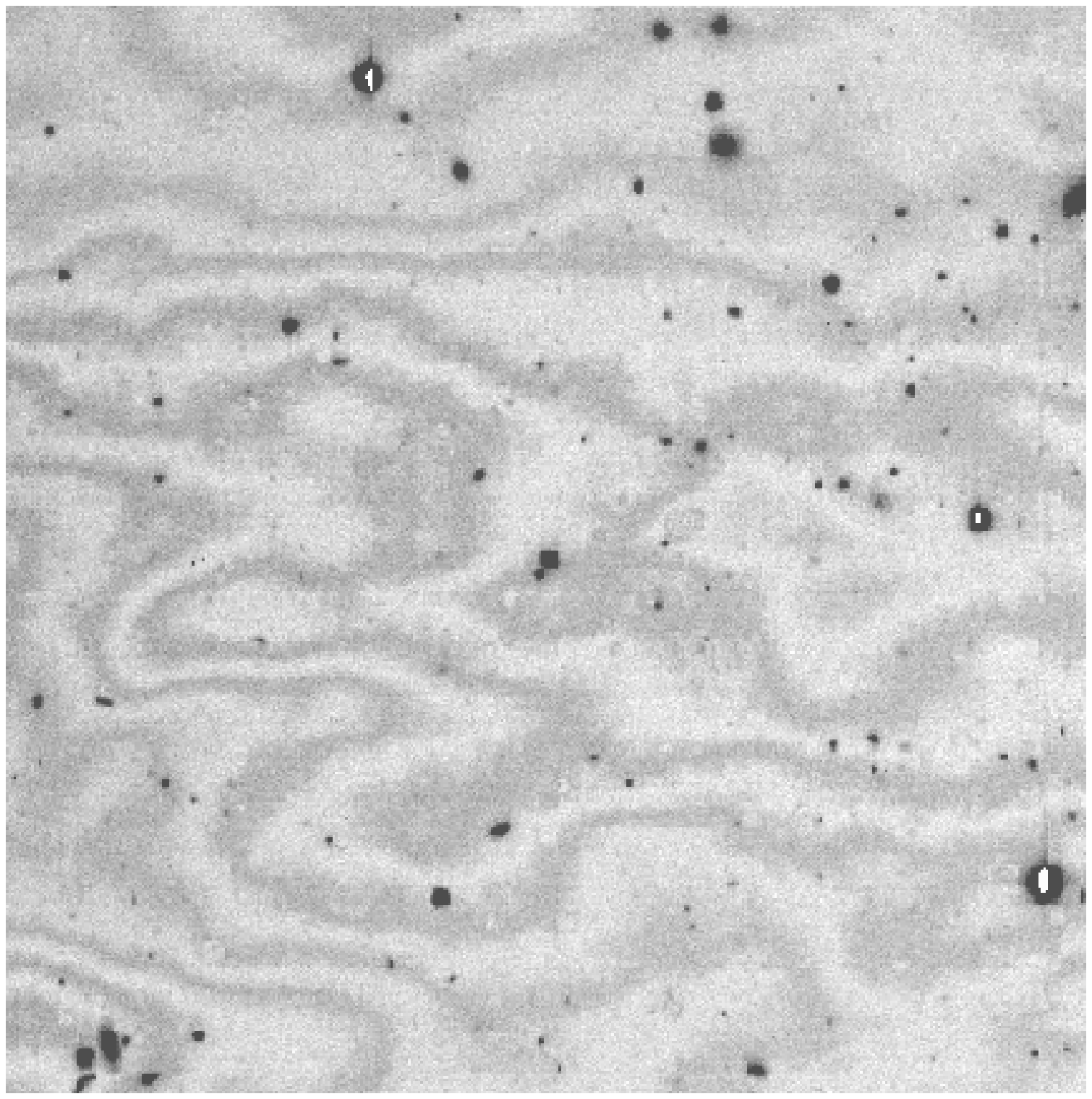,width=4.1cm,angle=0,clip=}\hspace{0.2cm}\psfig{file=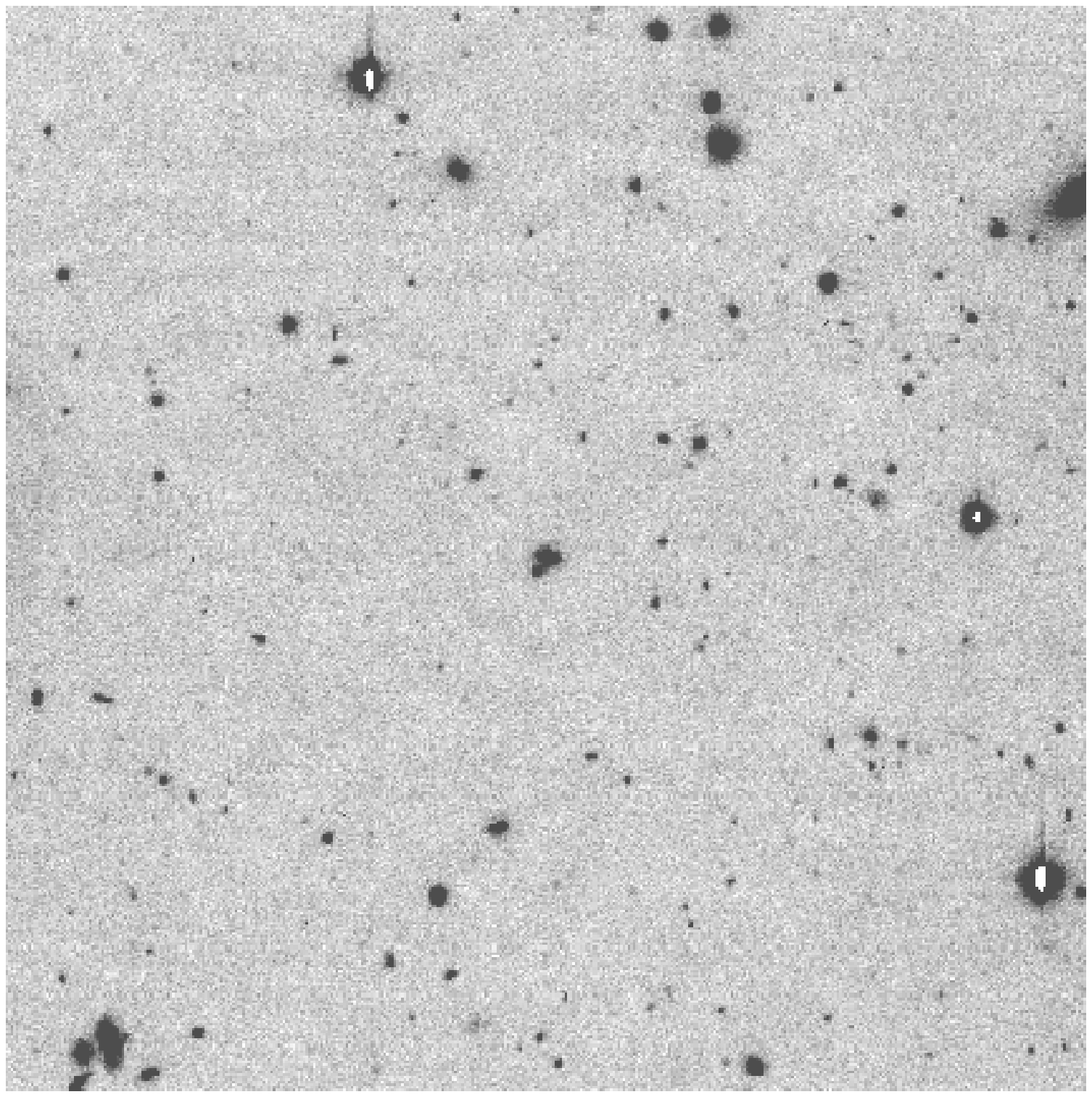,width=4.1cm,angle=0,clip=}}
\caption{A subsection of an $i^{\prime}$ image before (left) and after (right) undergoing de-fringing.}
\label{fig:fringe} 

\end{figure}

The $R$, $i^{\prime}$ and $Z$ band images were found to suffer from
fringing caused by the multiple reflection and interference of the
night-sky emission lines within the CCD.  Unfortunately
de-fringing is far from trivial as the relative intensities of the
emission lines are dependent on the atmospheric conditions, thus
fringing is a time varying effect. To de-fringe the data the bias
corrected, flat-fielded $R$, $i^{\prime}$ and $Z$ target frames from each night
were median combined. This resulted in master fringe frames for each band,
containing the residual sky fringe pattern. For each image,
the master fringe frame was subtracted, scaling the fringe level of
the master fringe frame to that of the fringing in the images. This
effectively removes the dominant zeroth order fringing component, an
example of which can be seen in figure~\ref{fig:fringe}.

For the WFC images, fringing effects were found to be $\sim 0.5\%$ of
the sky level for the $R$ band, $\sim 3\%$ for $i^{\prime}$ and $\sim
6\%$ for $Z$. After the de-fringing process the residual fringing
level was reduced to $< 0.1\%$ of the sky but the $Z$ band fringing
remained visible at $\sim 0.5\%$. Consequently, the acquisition of $Z$
band data using the INT had to be abandoned. $Z$ data for all observed
ODTS fields are now being obtained using the 8k camera on the 2.4m
Hiltner telescope at MDM ($3 \times 1200$ second exposures). This camera has thick CCDs and produces much
cleaner $Z$ band imaging.

Bad pixel masks were then created for each chip and the defective
pixels were effectively removed by interpolating across them in each
image

For each pointing in $BVRi^{\prime}$ three images were taken, offset
from each other by $5\arcsec$ to ensure that different areas of the
sky fell on the bad detector regions. The $U$ band data required 6
pointings to reach the required depth so these were taken in two
groups of three, offsetting in each group as for the other
bands. These were then aligned, median combined and trimmed, where the median combination was done using the {\tt IRAF} average sigma clipping algorithm, rejecting values deviating from the mean by $> \pm 3 \sigma$ to ensure the removal of any remaining bad pixels and cosmic rays.

To obtain the correct flux measurement for each of the sources the
contribution from the sky background also had to be considered. On
initial reduction of the ODTS data, significant sky gradients were
found across some images which SExtractor, the adopted source extraction
program \cite{sex} (see section~\ref{sec:sex}), had
difficulties correcting for. These gradients were found to vary between pointings, so were thought to be caused by diffuse
scattered light from bright stars in, or just outside, the observed fields, rather than being the result of vignetting within the instrument. An alternative background subtraction
algorithm was developed which computed the background value for every
pixel in the image by effectively centring a box (of width 15-25 pixels) on each pixel,
calculating the modal value of the background within the box
surrounding the central pixel and assigning that value to the central
pixel. A bi-cubic spline surface was then fit to the array of modal
values in order to create a smoothed sky background map, which was
then subtracted from the image.

\section{Photometry}

Initially, photometric zeropoints were estimated for each frame in every
band for each observing run using the standard star data acquired (see
section~\ref{sec:photo}). Using these zeropoints, SExtractor
\cite{sex} was then used to perform the photometry. After source
catalogues had been created for each image (see
section~\ref{sec:sex}), accurate  photometric
zeropoints were determined across the $V$ band by comparing objects in overlap regions (see
section~\ref{sec:overlap}), and adjusting their zeropoints relative to
a chosen calibrator chip. Finally, the other bands were adjusted relative to the $V$ band data
 by comparing the colours of the
stars in the ODTS images with those obtained using the Pickles (1998) library of stellar spectra, and applying zeropoint corrections needed to match these data sets (see section~\ref{sec:stell}).

\subsection{Initial Photometric Calibration using Standard Stars}
\label{sec:photo}

Extracting source counts from the data at this stage will result in a measure of the instrumental magnitude which must be converted to the apparent magnitude. In order to perform this flux calibration, short exposures of various fields containing several standard stars, calibrated by \scite{landolt}, were observed through the same filters as that night's target fields. Landolt fields SA92, SA95, SA101 and SA104 covered the ODTS fields well, with several standard stars falling in each frame. For each observing run, at least two of these standard fields were observed and were chosen to span a large airmass range, enabling us to monitor photometric quality throughout the night, and allowing for the subsequent estimation of extinction and colour terms.

As the survey progressed, the large number of non-photometric nights and the substantial
overheads associated with multiple-band observations of standard stars
prompted the use of an alternative procedure
to establish the multi-band photometry for each survey
region (see section~\ref{sec:overlap}). With this method, it is only the zeropointing of
the $V$ band calibrator chip that is of importance, as the zeropoints of the rest of the $V$ band data are determined by overlap matching (see section~\ref{sec:overlap},) and the other
bands are then corrected relative to the $V$ band, via stellar locus fitting (see section~\ref{sec:stell}). Data for the Andromeda $V$ band calibrator frame
and the corresponding standard star observations were obtained during photometric conditions, and the zeropoint was determined in the usual fashion. The uncertainty on the zeropoint of the $V$ band calibrator frame was found to be $0.07$ magnitudes.

\subsection{Object Extraction using SExtractor}
\label{sec:sex}

\begin{figure}
\centerline{\psfig{file=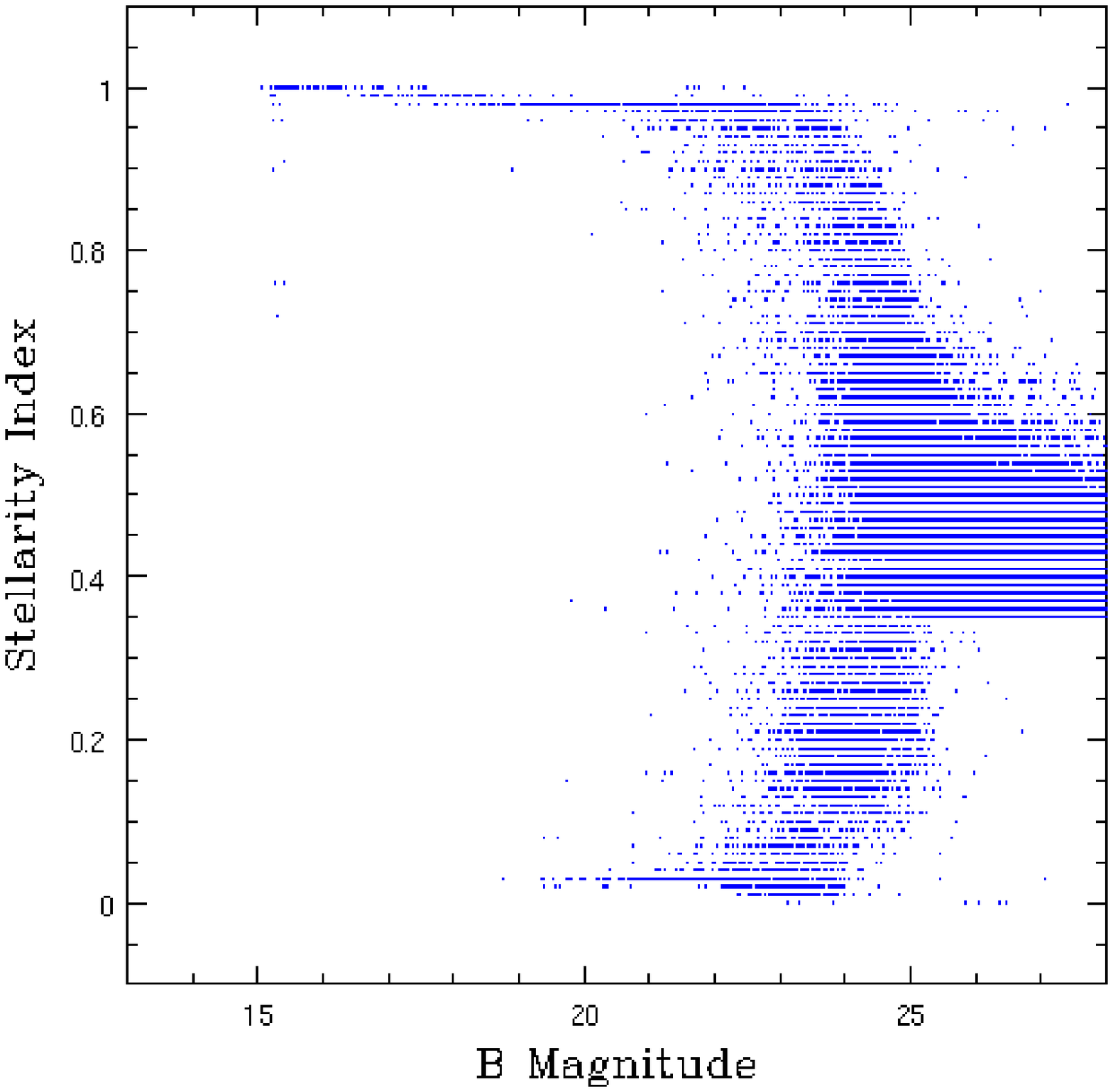,width=4.15cm,angle=0,clip=}\hspace{0.2cm}\psfig{file=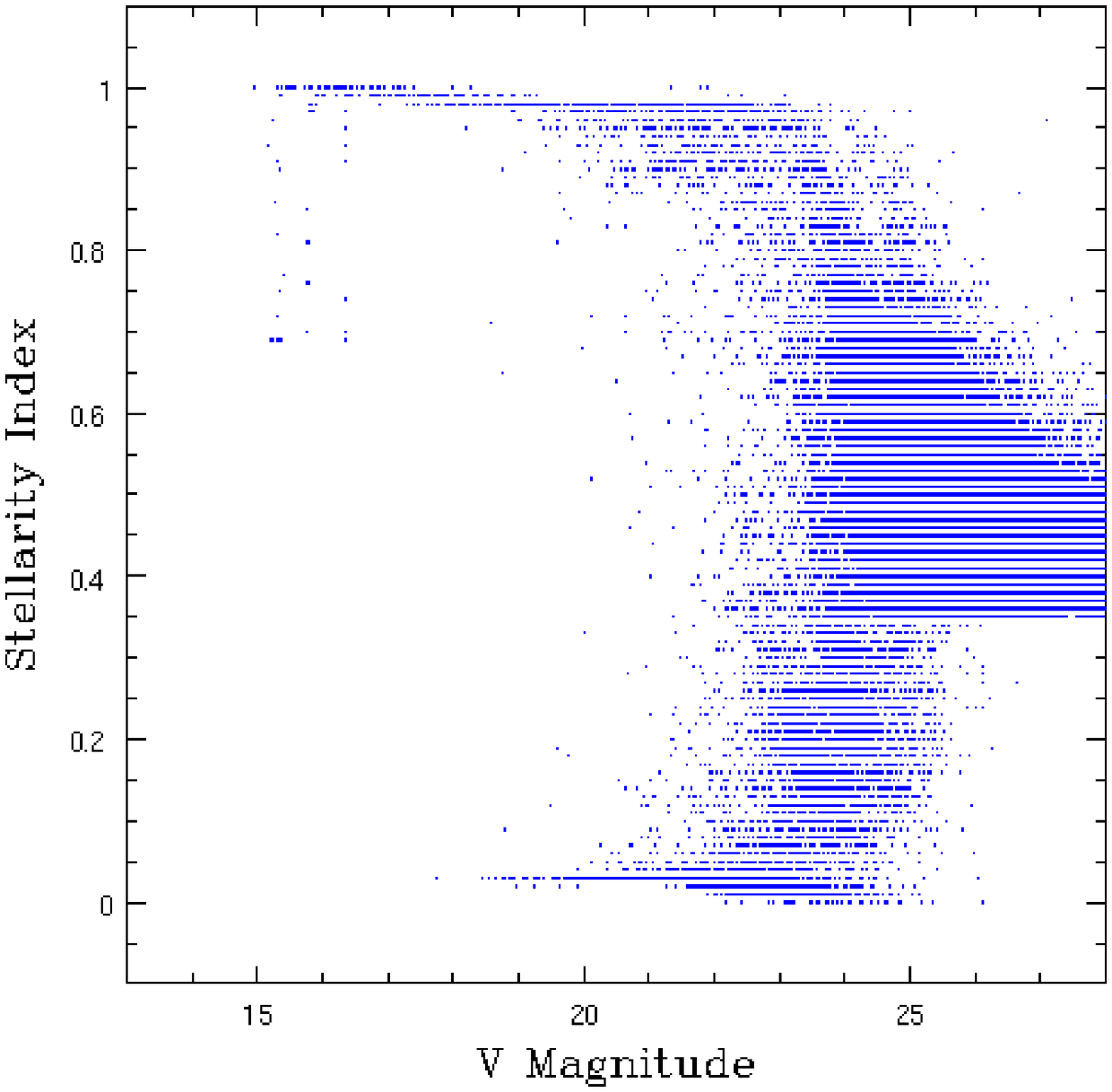,width=4.15cm,angle=0,clip=}}
\vspace{0.5cm}
\centerline{\psfig{file=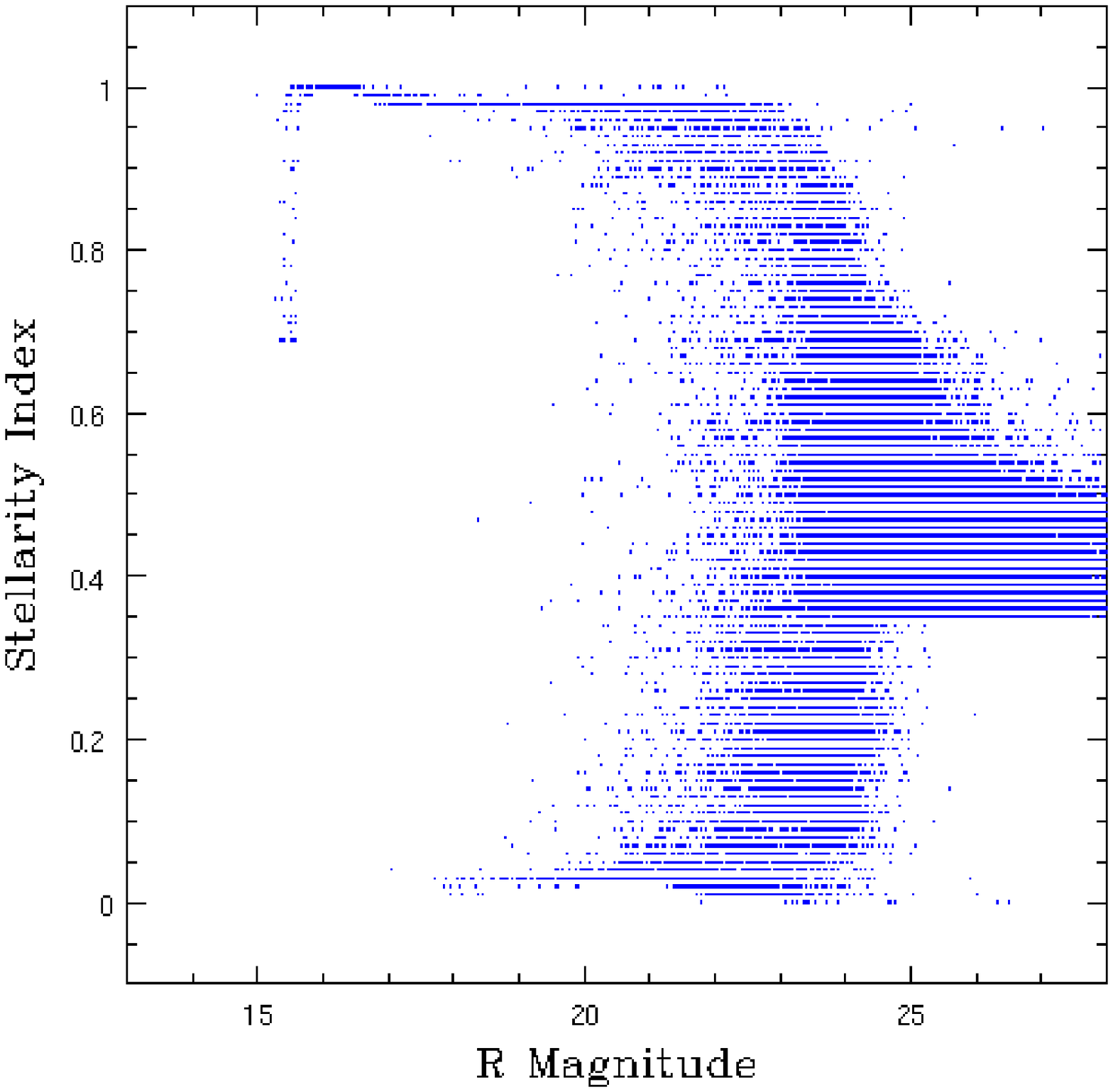,width=4.15cm,angle=0,clip=}\hspace{0.2cm}\psfig{file=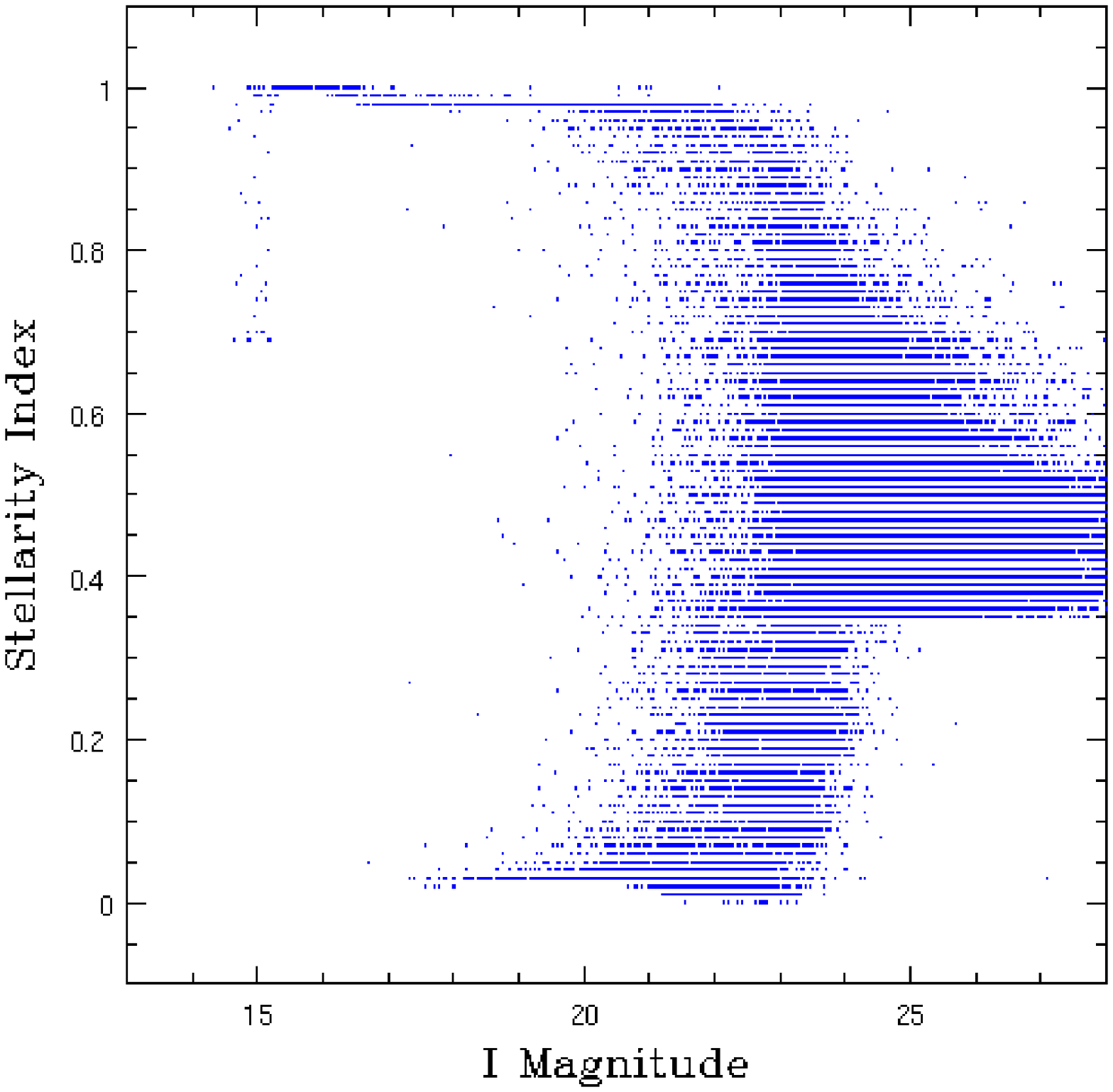,width=4.15cm,angle=0,clip=}}
\caption{SExtractor star/galaxy classifier (Stellarity Index), where 1 represents a star and 0 a galaxy, against magnitude for the one of the Andromeda pointings.}
\label{fig:star} 

\end{figure}

SExtractor \cite{sex}, an automated package, was used to perform the
image analysis and source extraction. It is designed for the detection
of faint objects in wide-field surveys and so is particularly
suited to the ODTS data. It also has the additional advantages of
speed, a de-blending algorithm and a neural network star-galaxy
classifier. Due to the afore mentioned problems with the background
subtraction, all images were background subtracted prior to SExtractor
object detection and the background subtraction algorithm used by
SExtractor was switched off.

Convolving the data with a Gaussian filter of full width half maximum
(FWHM) close to the seeing is known to optimise detections and reduce
noise levels during the source extraction. Hence, a Gaussian filter
described by a $3\times3$ matrix with an FWHM of $1.5-6.1$ pixels (= $0.5^{\prime\prime}-2^{\prime\prime}$) depending on the seeing, was utilised which had proved, after initial
tests, the most effective at faint source extraction. Objects meeting
the extraction criteria of having 8 connected pixels with flux $ >
0.4 \sigma$ above the local background level were analysed. This
corresponded to a minimum of $> 1.3 \sigma$ per object detection
\cite{booth}, however in practise only $> 5 \sigma$ detections are
included in the final catalogues.

Within SExtractor, the intensity profile of each source was automatically examined in order 
to ascertain whether it was a single source or a merged object, where the latter initiates the de-blending procedure within SExtractor. Simulations suggest that photometric errors for
objects de-blended by SExtractor are $< 0.2$ magnitudes and in most
cases are $< 0.1$ magnitudes. Astrometric errors due to de-blending
are typically $< 0.4$ pixels ($0.1^{\prime\prime}$) \cite{sex}.

On occasion, the comparatively low detection threshold used
resulted in spurious detections in the wings of both bright and
extended objects where the local background noise level was relatively
high. Hence the 'cleaning' procedure was implemented within SExtractor
whereby the contribution to the background from bright/extended
objects is estimated by fitting them with appropriate Gaussian
profiles. Local object intensities  remaining above the detection
threshold when the adjusted local background was subtracted were
accepted into the final catalogue. These spurious detections
typically accounted for $10-20\%$ of all detections and those few
which remained after cleaning were dealt with during the image masking
process detailed in section~\ref{sec:holes}.

SExtractor was used to calculate aperture and isophotal corrected magnitudes for every source. Aperture magnitudes were obtained by integrating the flux
within a fixed aperture and subtracting the contribution from the sky
background, its value estimated within an annulus outside the
aperture. For the ODTS, an aperture size of diameter $10$ pixels
($3.33^{\prime\prime}$) was used for all bands. Isophotal corrected
magnitudes were also computed whereby the flux within a specified
isophote, set at $2.5 \times$  the local background level, was integrated. To
retrieve flux existing outside the limiting isophote, a Gaussian
profile was then fit to the intensity distribution of the object and
an estimation of the omitted flux made and subtracted. All extracted
magnitudes have an associated rms error calculated within
SExtractor. This random error increases with faintness, but
remains small to the limiting depths of the ODTS compared with the
various calibration uncertainties (see section~\ref{sec:errors}).

Aperture magnitudes, although consistent, will tend to underestimate
the actual magnitude of an extended object due to the flux lost
outside the aperture. Simulations suggest that this is negligible for
seeing-limited objects fainter than $R \sim 18.5$ in the ODTS data
\cite{olding}. Isophotal corrected magnitudes work well for the brighter
more extended objects but become unstable at the faint end as they
involve assumptions about the shapes of objects which become highly
uncertain at faint magnitudes. In general the isophotal corrected
magnitudes were found to be less self-consistent than the aperture
magnitudes, ascertained by comparing the spread of the ODTS stellar
data around the main sequence and the scatter in the difference in
magnitudes for common objects found in the overlap regions.

Ultimately catalogues containing both aperture and isophotal corrected
source magnitudes for each frame at each pointing in every band were
archived, but aperture magnitudes were used in the final calibrations
performed in section~\ref{sec:photocalib}, as they give much more consistent colours. Archiving both magnitudes allows the user to choose the magnitude regime most appropriate for their work.

\begin{table}
\begin{center}
\begin{tabular}{l c c c c c }            \hline
\multicolumn{6}{|c|} {\bf Area in Square Degrees}\\
\bf Field& \bf $U$ & \bf $B$ & \bf $V$ & \bf $R$ & \bf $i^{\prime}$\ \\ \hline
Andromeda  & 0.82 & 2.16 & 2.01 & 2.53 & 2.23\\ 
Lynx  & n/a & 1.45  & 1.53 & 1.50 & 1.55\\ 
Hercules  & n/a & 0.99 & 0.58 & 0.84 & 1.07\\ \hline

\end{tabular}
\caption{Areas of the fully reduced, masked data for each band to date in the ODTS fields.} 
\label{tab:effareas}
\end{center}
\end{table}

\begin{figure*}
\begin{center}
\hspace{0.8cm}\epsfig{file=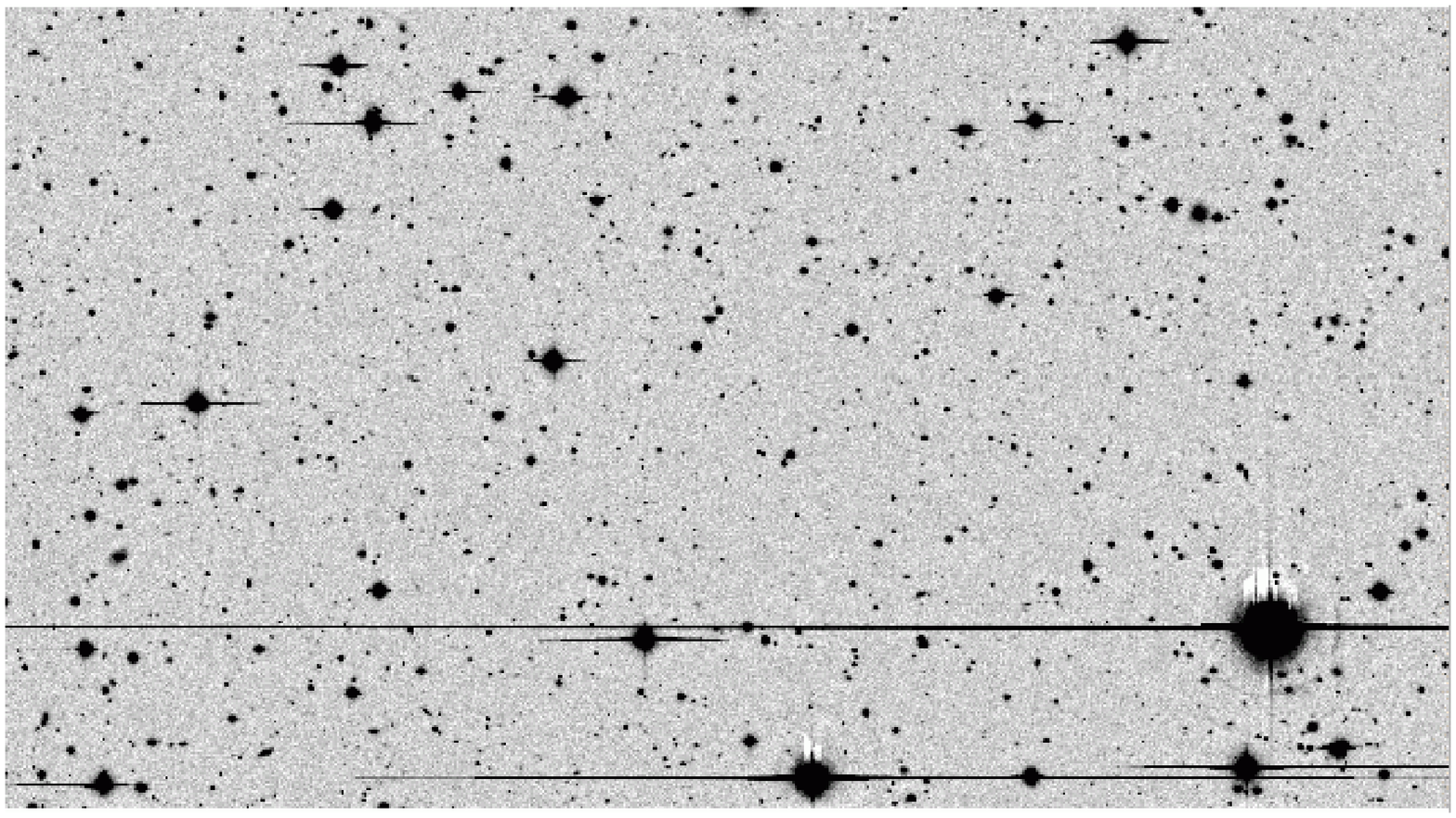, width=6.5cm, angle=-90}
\end{center}
\vspace{0.3cm}
\epsfig{file=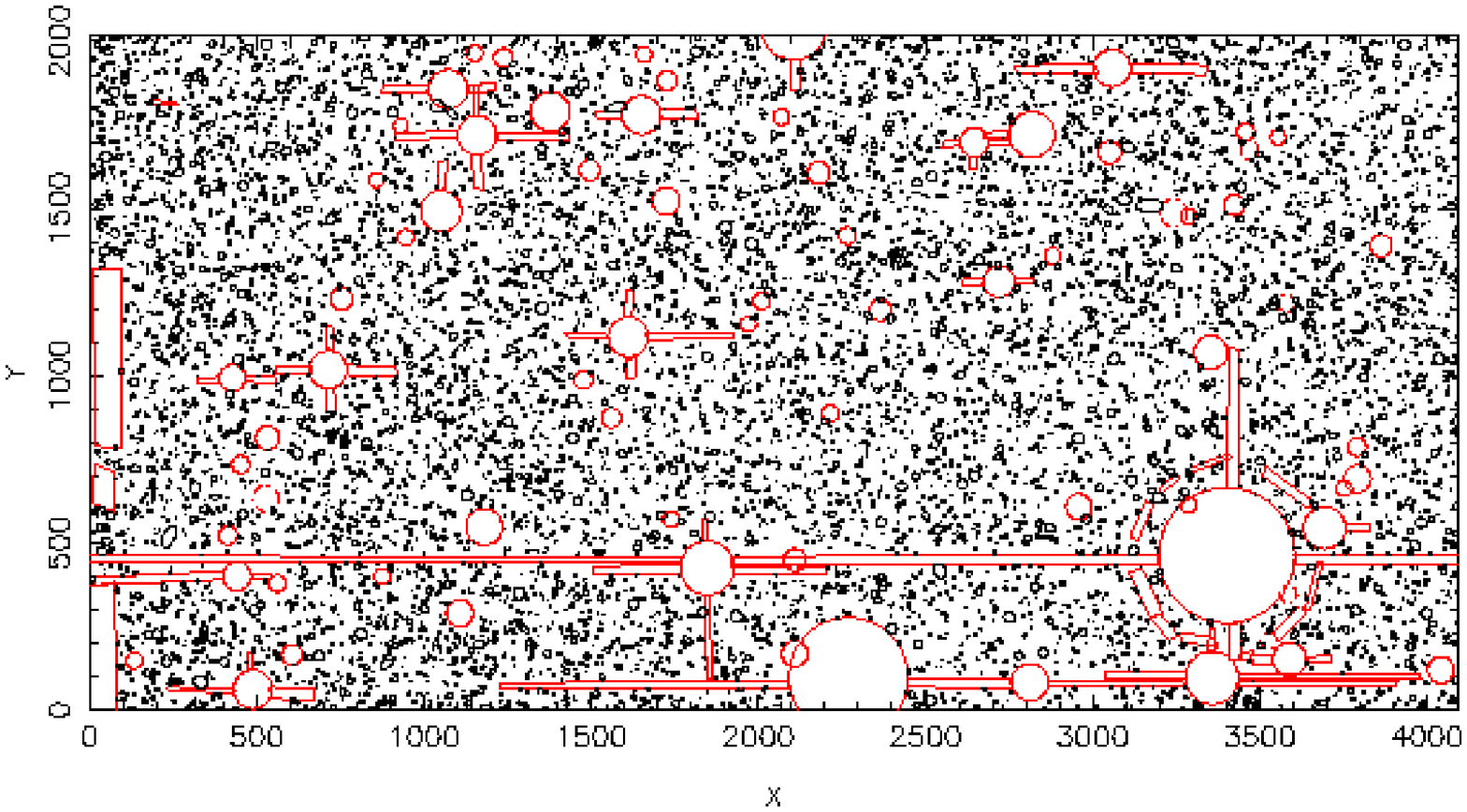, width=7.3cm, angle=-90} 
\caption{An example of the reduced data (an $R$ band image in Andromeda) and 
the corresponding image mask created
manually to ensure complete removal of all image defects. The black
circles in the lower figure represent the extracted sources from the
above image, and the red shapes represent the masking out of the
bright stars, diffraction spikes and spurious detections appearing at
the edge of the chip.}
\label{fig:holes}
\end{figure*}

\subsection{Star/Galaxy Separation}
\label{sec:stargal}

During the source extraction, SExtractor classifies each object as a
star or galaxy reflected in the value of the {\it stellarity index}
parameter.  A trained neural network was employed to determine the
stellarity index for each object dependent on the seeing, peak
intensity and a measurement of the isophotal area (see \scite{sex} for
details). The seeing for each pointing was obtained by taking the
median FWHM of all bright, unsaturated objects in the image,
identified from their profiles as stars. The stellarity index has a
value between 0 and 1, where 0 represents a galaxy, 1 a star and
intermediate values, by design, give an indication of the uncertainty
of the classification. \scite{sex} claim an algorithm success rate of
$\approx 95 \%$ to $R$ $\approx$ 22 when the seeing is $\approx 0.9^{\prime\prime}$,
however the seeing varies quite dramatically across the ODTS fields.
Figure~\ref{fig:star} depicts the behaviour of the stellarity index as a function
of magnitude for each band in the Andromeda field. As expected, at faint
magnitudes the stellarity index tends towards 0.5 as the distinction
between stars and galaxies becomes less pronounced due to object
profiles becoming seeing dominated. At very bright magnitudes, the
index tends to drop due to saturation effects. From
figure~\ref{fig:star}, it is apparent that the classifier begins to break down at magnitudes of $B > 22.5 $, $V > 22$, $R
> 21.5 $  and $I > 21.5 $. However, it should be noted that the depths to which the classifier is successful is a strong function of seeing and therefore varies between pointings. All objects with a stellarity index $ > $ 0.9
were considered to be stars. 

\begin{figure}
\begin{center}
\epsfig{file=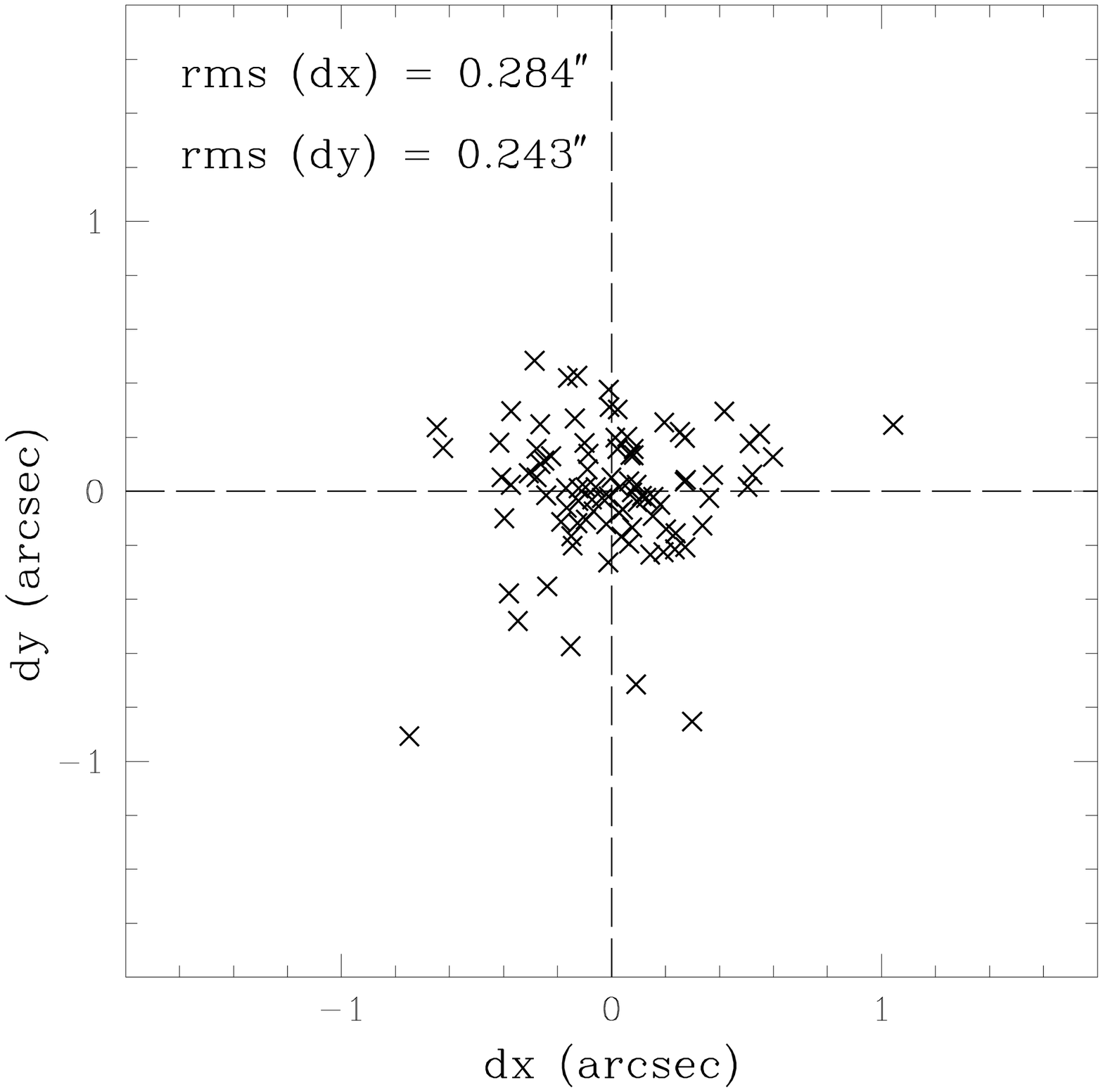, width=7.cm, angle=0}
\end{center}
\caption{Scatter diagram of the astrometric differences in arc seconds
between the matched ODT and USNO objects for an example field. The rms differences are small.
The pixel size is $0.33^{\prime\prime}$.}
\label{fig:astromscatter}
\end{figure}

\begin{figure}
\begin{center}
\vspace{-0.3cm} 
\epsfig{file=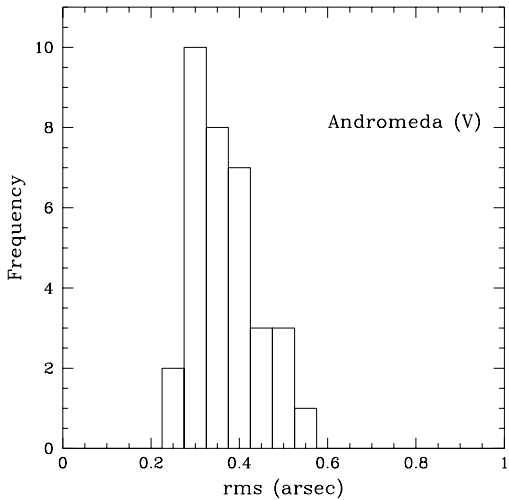, width=7.cm, angle=0}
\end{center}
\caption{Histogram of the rms astrometric residuals between the ODTS
  and USNO data found for each individual CCD frame after the final
  iteration of the astrometric matching procedure for the $V$ band
  Andromeda data.
}
\label{fig:astromhist}
\end{figure}

\begin{figure}
\begin{center}
\vspace{0.2cm} 
\epsfig{file=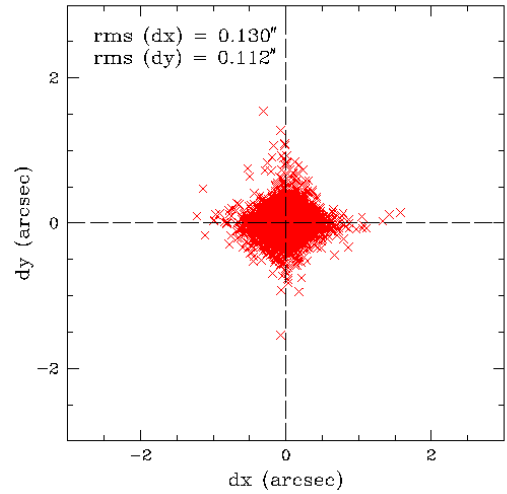, width=7.cm, angle=0}
\end{center}
\caption{Scatter diagram of the astrometric differences in arc seconds
between the matched objects in all the overlap regions of the reduced $V$ band Andromeda data. Objects with $V < 24.5$ have been used so as to avoid using data from overlapping frames which reach different depths.}
\label{fig:astromall}
\end{figure}

\begin{figure}
\begin{center}
\vspace{-0.52cm}
\epsfig{file=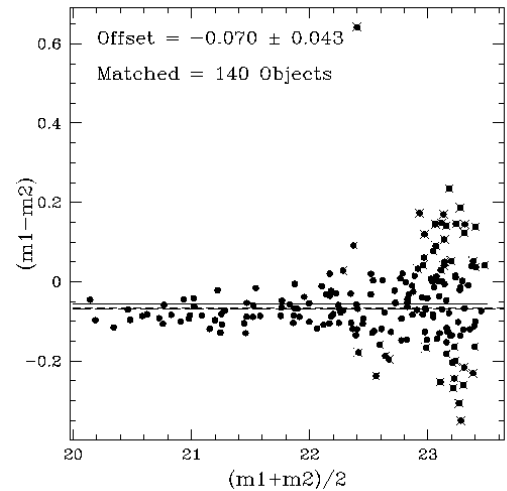, width=7.cm, angle=0}
\end{center}
\caption{The $V$ band magnitude (photometric) difference between objects in
  the overlap region of two adjacent frames plotted
  against the mean magnitude, where each point represents a matched
  pair of objects. The three lines represent the value of the mean
  determined after the first (solid line), second (short broken line)
  and third (long broken line) iterations of linear fitting to the
  data, where points deviating from the mean by $> 1.5\sigma$ are
  rejected after each fit (crossed out objects). The magnitude dependence of the photometric errors determined during source extraction is obvious.}
\label{fig:overlap} 
\end{figure}

\subsection{Image Masking}
\label{sec:holes}

False detections caused by the presence of asteroid and satellite
trails, excessive vignetting, low-level fringing, diffraction spikes,
and halos around bright stars had to be removed from,
or flagged in, the final catalogues. For each image a mask was created
manually which identified all the contaminated image areas. The method
employed consisted of drawing either rectangular or circular shaped
holes around the spurious structures in the data frame, where objects
appearing within the holes were then flagged appropriately, allowing
for subsequent rejection. An example of an image and its mask can be
seen in figure~\ref{fig:holes}.

The effective area of the survey was consequently reduced and the
final areas for each band in each field after masking are shown in
table~\ref{tab:effareas}.

\section{Astrometry}
\label{sec:astrom}

The conversion of the source pixel position to right ascension $\alpha$ and
declination $\delta$ was then carried out. The four
CCDs of the WFC maintain a fixed geometrical pattern relative to the
camera rotator centre however, the prime focus corrector of the INT
introduces a cubic radial distortion term to the plate scale of the
form 

\begin{equation}
r_{\rm true} = r + k r^3,
\end{equation}

\noindent where $r_{true}$ is the actual radial distance in radians
from the field centre, $r$ is the measured radial distance and $k$ is a
constant. For the WFC, $k$ was measured to be 220.0 radians$^{-2}$ 
(Irwin, private communication).

The astrometry is split into two stages. Initially, each individual
CCD frame is calibrated independently by roughly matching objects in
the images with objects in the corresponding Digitised Sky Survey
images \cite{Lasker}. Matched objects are then used to converge upon
an astrometric fit for each frame via an iterative process. This is
performed using the ASTROM package which relates the measured $x,y$ to
the true $\alpha,\delta$ coordinates by fitting a six coefficient
transformation, which includes orientation of the images, plate scale
and radial distortions. These fits gave rms residuals of the
order of $1^{\prime\prime}$ for each frame. The large residuals were
due to uncertainties in the exact position of the corrector axis
relative to the field rotation centre on the
sky.

 To compensate for this, the four CCD frames from each pointing were
considered as a whole and were matched to the more accurate data of
the United States Naval Observatory (USNO) A2.0 
astrometric catalogue
\cite{usno}. During this matching process, the field centre was
allowed to shift using the transformations derived in the initial
stage, until the astrometric differences between the ODT and USNO
catalogues (see figure~\ref{fig:astromscatter}) were minimised, thus
ascertaining a more accurate estimate of the true field centre. The
final residuals for a single CCD frame were found to be typically
$\sim 0.3^{\prime\prime}$ where figure~\ref{fig:astromhist} shows an
example of the rms residuals for the $V$ band astrometry.

\section{Photometric Calibration}
\label{sec:photocalib}

\begin{figure*}
\centerline{\psfig{file=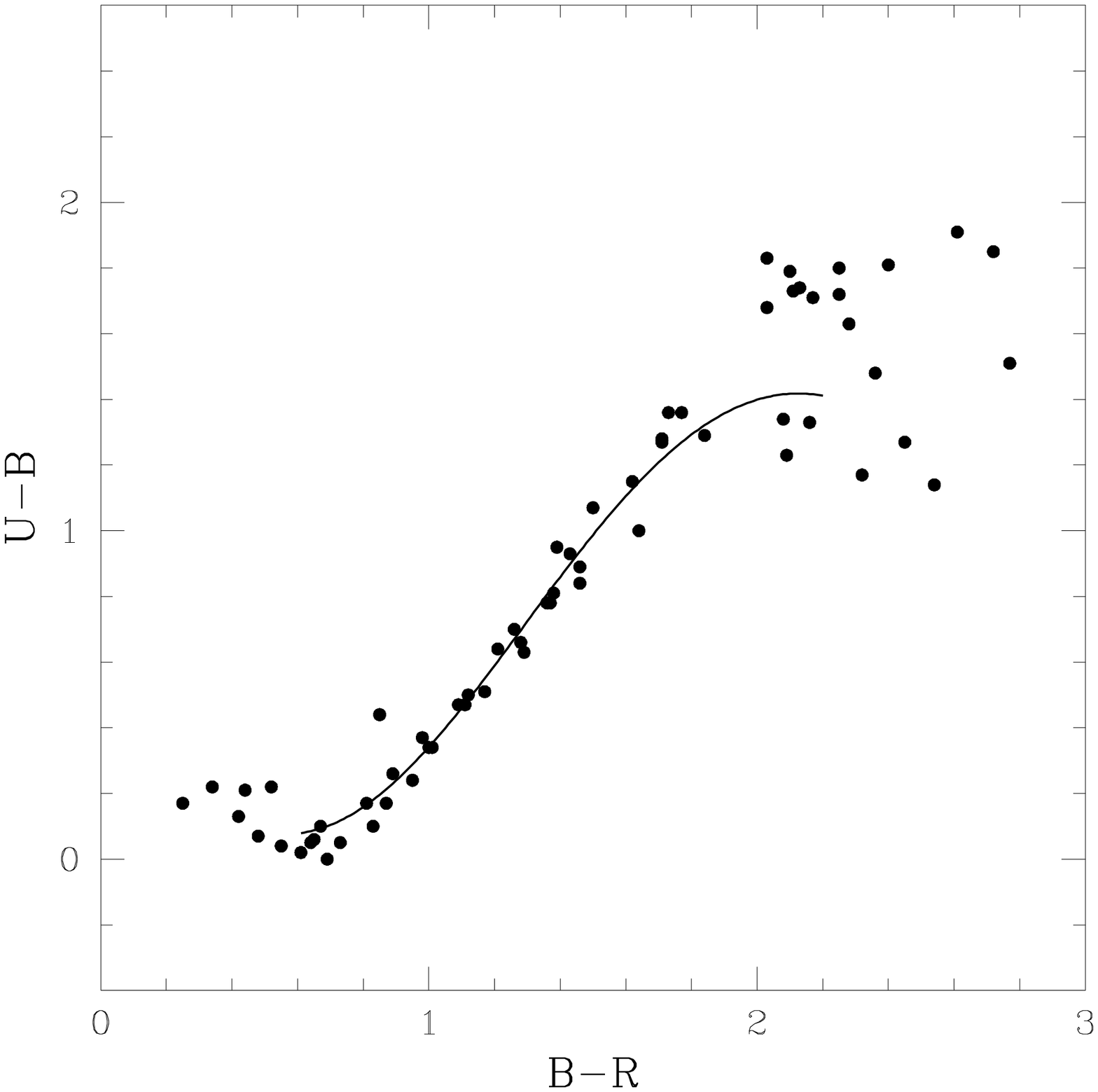,width=5.7cm,angle=0,clip=}\hspace{0.2cm}\psfig{file=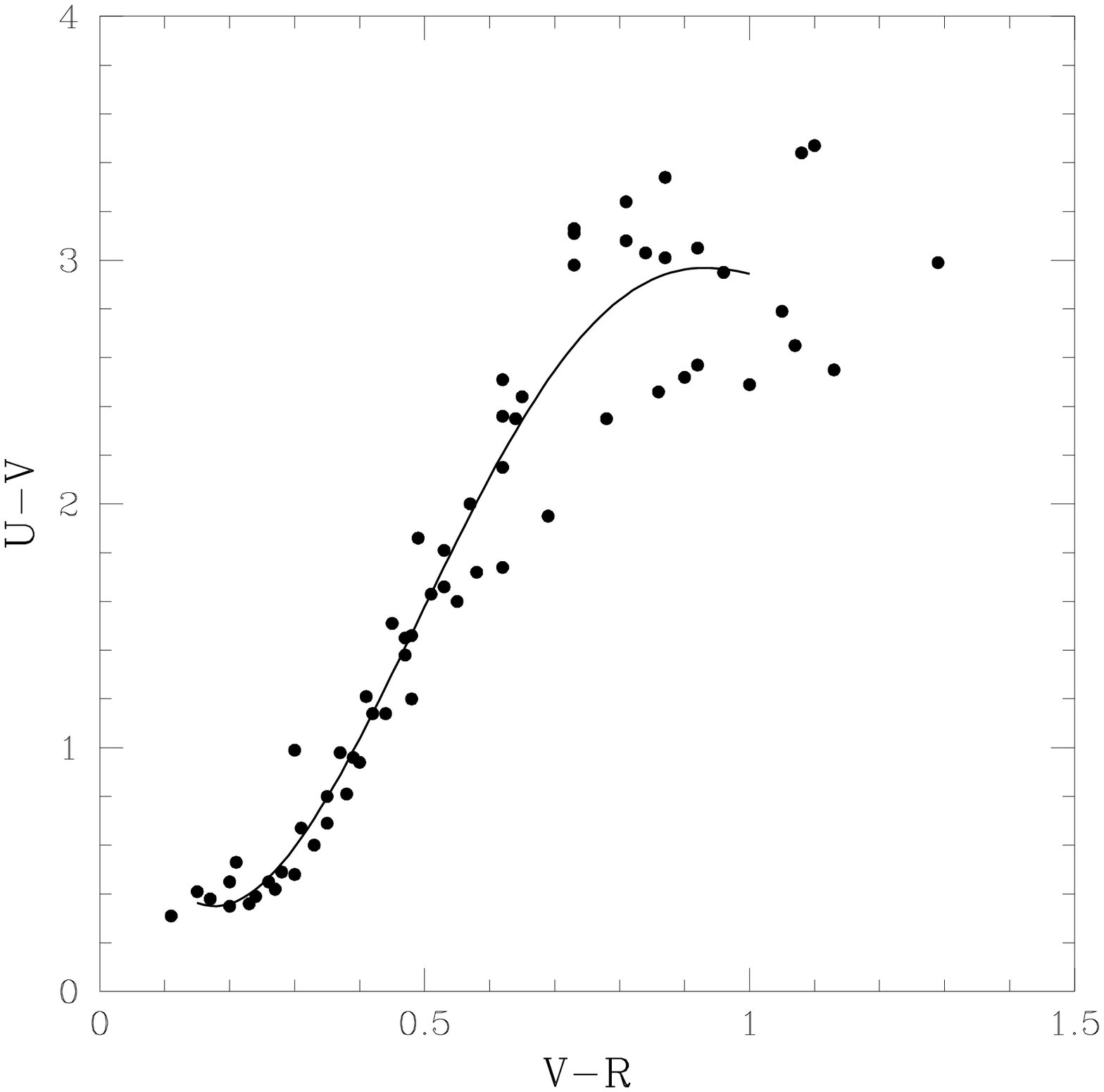,width=5.7cm,angle=0,clip=}\hspace{0.2cm}\psfig{file=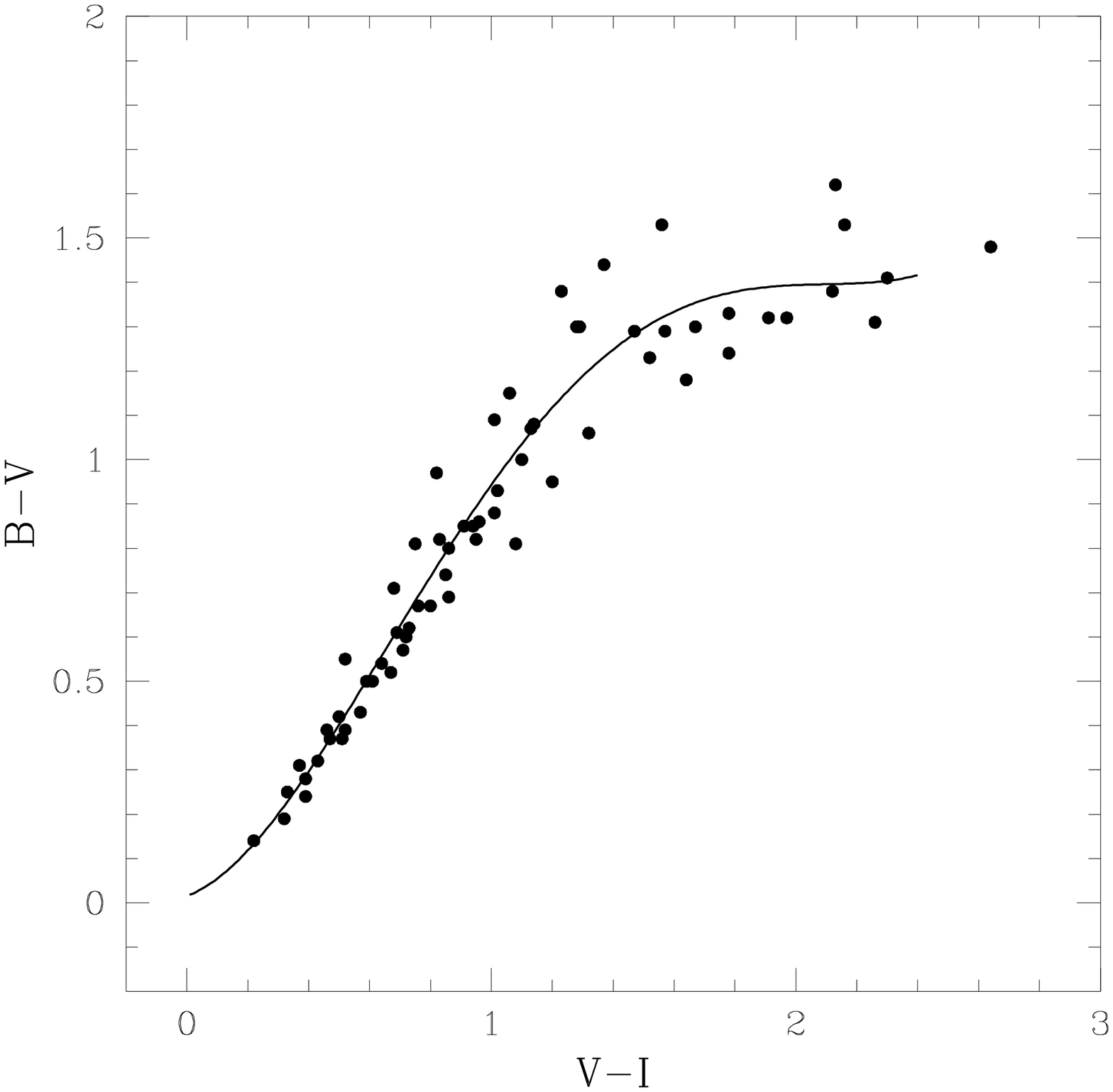,width=5.7cm,angle=0,clip=}}
\vspace{0.5cm}
\centerline{\psfig{file=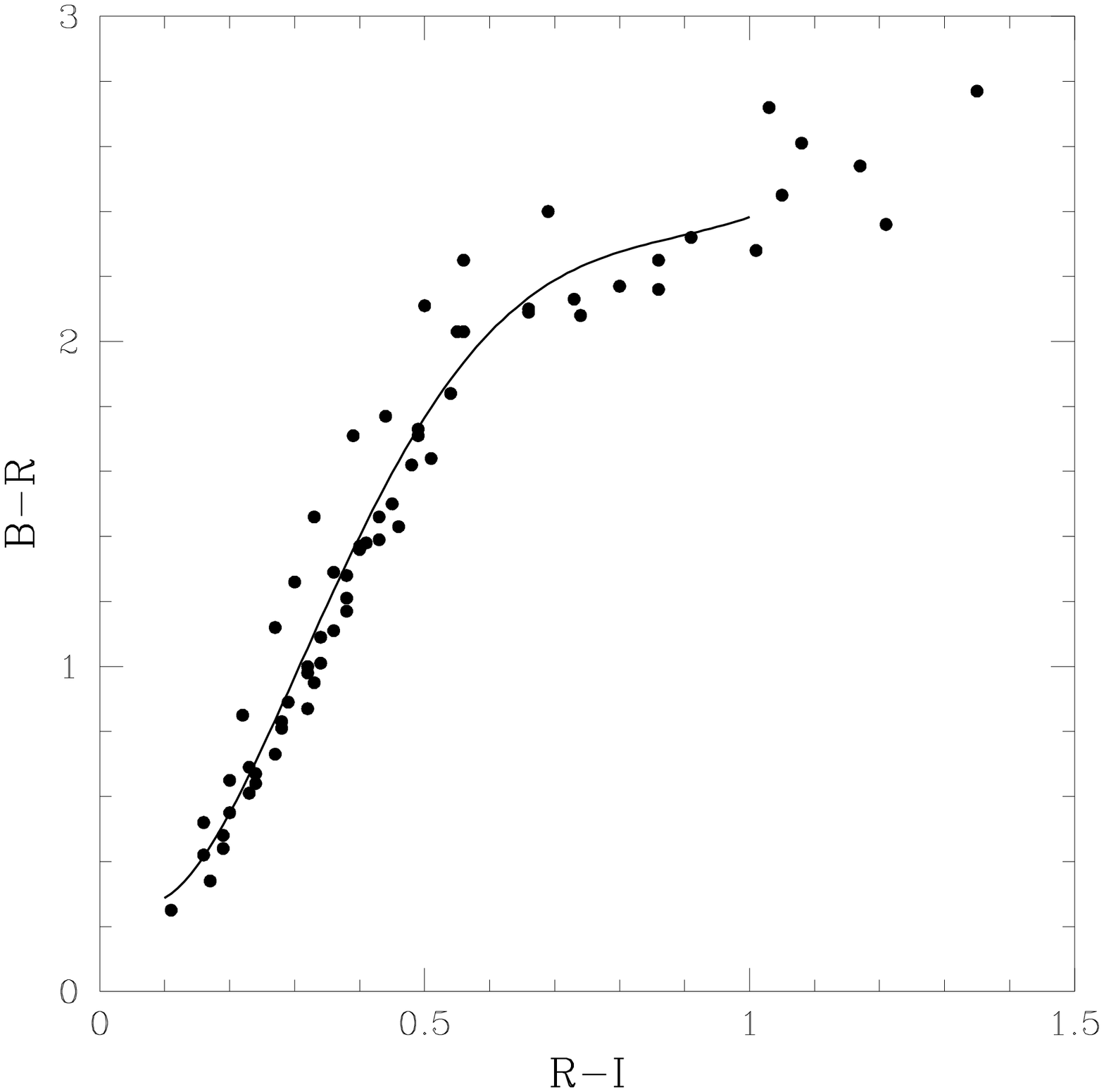,width=5.7cm,angle=0,clip=}\hspace{0.2cm}\psfig{file=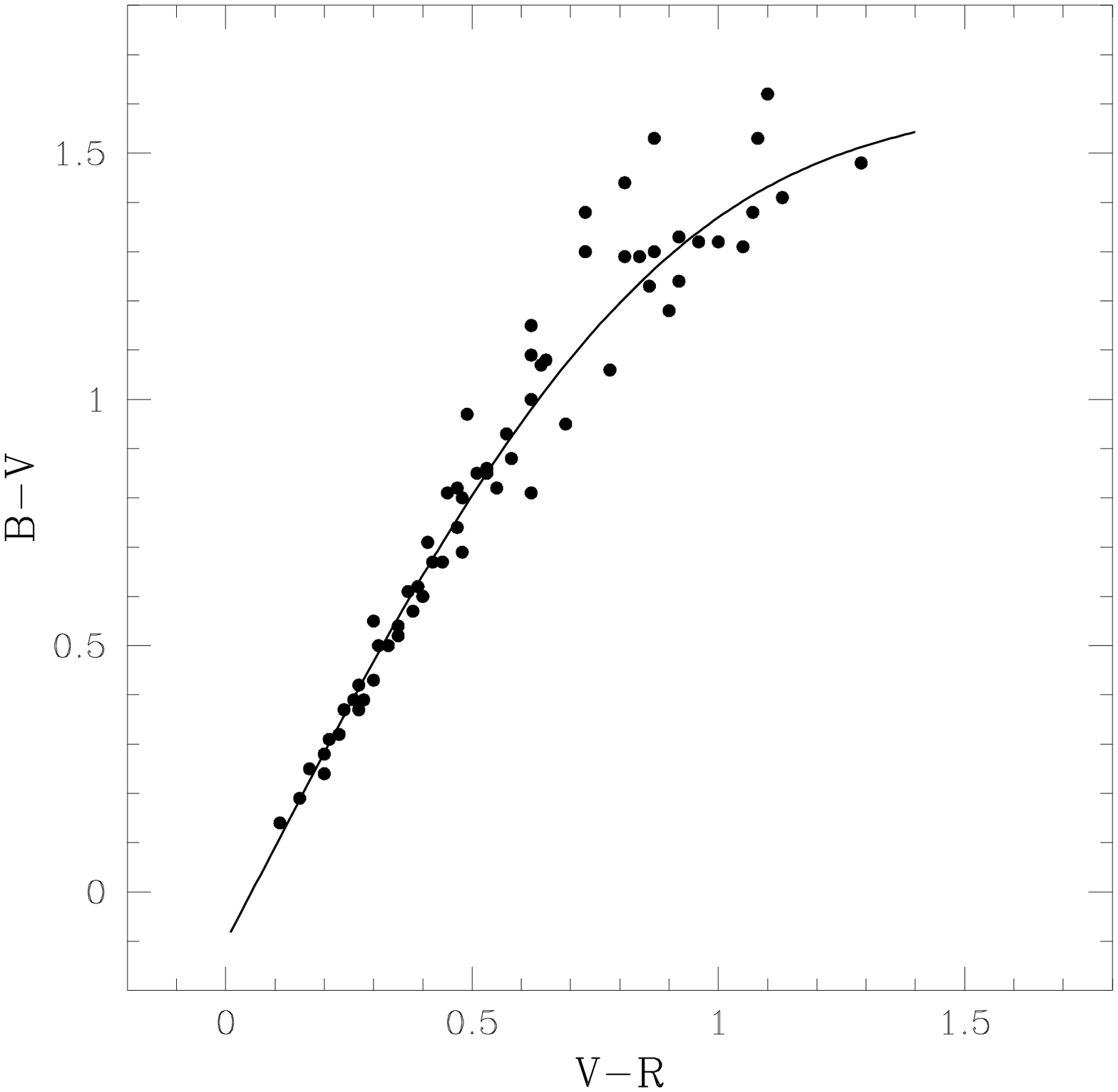,width=5.7cm,angle=0,clip=}\hspace{0.2cm}\psfig{file=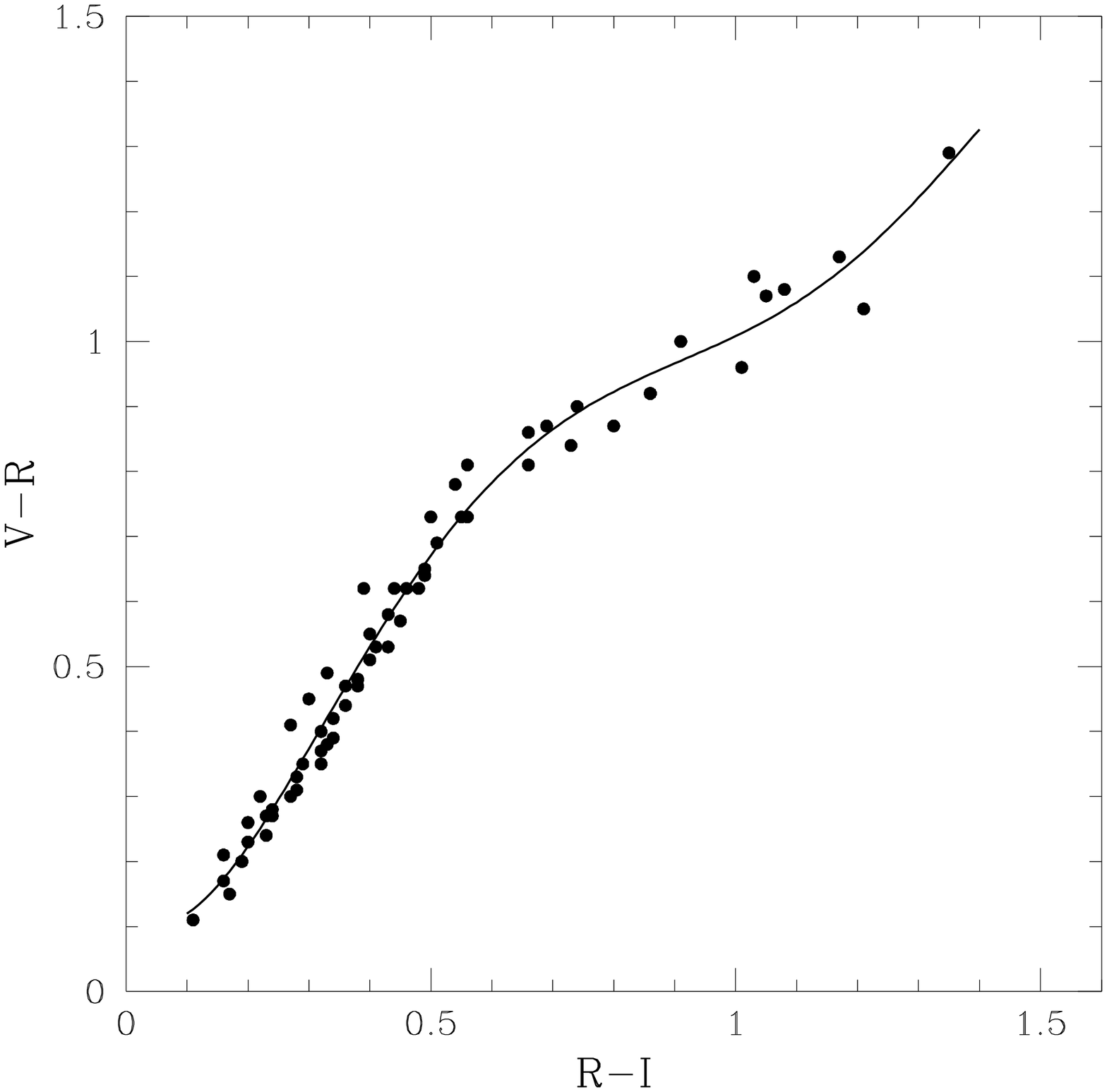,width=5.7cm,angle=0,clip=}}
\caption{The best fit polynomials (black lines) used in the photometric calibration of the ODTS data. Filled red circles represent the colours of Pickles (1998) stellar data in the ODTS filter system, as seen in six different colour-colour planes.}
\label{fig:poly} 
\end{figure*}

\begin{figure*}
\begin{center}
\centerline{\psfig{file=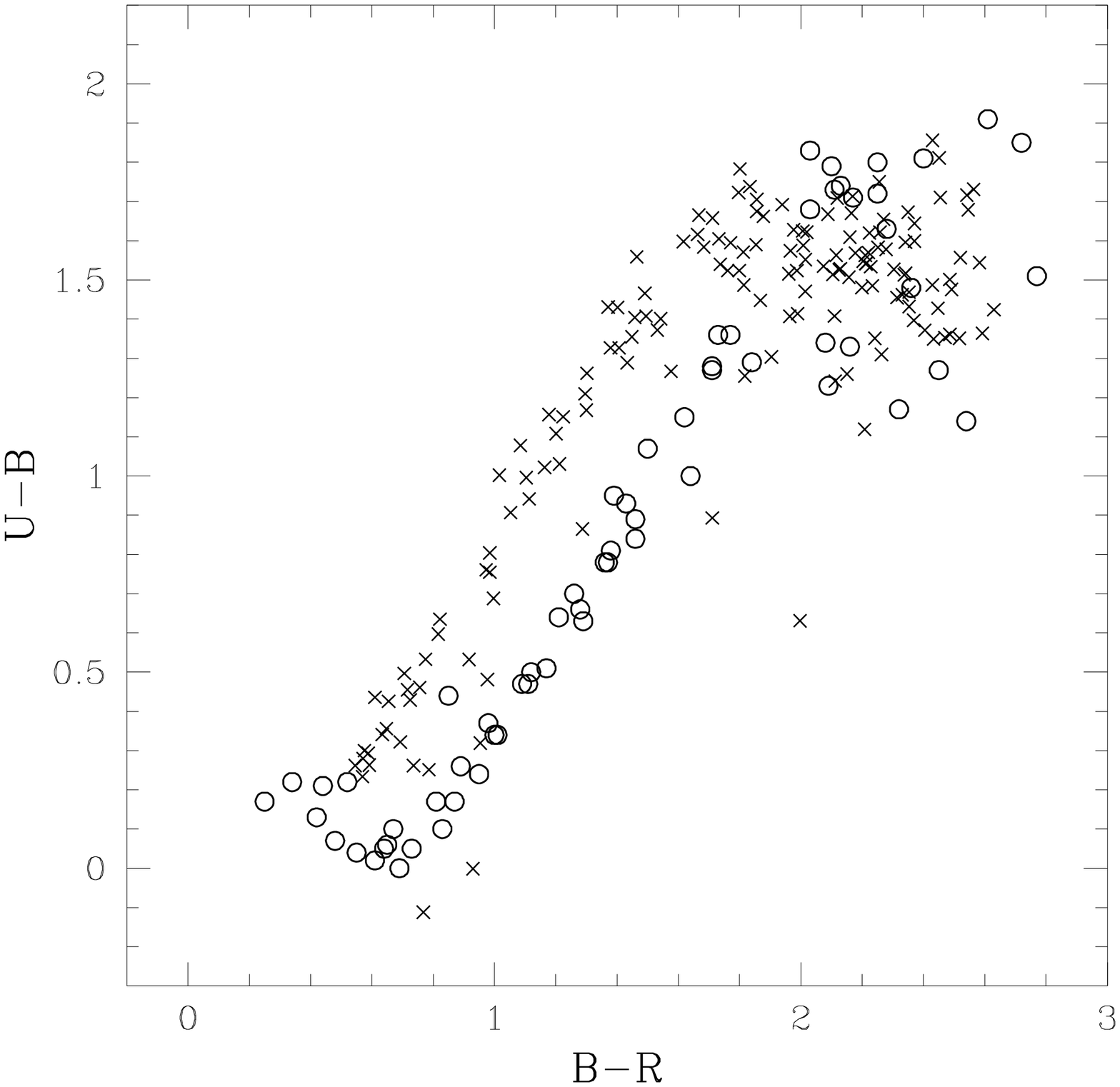,width=5.7cm,angle=0,clip=}\hspace{0.2cm}\psfig{file=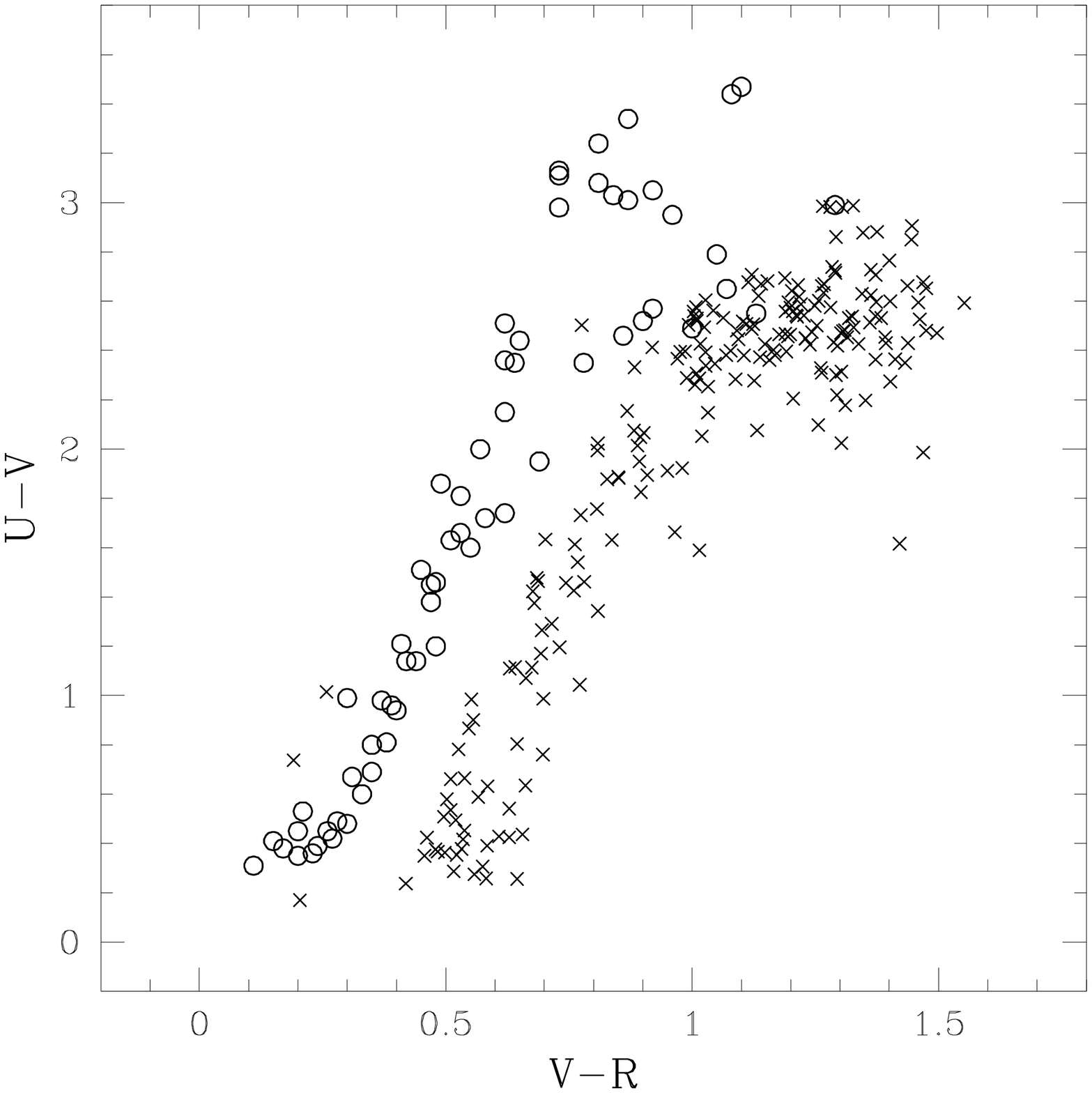,width=5.7cm,angle=0,clip=}\hspace{0.2cm}\psfig{file=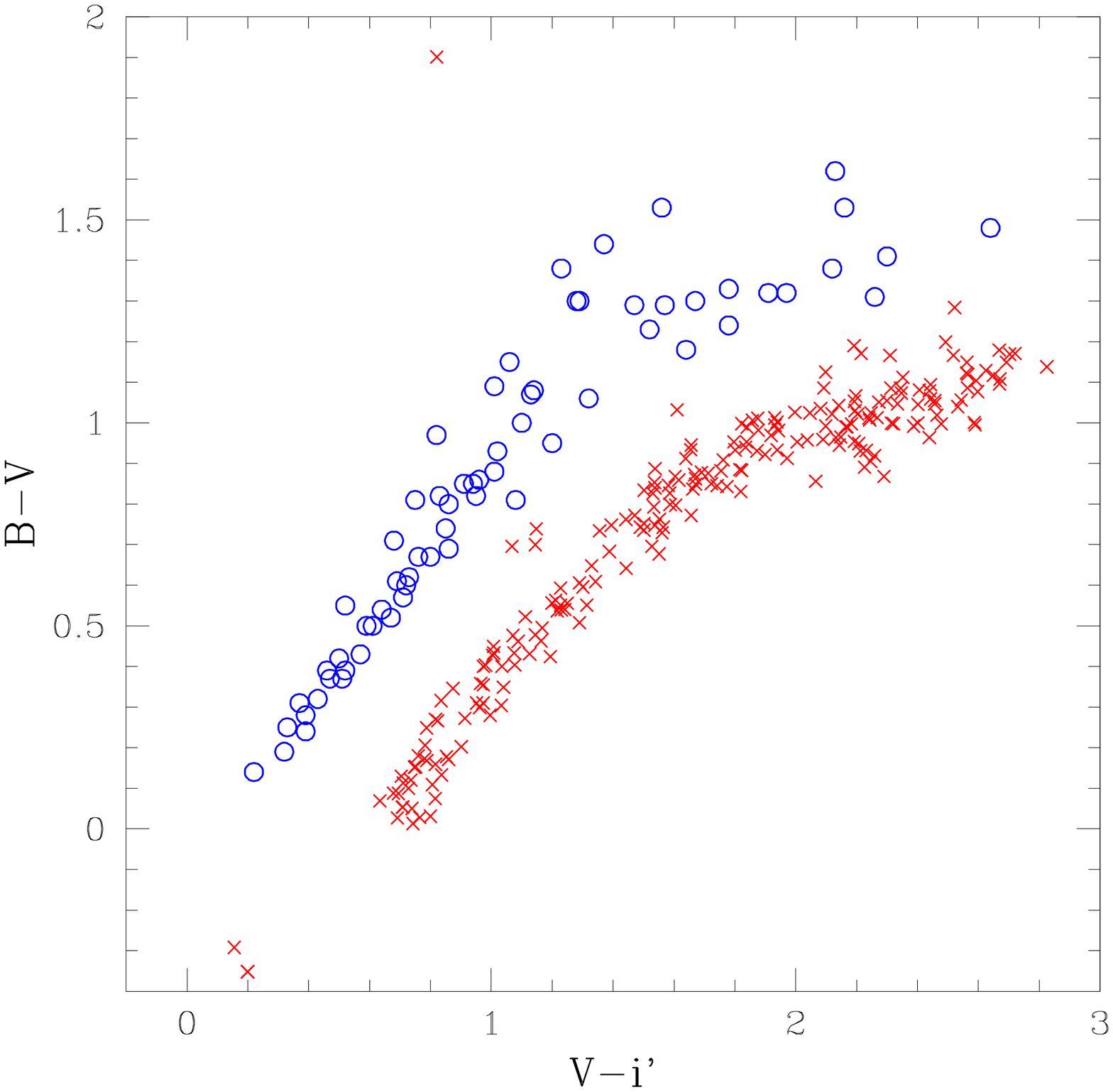,width=5.7cm,angle=0,clip=}}
\vspace{0.5cm}
\centerline{\psfig{file=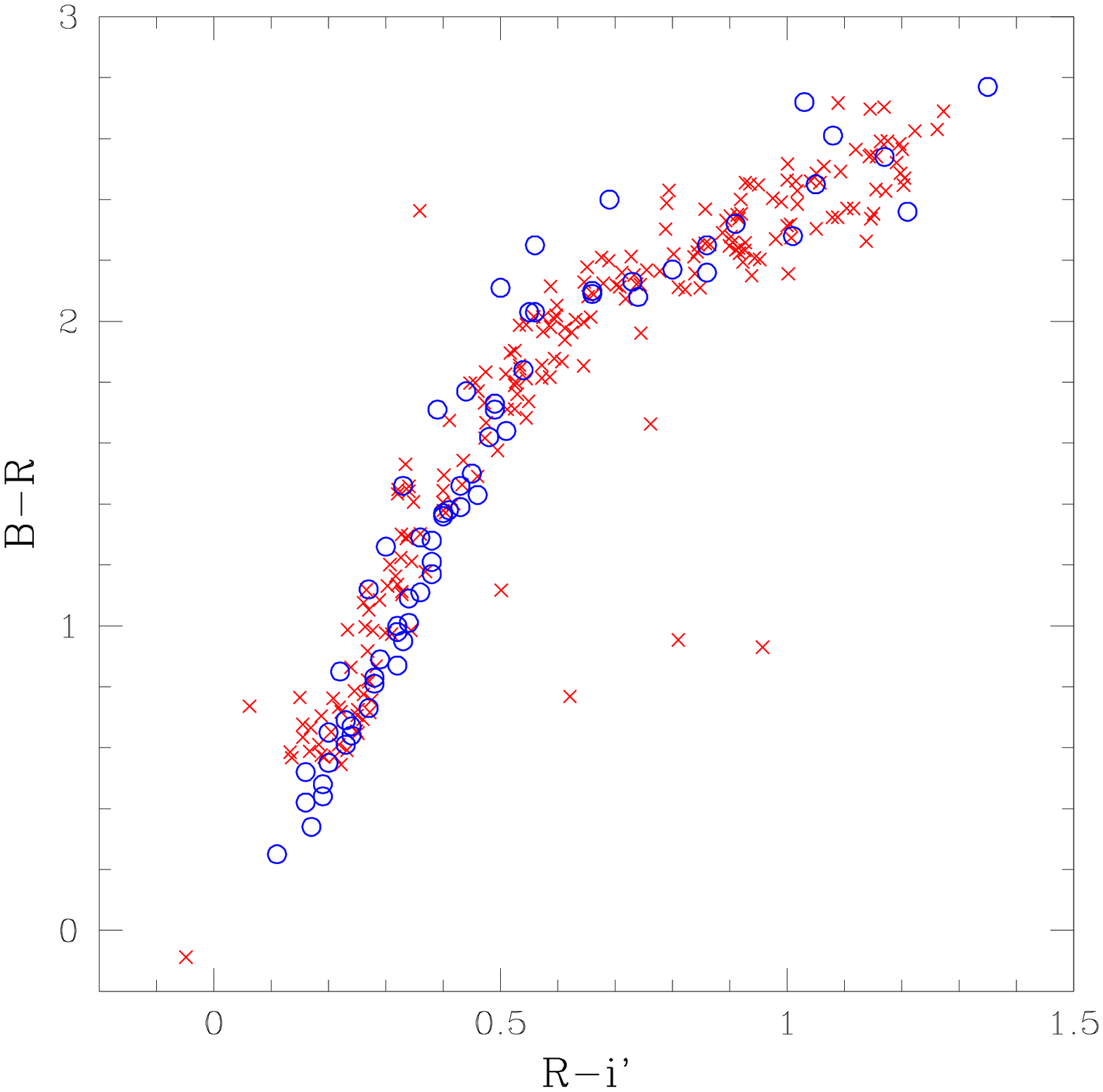,width=5.7cm,angle=0,clip=}\hspace{0.2cm}\psfig{file=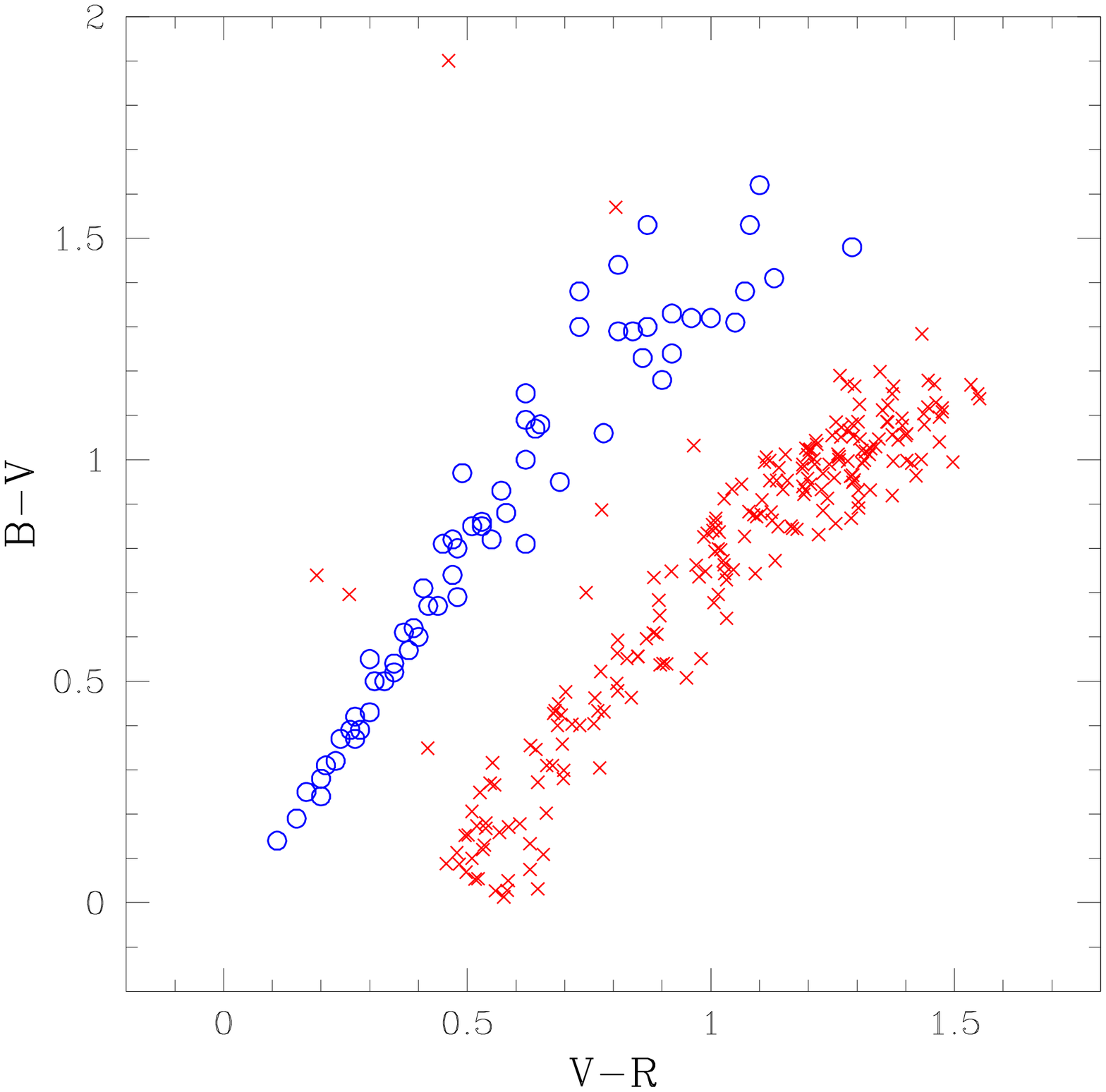,width=5.7cm,angle=0,clip=}\hspace{0.2cm}\psfig{file=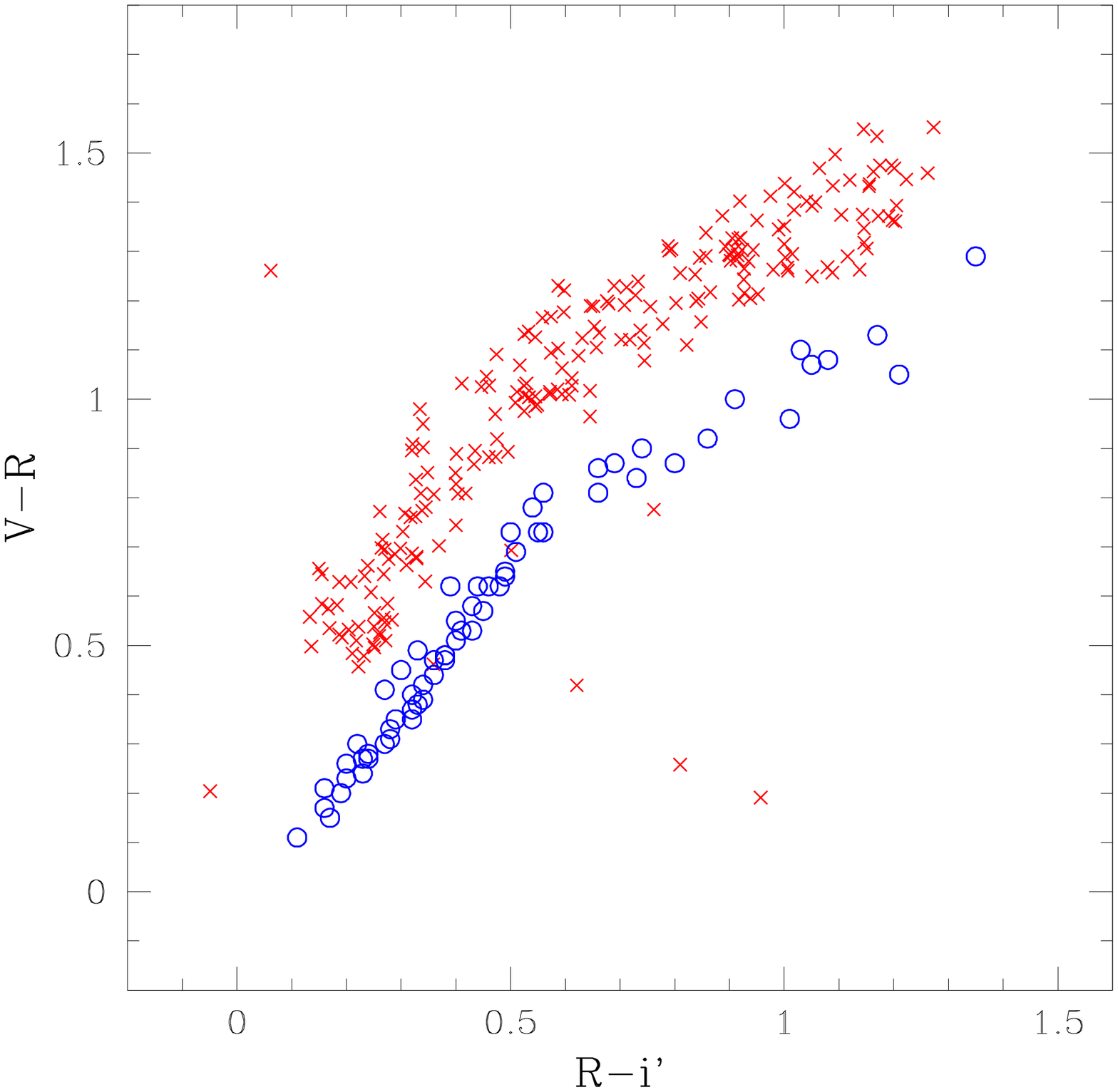,width=5.7cm,angle=0,clip=}}
\end{center}
\caption{Stellar colours with {\bf uncorrected} photometric zeropoints for an example frame. The open blue circles represent the colours of Pickles (1998) stellar data in the ODTS filter system and the red crosses show the ODTS stellar data. Large initial offsets were expected due to the differing observing conditions.}
\label{fig:before} 
\end{figure*}

\begin{figure*}
\begin{center}
\centerline{\epsfig{file=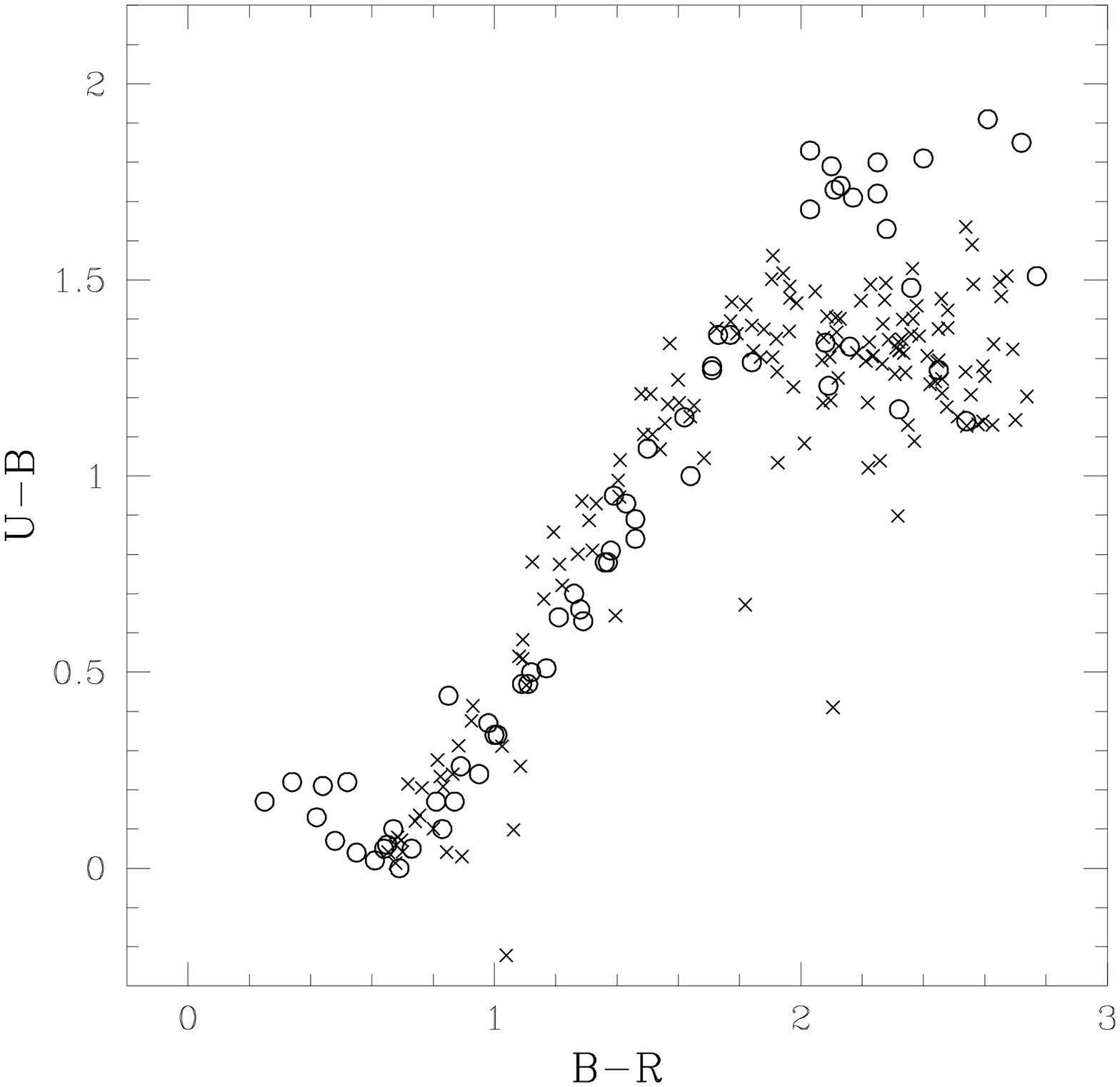,width=5.7cm,angle=0,clip=}\hspace{0.2cm}\epsfig{file=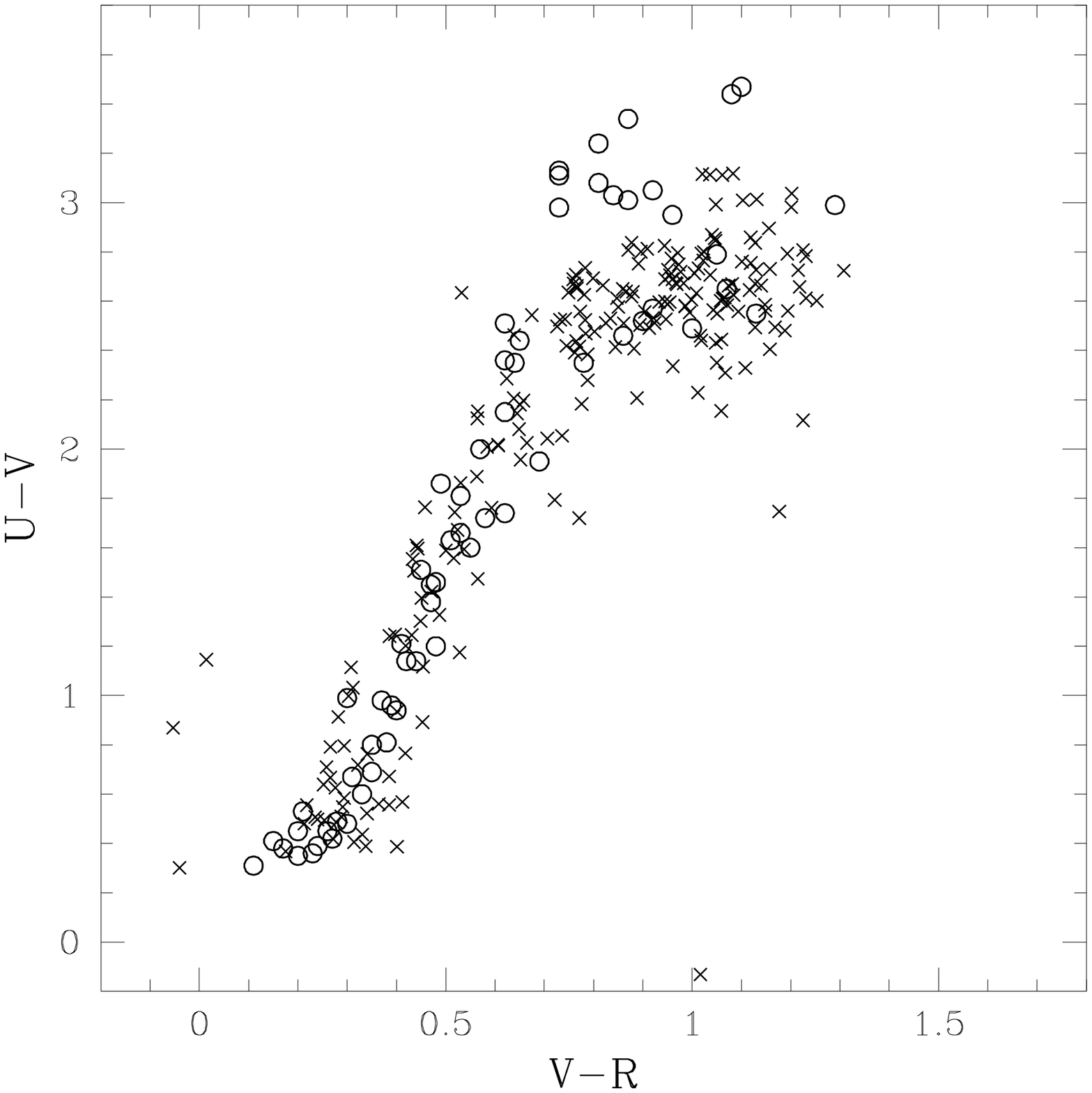,width=5.7cm,angle=0,clip=}\hspace{0.2cm}\epsfig{file=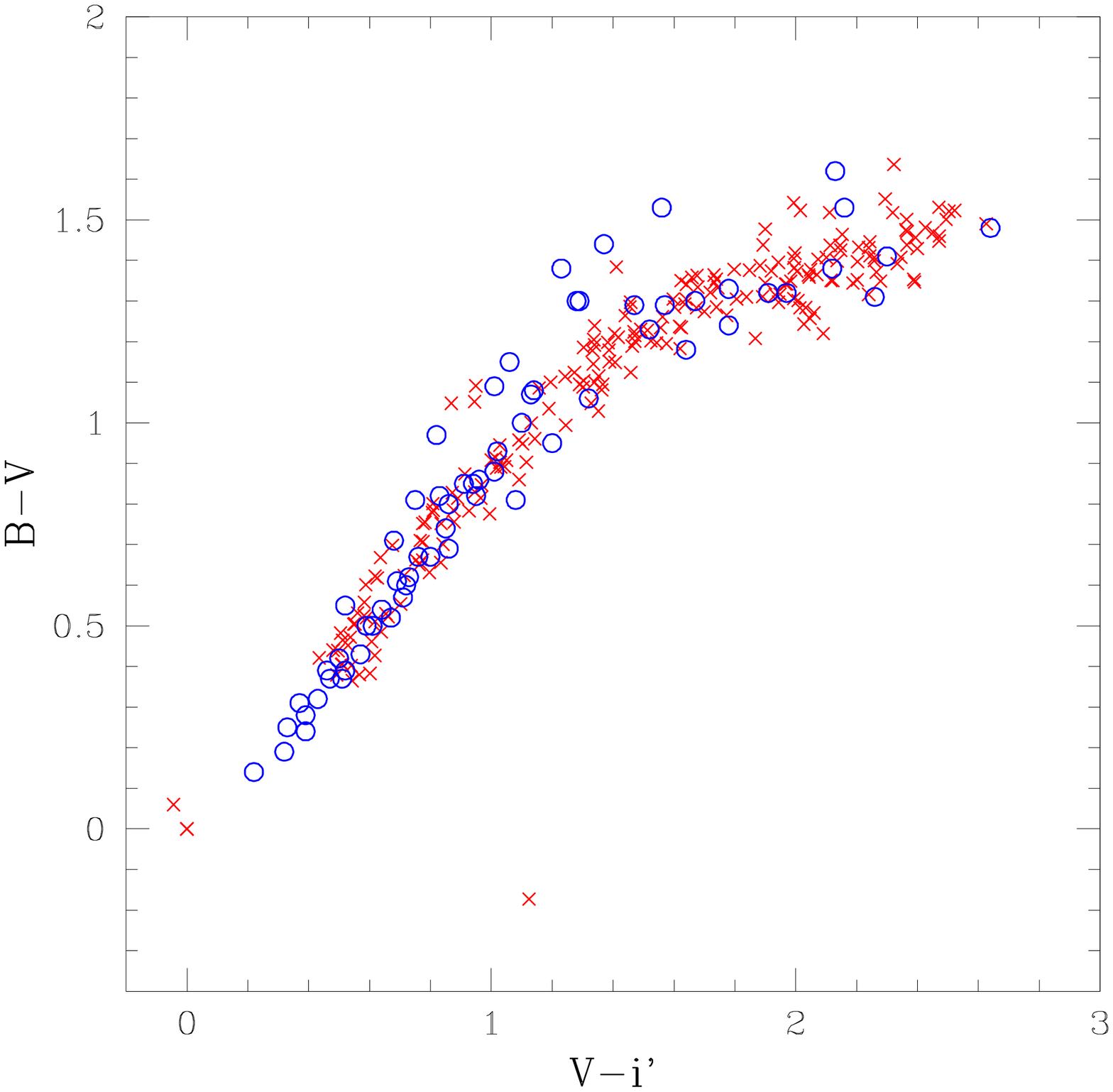,width=5.7cm,angle=0,clip=}}
\vspace{0.5cm}
\centerline{\epsfig{file=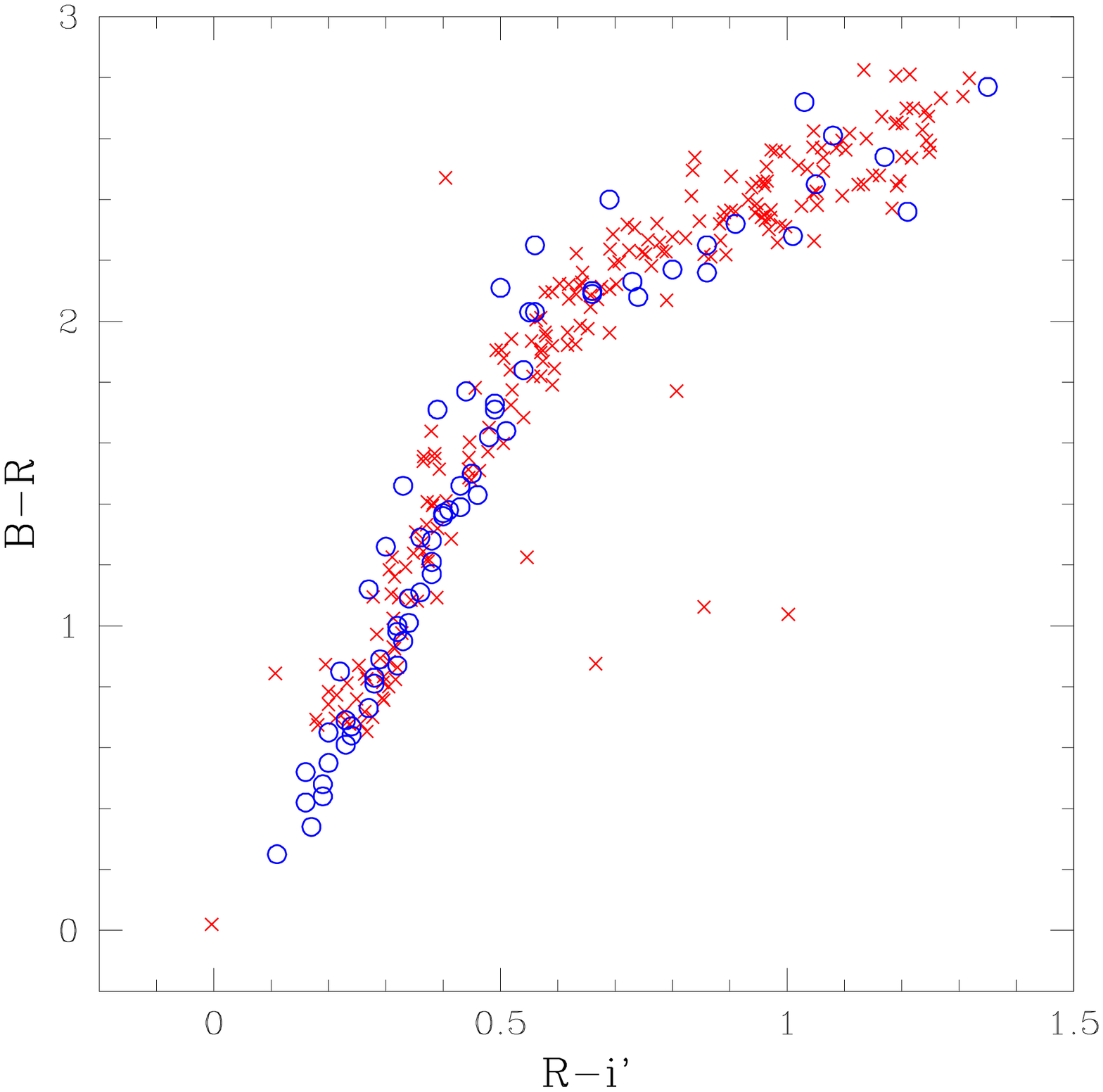,width=5.7cm,angle=0,clip=}\hspace{0.2cm}\epsfig{file=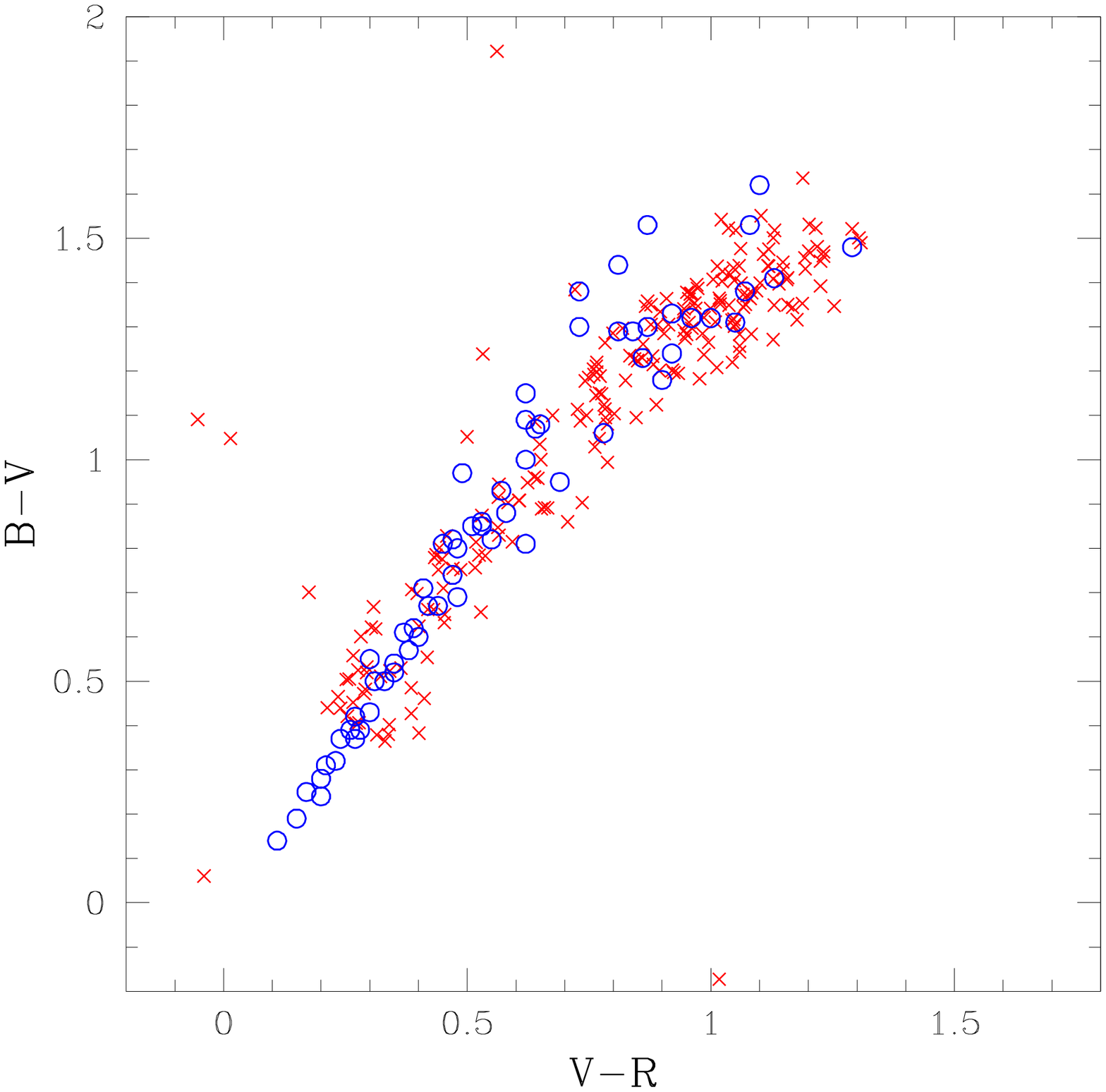,width=5.7cm,angle=0,clip=}\hspace{0.2cm}\psfig{file=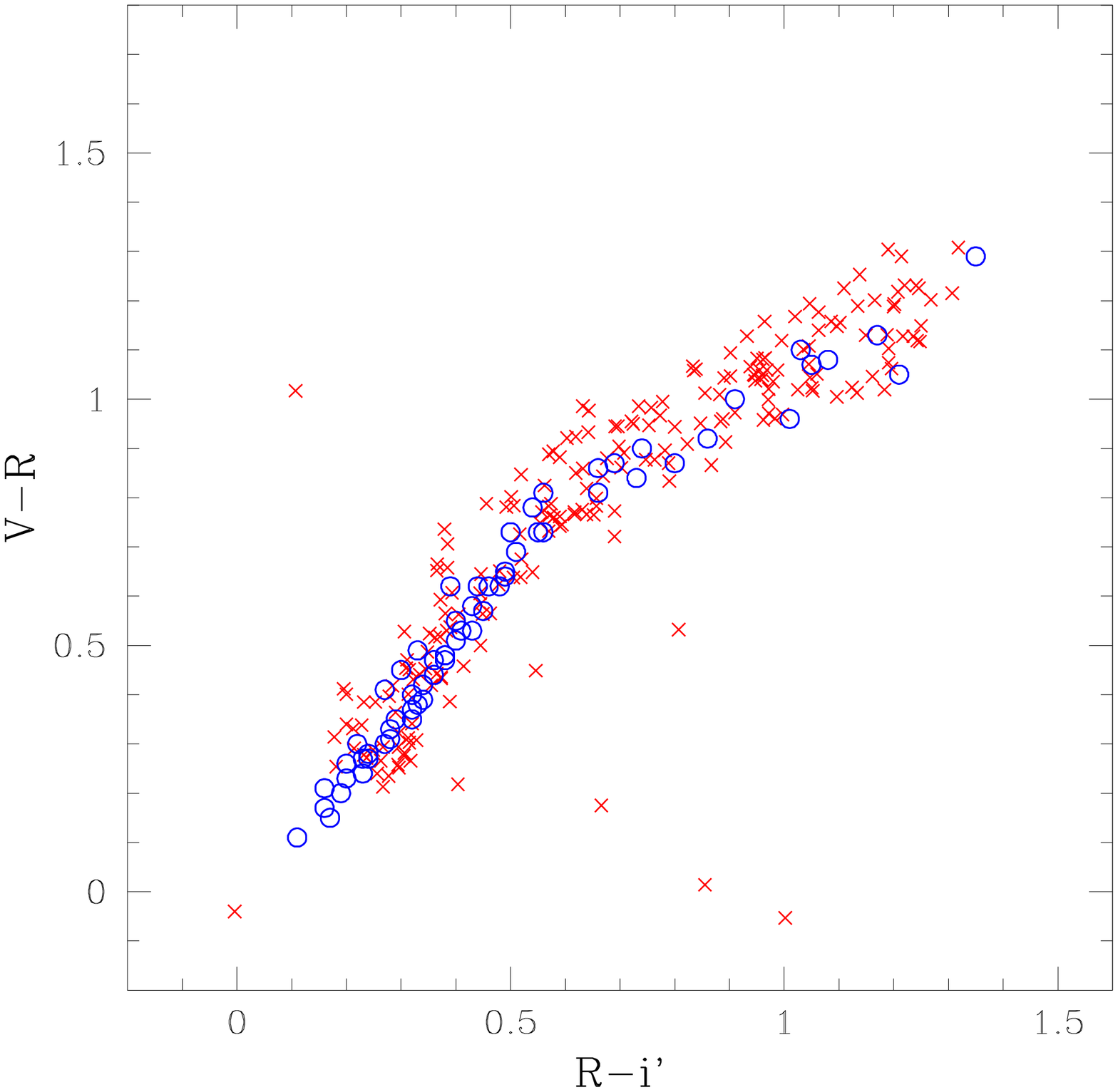,width=5.7cm,angle=0,clip=}}
\end{center}
\vspace{-0.3cm}
\caption{Stellar colours with {\bf corrected} photometric zeropoints for an example frame. The open blue circles represent the colours of Pickles (1998) stellar data in the ODTS filter system and the red crosses show the ODTS stellar data.}
\label{fig:after} 
\end{figure*}

Producing a final catalogue with consistent photometry throughout
requires each contiguous field to have a common zeropoint. At this
stage, the individual zeropoints derived in section~\ref{sec:photo}
will each have a residual uncertainty due to differing observing
conditions and airmass and extinction variations between
observations. Many observations were taken in non-photometric conditions. In order to achieve homogeneity in the photometric
calibration, the zeropoints of the individual frames must be adjusted
relative to a common zeropoint adopted for the whole survey. This was
carried out in two stages, where the first involved using the $V$ band
as the calibrator band (see section~\ref{sec:overlap}) and considering
the magnitude differences between common objects in the overlap
regions of the $V$ band pointings. The photometric zeropoints could then be
corrected relative to a chosen calibrator frame, effectively creating
a common zeropoint across the $V$ band data. Following this, the
zeropoints of all the other bands were corrected relative to the $V$
band via stellar locus fitting (see section~\ref{sec:stell}). This was
done by examining the colours of stars in each frame and adjusting
their zeropoints until the stars had the expected colours of the
stellar main sequence.

\subsection{Overlap Matching}
\label{sec:overlap} 

To ensure a common photometric zeropoint for the $V$ band, the approach introduced by \scite{glazebrook} was adopted whereby 
the magnitudes of objects common to  overlapping images are compared in order to 
ascertain the difference in their zeropoints. Objects in the overlap regions of adjacent fields were matched with a tolerance of $\sim 1^{\prime\prime}$, also a useful check of the astrometric accuracy. Figure~\ref{fig:astromall} shows the astrometric differences for all the matched objects in the $V$ band overlap regions, where rms residuals were found to be $< 0.16^{\prime\prime}$ (with equivalent results for the other bands). If $m_{i}$ and $m_{j}$ denote the magnitudes of the matched objects in the overlap regions of frames $i$ and
$j$, then the magnitude offset between overlapping frames, $T_{ij}$, can be determined by plotting the magnitude differences, $(m_{i}-m_{j})$, against the mean magnitudes, $(m_{i}+m_{j})/2$, for all the objects in the overlap region and making a linear fit to the data to find the average magnitude difference. For the ODTS, the mean magnitude offset was calculated by performing three iterations, after each of which objects deviating by $> 1.5\sigma$ from the mean were rejected. A typical example of this is shown in figure~\ref{fig:overlap} where 140 objects have been matched, 48 being rejected during the iterative process used to determine the mean. In this particular case, the value of $T_{ij}$ is -0.07 magnitudes with a rms of 0.04.

If frame $i$ differs from the 'true' zeropoint, $m_{0}$, so that
$m_{0}=m_{i}+C_{i}$, where $C_{i}$ is the correction factor for frame
$i$, then the zeropoint offset between individual frames can be written

\begin{equation}
T_{ij}=m_{i}-m_{j}=C_{j}-C_{i}=-T_{ji}
\label{eqn:b}
\end{equation}

For each frame there will
be up to four correction factors corresponding to each of its
immediate neighbours. In order to find the best correction, the minimisation of the following summation is required, 

\begin{equation}
S=\sum_{i=1}^{n}\sum_{j=1}^{n}w_{ij}\theta_{ij}(T_{ij}+C_{i}-C_{j})^{2}
\label{eqn:sum}
\end{equation}

where n is the number of frames, $w_{ij}$ are the weighting factors $\propto1/var(T_{ij})$
thus favouring overlaps with small variance and $\theta_{ij}$ is the overlap function given by

\begin{center}
\[\theta_{ij}=\left\{\begin{array}{ll}
1 & \mbox{if i and j overlap} \\
0 & \mbox{if i and j don't overlap} \\
1 & \mbox{if i=j}
\end{array}  \right. \]
\end{center}

Of the $n$ frames involved, $(1....m)$ are, by design, uncalibrated
and $(m+1,....n)$ are calibrated. For the ODTS, only one frame in the
central region of the field was adopted as the calibration frame based
on its position, good seeing, photometric conditions during
observation, and galaxy number counts.
The systematic errors in the plate
to plate variations should be (and were) $\leq 0.05$ magnitudes to avoid introducing artificial large-scale structure \cite{geller}.

Initially, overlap matching to find a common photometric zeropoint
was carried out for all bands and it was the $V$ band that was found
to possess the smallest scatter in zeropoint corrections. This,
combined with the $V$ band having the largest coverage, being less
likely than $B$ to be affected by variations in both Galactic and
atmospheric extinction and being free from fringing effects present in
$R$ and $i^{\prime}$ data justifies the choice of $V$ as the calibrator band.

\begin{figure*}  
\begin{center}
\leavevmode      
\begin{tabular} {cl}    

\includegraphics[scale=0.58,angle=0.0]{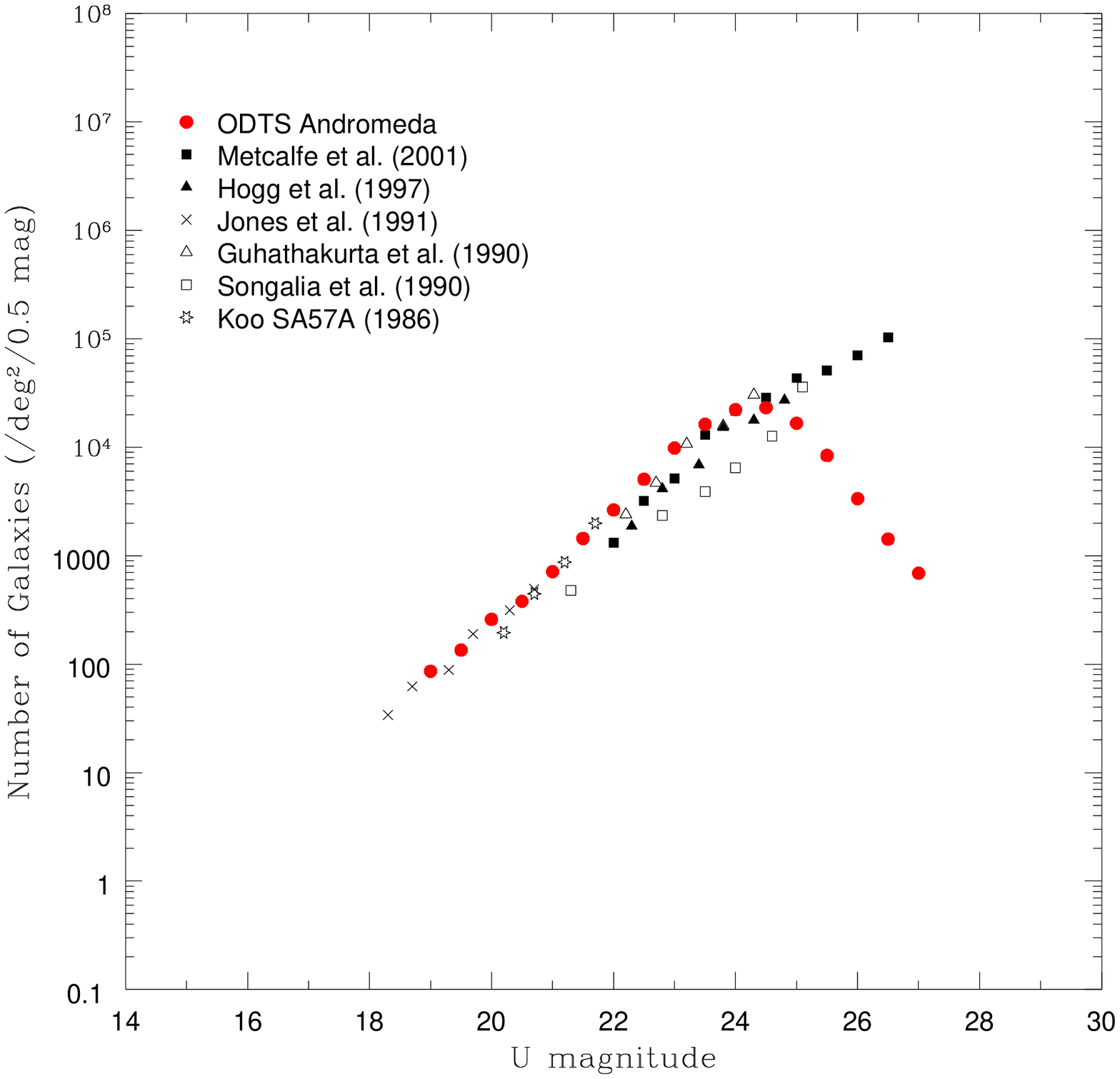} &

\begin{tabular}[b]{ccc} \\
\hline\\
{\bf Magnitude} &  \multicolumn{2}{|l|} {\bf Galaxies 0.5 mag$^{\bf -1}$}\\ 
{\bf U} & {\bf Total} & {\bf deg$^{\bf -2}$} \\ \hline

18.75-19.25 &    32  &     86  \\
19.25-19.75 &    50  &    135  \\
19.75-20.25 &    96  &    259  \\
20.25-20.75 &   141  &    380  \\
20.75-21.25 &   265  &    714  \\
21.25-21.75 &   537  &   1446  \\
21.75-22.25 &   981  &   2642  \\
22.25-22.75 &  1880  &   5064   \\
22.75-23.25 &  3651 &    9834  \\
23.25-23.75 &  6049 &   16290  \\ 
23.75-24.25 &  8246 &   22210  \\ 
24.25-24.75 &  8607 &  (23180)   \\
24.75-25.25 &  6189 &   (16670)  \\
25.25-25.75 &  3115 &    (8390)  \\ 
25.75-26.25 &  1247 &    (3359)  \\ 
26.25-26.75 &   527 &    (1419)  \\ 
26.75-27.25 &   256 &     (690)  \\ \hline
& & \\
& & \\
& & \\
& & \\
\label{tab:ugal}
\end{tabular}              \\ 
	        
\end{tabular}
\vspace{-0.3cm}
\caption{\label{fig:ucounts}{\bf U} band galaxy number counts from the ODTS compared with a complication of results taken from various sources. All counts have been converted to the standard photoelectric system. Conversion of the ODTS $U$ data to the standard {\bf U} filter (using equation~\ref{eqn:u}) requires matches in $U$ and $B$, and consequently the completeness of the {\bf U} data will be dependent of the shallowest depths reached by the matched ODTS $U$ and $B$ band data. Here, {\bf U} is found to be complete to a depth of 24.4.}
\end{center}
\end{figure*}

\subsection{Stellar Sequence Fitting}
\label{sec:stell}

To calibrate the $UBRi'$ data, the colours of the stellar objects in the ODTS were compared with the expected colours of stars, obtained from the \scite{pickles} stellar library. The model stellar colours were determined by convolving the ODTS spectral response curves (seen in figure~\ref{fig:filters}) with the SEDs of the \scite{pickles} sources. Model stellar sequences were derived in six different colour-colour regimes using $6^{th}$ order polynomials of the form\\
  
\begin{center}
\begin{equation}
y=\sum_{j=0}^{n}a_{j}x^{j}
\label{eqn:poly}
\end{equation}
\end{center}

where $x$ and $y$ are the chosen colours and $a_{j}$ are the polynomial coefficients. The best fit polynomials are shown in figure~\ref{fig:poly}.

Stellar objects were selected from each ODTS frame by choosing stars with $17 < R < 21$ and a star-galaxy classification of $> 0.9$ (see section~\ref{sec:stargal}). This made use of data with small photometric measurement errors and reliable SExtractor star-galaxy classification. Any stellar objects lying more that 0.2 magnitudes from the main sequence of objects (in the colour-colour diagrams) were excluded as outliers. 

For each frame, in each colour-colour plane shown in figure~\ref{fig:poly}, the zeropoints were allowed to vary for each of the three colours being examined, with the exception of the $V$ band zeropoint. This remains fixed as $V$ is the chosen calibrator band and the zeropoint correction factors for other bands are calculated relative to this. The $\chi^{2}$ statistic was then computed via\\

\begin{center}
\begin{equation}
\chi^{2}=\sum_{i=1}^{N}(y_{i} - Y(x_{i}))^{2}
\label{eqn:chi}
\end{equation}
\end{center}

where there are $N$ data points with positions $x_{i}$ and $y_{i}$, and $Y(x_{i})$ is the value of the best fitting polynomial (from equation~\ref{eqn:poly}) evaluated at $x_{i}$. The shifts in $x$ and $y$ which minimise this statistic were computed, as was the final change in colours required for the stellar loci of the ODTS data to match those of \scite{pickles}. This process was carried out for all six colour-colour regimes (or 4 if only $BVRi^{\prime}$ data was available). 

If the zeropoint for a given frame in a certain band differs from the 'true' zeropoint by a correction factor, $C_{i}$, then the measured shifts in colour correspond to the difference between the zeropoint correction factors of two different bands, $C_{i}$ and $C_{j}$. Using the same formalisation introduced in the previous section, the $T_{ij}$ matrix which can be solved to find the $C_{i}$ values for each frame. As the combination of colour-colour regimes constrain the zeropoint corrections for each band, the rms scatter in each zeropoint determination can be computed from\\
 
\begin{center}
\begin{equation}
rms=\frac{\sum_{i=1}^{n}\sum_{j=1}^{n}w_{ij}\theta_{ij}(T_{ij}+C{j}-C_{i})^{2}}{\sum_{i=1}^{n}\sum_{j=1}^{n}w_{ij}\theta_{ij}}
\label{eqn:rms}
\end{equation}
\end{center}

Typical rms values were found to be of the order 0.02 magnitudes. An example of the six colour-colour planes showing data before and after correction is shown in figures~\ref{fig:before} and ~\ref{fig:after}. Comparing the $UBRi'$overlap matches after zeropoint corrections were applied gave offsets $< 0.05$ magnitudes suggesting that the photometry is internally consistent at this level.

\subsection{Photometric Errors}
\label{sec:errors}
\begin{figure}
\begin{center}
\epsfig{file=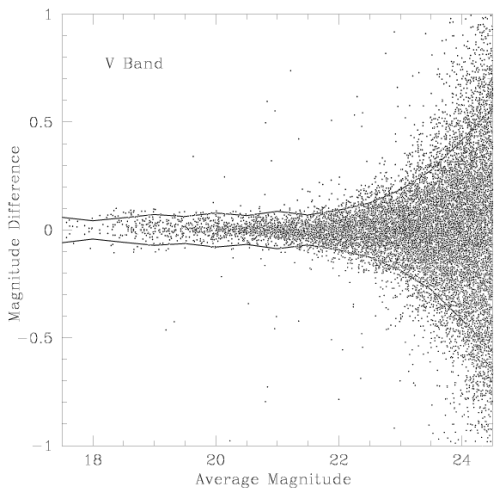, width=6.8cm, angle=0}
\end{center}
\caption{$V$ band magnitude difference between common objects in overlap regions against the mean magnitude for the reduced Andromeda data after photometric calibration. $2\sigma$ error curves have been plotted and objects with $V < 24.5$ have been used so as to avoid using data from overlapping frames which reach  different depths.}
\label{fig:magerrorsall}
\end{figure}

A number of sources of error can affect the photometry. Random errors
are estimated during the extraction and measurement of photon
counts for an object, and its subsequent conversion to a
magnitude. These measurement errors are estimated by SExtractor for
each source extraction. For a typical V-band frame, these errors are
$< 0.025$ mag at $V \leq 22$, rising to $\sim 0.075$ mag by $V\sim 23$
  and $\sim 0.3$ mag at $V \sim 24.5$. An alternative error estimate was made by plotting the magnitude difference between objects in the overlap region, after photometric calibration, against the average magnitude. This is shown for the $V$ band in figure~\ref{fig:magerrorsall}, where $2\sigma$ error curves have been superimposed. The error estimates given by SExtractor appear to be systematically lower (by $\approx 0.03$ mags) than the errors obtained by examining the photometry of common objects in overlap regions after photometric calibration. This is the case for all the bands.

\begin{figure*}  
\begin{center}
\leavevmode      
\begin{tabular} {cc}    

\includegraphics[scale=0.58,angle=0.0]{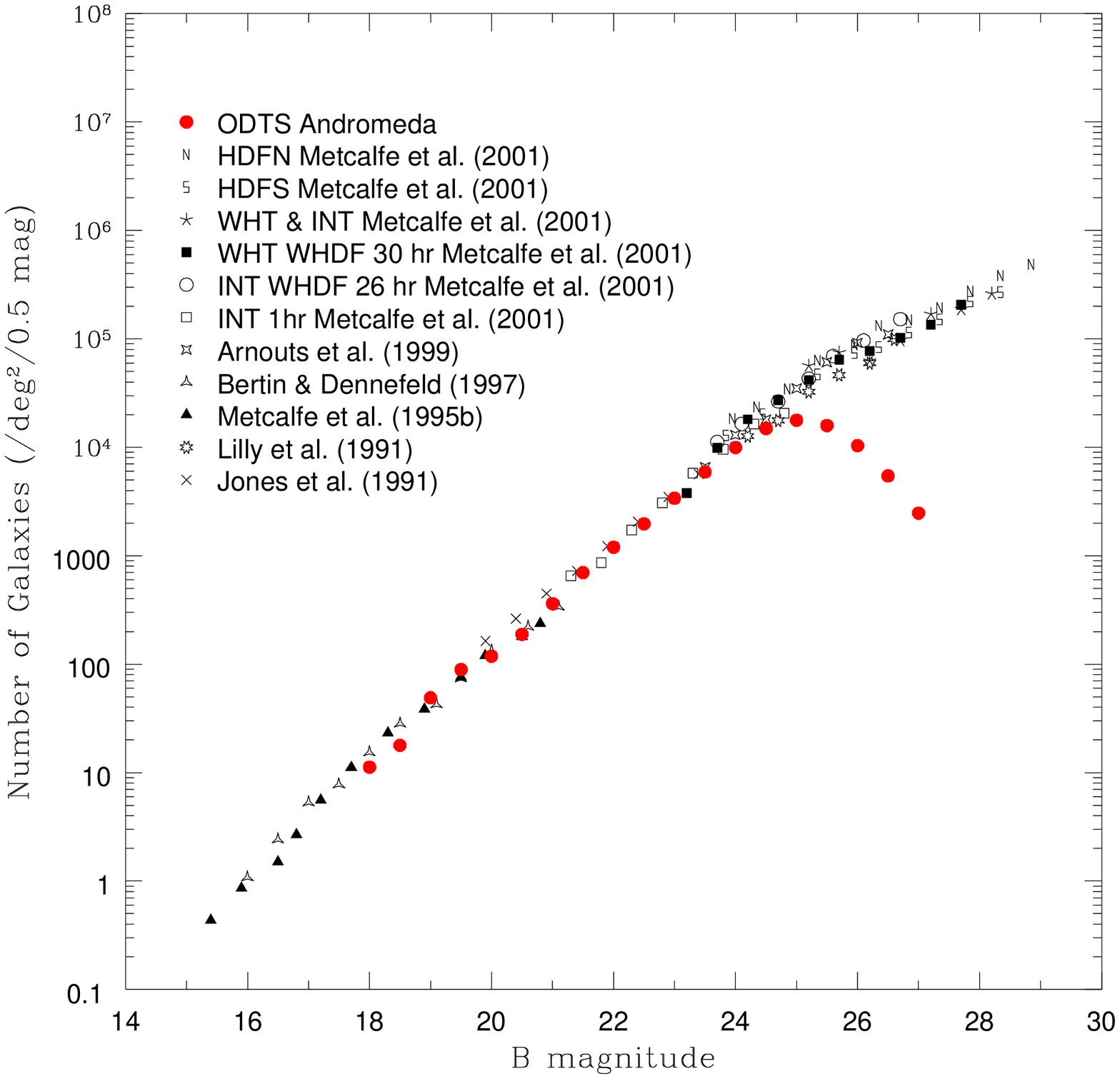} &

\begin{tabular}[b]{ccc} \\
\hline\\
{\bf Magnitude} &  \multicolumn{2}{|c|} {\bf Galaxies 0.5 mag$^{\bf -1}$}\\ 
{\bf B} & {\bf Total} & {\bf deg$^{\bf -2}$} \\ \hline
17.75-18.25 &      17   &    11  \\
18.25-18.75 &      27   &    18  \\
18.75-19.25 &      74   &    49  \\
19.25-19.75 &     135   &    89  \\
19.75-20.25 &     204   &    198  \\
20.25-20.75 &    286  &   189  \\
20.75-21.25 &    548  &   362  \\
21.25-21.75 &   1058  &   699  \\
21.75-22.25 &   1812  &    1198  \\
22.25-22.75 &   2977  &    1968  \\
22.75-23.25 &   5131  &    3392  \\
23.25-23.75 &   8935  &    5906  \\ 
23.75-24.25 &  15060  &    9956  \\ 
24.25-24.75 &  22610  &   14940  \\ 
24.75-25.25 &  26890  &   (17770)  \\ 
25.25-25.75 &  24050  &   (15890)  \\ 
25.75-26.25 &  15680  &   (10370)  \\ 
26.25-26.75 &   8246  &   (5451)   \\
26.75-27.25 &   3745  &   (2476)  \\ \hline
& & \\
& & \\
& & \\
& & \\
\label{tab:bgal}
\end{tabular}              \\ 

\end{tabular}
\vspace{-0.3cm}
\caption{\label{fig:bcounts}{\bf B} band galaxy number counts from the ODTS compared with a complication of results taken from various sources. All counts have been converted to the standard photoelectric system. Conversion of the ODTS $B$ data to the standard {\bf B} filter (using equation~\ref{eqn:bb}) requires matches in $B$ and $V$, and consequently the completeness of the {\bf B} data will be dependent of the shallowest depths reached by the matched ODTS $B$ and $V$ band data. Here, {\bf B} is found to be complete to a depth of 25.1.}
\end{center}
\end{figure*}

Two sources of calibration error are introduced during the photometric
calibration process; one from the zeropointing of the $V$ band mosaic,
typically good to $\sim 0.03$ magnitudes, and the other from the zeropoint
calibration of the other bands relative to $V$ using the stellar locus
fitting technique, found to introduce errors of $\sim 0.03$
magnitudes. Adding these errors in quadrature confirmed that the
photometry is {\it internally consistent} in all bands at the $< 0.05$
magnitude level. A final source of uncertainty comes from the
calibration of the $V$ band data using the photometric standards. The
frame used as the calibrator had a zeropoint uncertainty of 0.07
magnitudes, resulting in overall photometric uncertainties of $< 0.1$ magnitudes.

\begin{figure*}  
\begin{center}
\leavevmode      
\begin{tabular} {cc} 

\includegraphics[scale=0.58,angle=0.0]{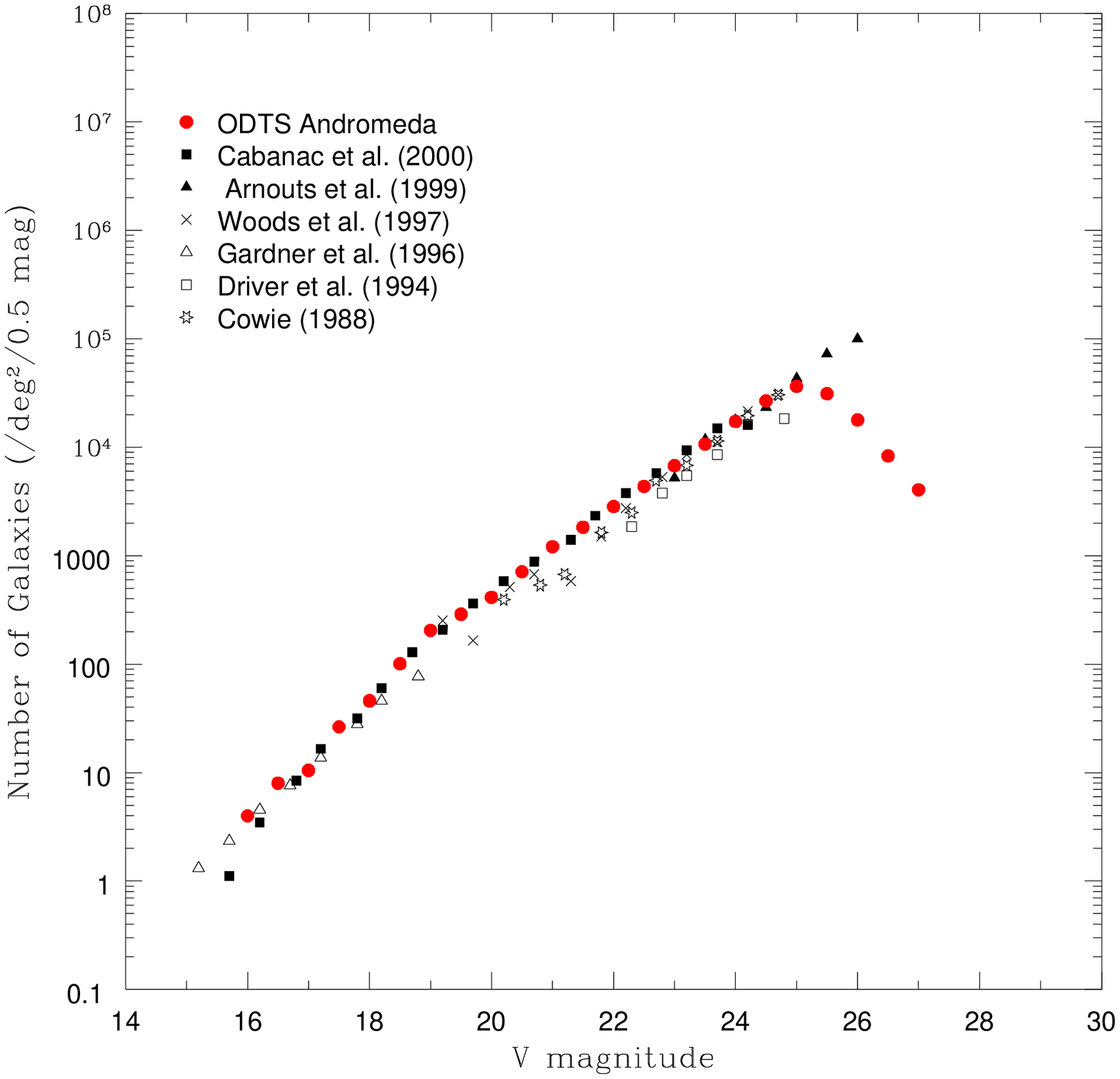} &

\begin{tabular}[b]{ccc} \\
\hline\\
{\bf Magnitude} &  \multicolumn{2}{|c|} {\bf Galaxies 0.5 mag$^{\bf -1}$}\\ 
{\bf V} & {\bf Total} & {\bf deg$^{\bf -2}$} \\ \hline

15.75-16.25 &      8   &      4  \\
16.25-16.75 &     16   &      8  \\
16.75-17.25 &     21   &      11  \\
17.25-17.75 &     53   &      26  \\
17.75-18.25 &     92   &      46  \\
18.25-18.75 &    202   &      101  \\
18.75-19.25 &    411   &      205  \\
19.25-19.75 &    579   &      289 \\
19.75-20.25 &   830    &      414  \\
20.25-20.75 &  1426    &      712  \\
20.75-21.25 &  2430    &   1213  \\
21.25-21.75 &  3665    &   1829  \\
21.75-22.25 &  5697    &   2843  \\
22.25-22.75 &  8745    &   4364   \\
22.75-23.25 &  13540  &    6756  \\
23.25-23.75 &  21470  &   10720  \\ 
23.75-24.25 &  34650  &   17290  \\ 
24.25-24.75 &  53670  &   26780   \\
24.75-25.25 &  72940  &   36400  \\ 
25.25-25.75 &  6240   &   (31140)  \\ 
25.75-26.25 &  35850  &   (17890)  \\ 
26.25-26.75 &  16660  &   (8316)  \\ 
26.75-27.25 &   8141  &   (4063)  \\ \hline
& & \\
& & \\
& & \\
& & \\
\label{tab:vgal}
\end{tabular}              \\ 

\end{tabular}
%\vspace{-0.3cm}
\caption{\label{fig:vcounts}{\bf V} band galaxy number counts from the ODTS compared with a complication of results taken from various sources. All counts have been converted to the standard photoelectric system. As the Harris $V$ filter used in the ODTS is approximately the same as the standard {\bf V} filter, no conversion of the ODTS data was required. Consequently, the completeness of the {\bf V} data will only be dependent on the shallowest depths reached by the ODTS $V$ data. Here, {\bf V} is found to be complete to a depth of 24.8.}

\end{center}
\end{figure*}

\section{Matched Catalogue Generation}
\label{scmatcat}

The final catalogues were created by matching the detections within the single colour catalogues to produce one master catalogue for each field, containing magnitudes and errors in all bands, and the corresponding extraction flags.

\begin{table}
\begin{center}
\begin{tabular}{cl} \\
Flag & \multicolumn{1}{c}{Description}  \\
Value &    \\\hline
$0+$ & Detected object given its SExtractor extraction flag. \\
$-1$ & No object detected (but region has been observed). \\
$-2$ & Object lies in a drilled region (bright star or bad area \\
     & of the chip. \\
$-3$ & Object lies in a drilled region (bright extended source). \\
$-4$ & Inconsistent matching, would need follow-up by eye. \\
$-8$ & Region yet to be observed. \\
\end{tabular}
\vspace{-0.2cm}
\caption{Flags used in the matched catalogues determining
the reliability of objects or reasons for not matching. The catalogue
stores one flag for each filter for every catalogue object.
}
\label{tab:flags}
\end{center}
\end{table}

Routines were constructed to match the single-colour ODTS data using the following  
approach. All of the astrometric solutions and hole positions were read, along with each
of the single colour catalogues. The catalogues were then sorted in turn by right ascension, and matched to the other catalogues using an astrometric tolerance of $3^{\prime\prime}$.  A large array was subsequently
produced which contained all the matching information for every object
in each filter, and a record of which objects in the other bands they
matched. Any inconsistencies arising after the first iteration of the matching routine, e.g. an object de-blended by SExtractor in one filter but not in
others due to complex morphology, were isolated and flagged appropriately. The matching process was then repeated taking account of the new flags. Within the programme, checks were made
to establish whether observations in the filter being matched existed,
if the object being matched lay within a hole, or if there simply
was not a detection in that band. Table~\ref{tab:flags} gives a list of
the flags that were assigned to each detection in the final matched
catalogue. The $-2$ and $-3$ flags took precedence over the SExtractor
extraction flags unless the SExtractor flag was $
> 4$, which indicates saturation or some sort of extraction
problem. Parameters stored for each object included the ODTS identification
number, unique to each object and containing information about the
field, filter, chip number and pixel position, the RA and
Dec. coordinates, the SExtractor star-galaxy classifier, the fluxes,
magnitudes, errors in each band and the flags for each band. The final
value of the star-galaxy classifier assigned to each object was taken from the data according to the hierarchy {\it{RBV$i^{\prime}$KU}} which was so ordered based on the quality, depth and range of the data available. The final matched catalogue produced for the Andromeda field contained $\sim 1.3$ million objects.

\section{Photometric redshifts}

\begin{figure}
\begin{center}
 \vspace{-0.4cm}
{\leavevmode \epsfxsize=8.cm \epsfysize=6.25cm \epsfbox{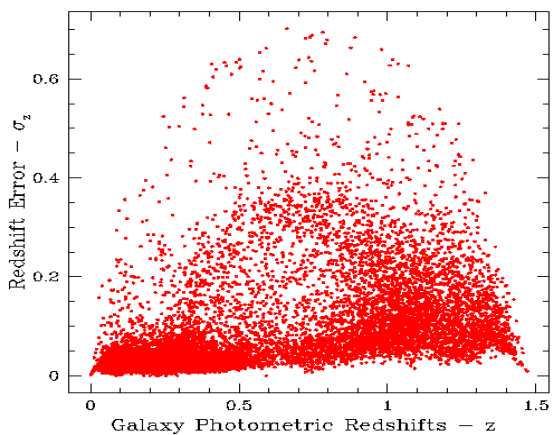}}
\end{center} 
\caption{Determined photometric redshifts versus the error in the redshift for a subsection of the matched data from the Andromeda field. All galaxies have $R < 23 $ to avoid including fainter galaxies with large photometric errors.}
\label{fig:photoz}
\end{figure}

\begin{figure*}  
\begin{center}
\leavevmode      
\begin{tabular} {cc}    

\includegraphics[scale=0.58,angle=0.0]{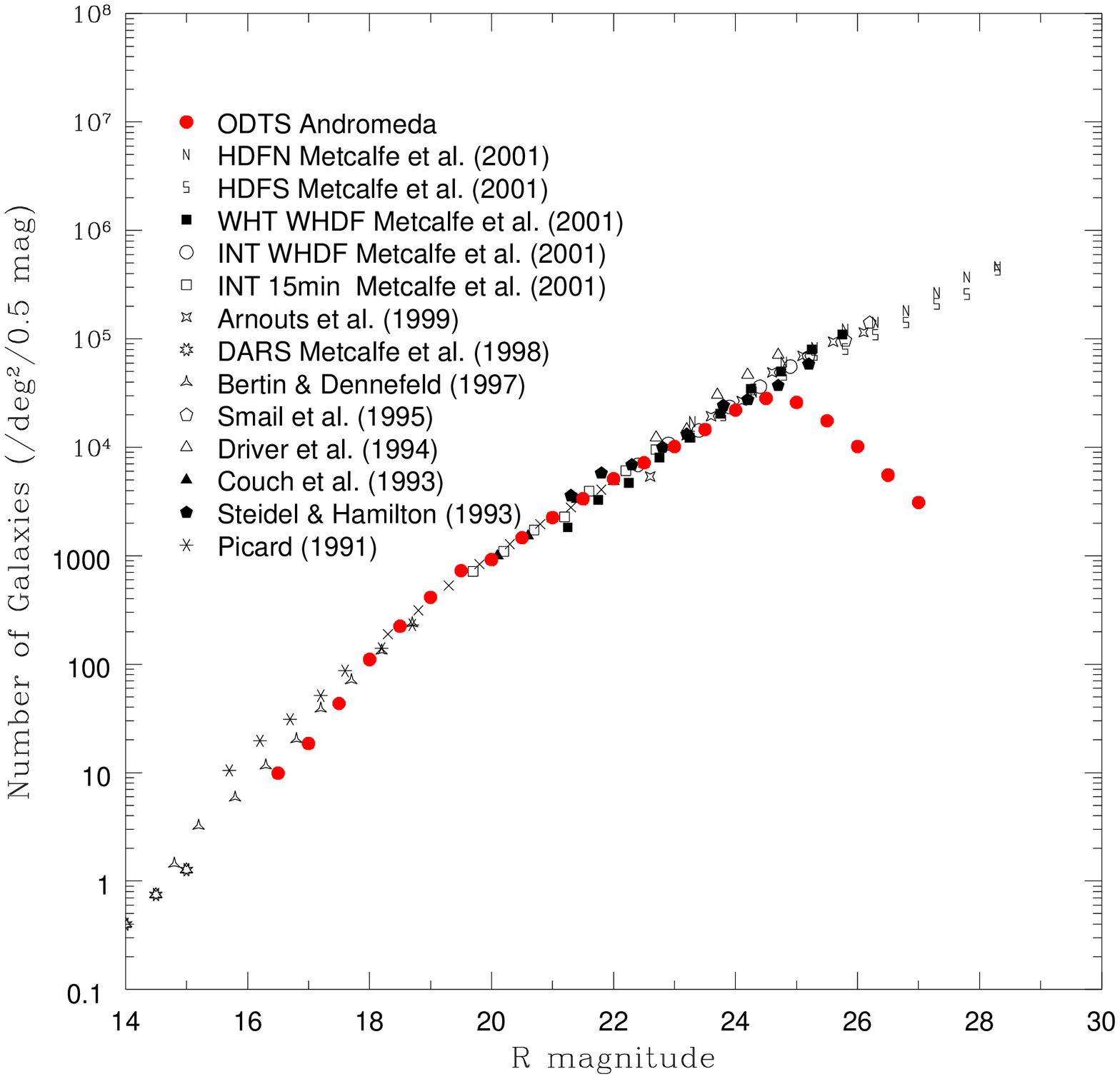} &

\begin{tabular}[b]{ccc} \\
\hline\\
{\bf Magnitude} &  \multicolumn{2}{|c|} {\bf Galaxies 0.5 mag$^{\bf -1}$}\\ 
{\bf R} & {\bf Total} & {\bf deg$^{\bf -2}$} \\ \hline

16.25-16.75 &       25  &        10     \\
16.75-17.25 &       47  &        19     \\
17.25-17.75 &      110  &        43     \\
17.75-18.25 &      280  &        111     \\
18.25-18.75 &      566  &        223     \\
18.75-19.25 &     1049  &        414     \\  
19.25-19.75 &     1846  &        729     \\  
19.75-20.25 &       2329   &     919 \\
20.25-20.75 &       3706   &    1463  \\
20.75-21.25 &       5700   &    2250  \\
21.25-21.75 &       8473   &    3345  \\
21.75-22.25 &  12950   &    5112  \\
22.25-22.75 &  18250   &    7202  \\
22.75-23.25 &  25720   &   10150 \\
23.25-23.75 &  37040   &   14620  \\
23.75-24.25 &  55750   &   22010  \\
24.25-24.75 &  71620   &   (28270)  \\
24.75-25.25 &  65680   &   (25930)  \\
25.25-25.75 &  44420   &   (17540)  \\
25.75-26.25 &  25790   &   (10180)  \\
26.25-26.75 &  14050   &   (5544)  \\ 
26.75-27.25 &  7853   &     (3100)  \\ \hline
& & \\
& & \\
& & \\
& & \\
\label{tab:rgal}
\end{tabular}              \\ 
\end{tabular}
\vspace{-0.3cm}
\caption{\label{fig:rcounts}{\bf R} band galaxy number counts from the ODTS compared with a complication of results taken from various sources. All counts have been converted to the standard photoelectric system. As the Harris $R$ filter used in the ODTS is approximately the same as the standard {\bf R} filter, no conversion of the ODTS data was required. Consequently, the completeness of the {\bf R} data will only be dependent on the shallowest depths reached by the ODTS $R$ data. Here, {\bf R} is found to be complete to a depth of 24.3.}
\end{center}
\end{figure*}

\begin{figure*}  
\begin{center}
\leavevmode      
\begin{tabular} {cc} 
	        
\includegraphics[scale=0.58,angle=0.0]{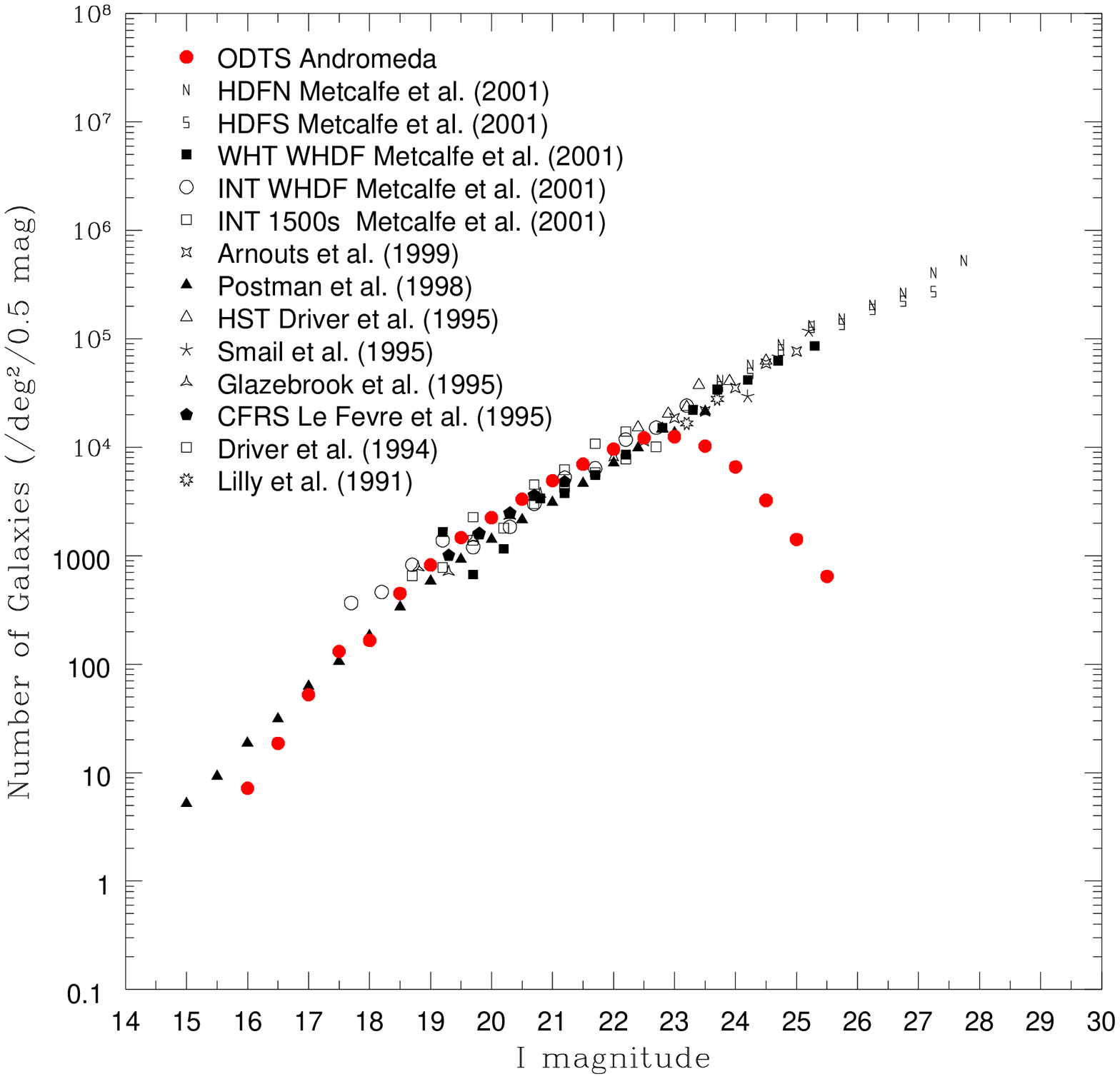} &

\begin{tabular}[b]{ccc} \\
\hline\\
{\bf Magnitude} &  \multicolumn{2}{|c|} {\bf Galaxies 0.5 mag$^{\bf -1}$}\\ 
{\bf I} & {\bf Total} & {\bf deg$^{\bf -2}$} \\ \hline
15.75-16.25 &     10   &    7    \\
16.25-16.75 &     26   &    19   \\
16.75-17.25 &     73   &    52   \\
17.25-17.75 &    183   &    131    \\
17.75-18.25 &    231   &    166    \\
18.25-18.75 &    628   &    450   \\ 
18.75-19.25 &   1145   &    820   \\ 
19.75-20.25 &   2050   &    1468   \\ 
20.25-20.75 &   3129   &    2241   \\ 
20.75-21.25 &   4634   &    3319    \\ 
21.25-21.75 &   6871   &    4922   \\ 
21.75-22.25 &   9741   &    6978   \\ 
22.25-22.75 & 13440   &    9626   \\ 
22.75-23.25 & 17030   &    12200   \\  
23.25-23.75 & 17510   &    (12540)   \\ 
23.75-24.25 & 14330   &    (10270)    \\
24.25-24.75 &  9216   &    (6601)    \\
24.75-25.25 &  4523   &    (3240)   \\ 
25.25-25.75 &  1978   &    (1417)   \\ \hline 
& & \\
& & \\
& & \\
& & \\
\label{tab:igal}
\end{tabular}              \\ 

\end{tabular}
\vspace{-0.3cm}
\caption{\label{fig:icounts}{\bf I} band galaxy number counts from the ODTS compared with a complication of results taken from various sources. All counts have been converted to the standard photoelectric system. Conversion of the ODTS $i^{\prime}$ data to the standard {\bf I} filter (using equation~\ref{eqn:i}) requires matches in $R$ and $i^{\prime}$, and consequently the completeness of the {\bf I} data will be dependent of the shallowest depths reached by the matched ODTS $R$ and $i^{\prime}$ band data. Here, {\bf I} is found to be complete to a depth of 22.8.}
\end{center}
\end{figure*}

Photometric redshifts were obtained by comparing the ODTS photometric
 data with an extensive set of model spectra taken from \scite{kinney}
 galaxy templates, PEGASE stellar synthesis code \cite{pegase}, SDSS
 QSO libraries \cite{sdssqso} and the \scite{pickles} stellar templates. 
Galaxy and QSO models were translated to longer
 wavelengths, effectively providing comparison spectra over a range of
 redshifts, and the complete set of templates were then multiplied with
 the ODTS spectral response functions to obtain potential $U-B$, $B-V$,
 $V-R$ and $R-I$ colours. The photometric data for each object were then
 compared with the possible model template colours and likelihoods
 obtained for each fit. Each object was then classified as a star,
 galaxy, or QSO, depending on the calculated likelihoods and on the
 expected number of the three types based on the object's I-band
 magnitude. The redshift probability distribution was found by summing
 up the the maximum likelihoods obtained for all fits to the model
 SEDs. The redshift error, $\sigma_{z}$, is effectively the standard deviation of the
 distribution (detailed in forthcoming Edmondson et al (in prep) paper). If the source was subsequently classified as a galaxy or QSO then the maximum
 likelihood redshift determined was recorded, along with the width of  the likelihood peak in redshift.\\
\indent Photometric redshifts were determined for the Andromeda pointings with matched $UBVR{\emph i^{\prime}}$ data and figure~\ref{fig:photoz} shows the determined redshifts against $\sigma_{z}$ for a subsection of this data. Objects were chosen to have $R < 23$ in order to avoid including the fainter objects which have larger associated photometric errors. Within this representative sample, $\approx 8000$ ($81\%$) were found to have $\sigma_{z} < 0.1$ and $\approx 6000$ to have $\sigma_{z} < 0.05$ ($60\%$).\\
\indent An estimate for the median redshift of the ODTS was obtained using the
redshift distributions of
\scite{Fsoto}, \scite{Cohen}, \scite{goods} and \scite{goods2}. From \scite{BaughEfst},
 a magnitude
dependent redshift distribution can be defined using

\begin{center}
\begin{equation}
n(z,R) \propto z^2 \exp -\left [ \frac{z}{z_0(R)}\right ] ^{1.5} 
\label{eqn:photoz}
\end {equation}
\end{center}

 where $z_0(R) = z_m(R)/1.4$ and $z_m(R)$ is the
median redshift as a function of R band magnitude. The
$z_m(R)$ is estimated from the redshift distribution of galaxies with known
spectroscopic or photometric redshifts sampled in R
band magnitude bins of width 0.5 magnitudes.  Summing the magnitude
dependent model redshift distributions, weighted according
to the magnitude distribution of the ODTS, results in an estimate
of the redshift distribution for the ODTS, found to be $z_{m} \approx 0.7$. This was in good
agreement with the $z_{m}$ determined from the median magnitude/median redshift 
relationship, derived from the COMBO-17 survey, documented in \pcite{MLB02}, where the
ODTS median R band magnitude of $R_{m} = 23.2$, corresponds to $z_m \approx 0.7$.

\section{Galaxy Number Counts}
\label{sec:galnocounts}

All of the ODTS magnitudes were converted to the standard
Johnson-Morgan-Cousins photoelectric filter system, denoted here as
{\bf UBVRI}, using the following colour transformation
equations\footnote{See http://www.ast.cam.ac.uk/$\sim$wfcsur/colours.php}\\

\begin{center}
\begin{equation}
{\bf U}=\frac{(U_{\rm RGO}-B_{\rm KPNO})}{12.167} + U_{\rm RGO} - 0.119
\label{eqn:u}
\end {equation}
\end{center}

\vspace{0.3cm}

\begin{center}
\begin{equation}
{\bf B}=\frac{(B_{\rm KPNO}-V_{\rm Har})}{7.008} + B_{\rm KPNO} - 0.001
\label{eqn:bb}
\end {equation}
\end{center}

\vspace{0.3cm}

\begin{center}
\begin{equation}
{\bf I}=i'_{\rm Sln} - 0.2006(R_{\rm Har} - i'_{\rm Sln}-0.004)
\label{eqn:i}
\end {equation}
\end{center}

\vspace{0.5cm}

The Harris $V$ and $R$ filters used in the ODTS were deemed close
enough to the standard photoelectric bands to warrant no
conversion. Galaxy number counts for the ODTS are shown in figures~\ref{fig:ucounts},~\ref{fig:bcounts},~\ref{fig:vcounts},~\ref{fig:rcounts}
and ~\ref{fig:icounts} and are found to be in excellent agreement
with previous results. It should be noted that the results shown use {\it all} of the data from the Andromeda field, thus represent the average depths reached. For a few pointings, where data was acquired in the best conditions, i.e. when the seeing was $\leq 1$, the target depths given in table~\ref{tab:filters} were reached. For the {\bf V} and {\bf R} bands, where no filter conversions are needed, the depths reached in figures~\ref{fig:vcounts} and ~\ref{fig:rcounts} depend only on on the depths of the ODTS $V$ and $R$ data respectively. However, to enable the conversion of the $U,V$ and $i^{\prime}$ ODTS data (as in equations~\ref{eqn:u},~\ref{eqn:bb} and ~\ref{eqn:i}), only galaxies from the matched catalogue with the necessary multi-colour data 
could be used to derive the galaxy number counts
shown. Hence the depths reached in
figures~\ref{fig:ucounts},~\ref{fig:bcounts},
and ~\ref{fig:icounts} will depend on the most shallow of the multi-band data used. As a result of this, the depths reached are {\bf U}= 24.4, {\bf B}= 25.1, {\bf I}= 22.8, slightly shallower than the average depths stated in table~\ref{tab:seeing} which is as expected, {\bf V}=24.8 and {\bf R}=24.3, the same depths as those quoted in table~\ref{tab:seeing}.

\section{Summary}

The Oxford-Dartmouth Thirty Degree Survey (ODTS) is a deep, wide,
multi-band imaging survey. The Wide Field Camera on the 2.5m Isaac
Newton Telescope on La Palma has been used to obtain $UBVRi^{\prime}$
data and the $Z$ band data are being acquired using the 2.4m Hiltner
Telescope at the MDM observatory on Kitt Peak. A complementary $K$
band survey is currently being carried out using the 1.3m McGraw-Hill
Telescope at MDM. Four survey regions of 5-10 deg$^{2}$, centred at 
00:18:24 +34:52 (Andromeda),
09:09:45 +40:50 (Lynx), 13:40:00 +02:30 (Virgo) and 16:39:30 +45:24
(Hercules) have been covered to average $5\sigma$ limiting depths (Vega) of $B$ = 25.5, $V$ = 25.1, $R$ = 24.6, and $i^{\prime}$ = 23.5, with $U$ and $K$ subsets covered to depths of 25.1 and 18.5 respectively. Initial data analysis indicates that the ODTS reaches  depths
$\approx 0.5$ magnitudes shallower than previously anticipated,
attributed to less than optimal seeing during observations. On completion of 
INT observations for the ODTS, approximately 23 square degrees have been
covered in $BVRi^{\prime}$, with a subset of 1.5 square degrees in $U$.

This paper details the process from data acquisition, through data reduction and calibration, to the resultant multi-colour catalogues.

Photometric redshifts were calculated for a representative sample of objects in the Andromeda field, and were found to have $\sigma_{z} < 0.1 $ for $\sim 80\%$ of the data and $\sigma_{z} < 0.05 $ for $\sim 60\%$ of the data. The median redshift of the survey to date was estimated to be z $\approx 0.7$. 

Galaxy number counts were determined and were found to compare well with previous survey results.

Preliminary evaluation of the ODTS data shows that the overall quality and quantity of the data is sufficient to meet the initial science objectives of the ODTS. Research undertaken using the ODTS and the results obtained will be documented in the forthcoming papers of this series.

\section{Acknowledgements} 
 
The Isaac Newton Telescope is operated on the island of La Palma by the Isaac Newton Group in the Spanish Observatorio del Roque de los Muchachos of the Instituto de Astrofisica de Canaries. Kitt Peak National Observatory, National Optical Astronomy Observatory, is operated by the Association of Universities for Research in Astronomy, Inc. (AURA) under cooperative agreement with the National Science Foundation.  ECM thanks the C.K. Marr Educational Trust. PDA, CEH, EME, and CAB acknowledge the support of PPARC Studentships. This work was supported by the PPARC Rolling Grant PPA/G/O/2001/00017 at the University of Oxford.

\nocite{1991MNRAS.249..481J}
\nocite{1986ApJ...311..651K}
\nocite{1990ApJ...357L...9G}
\nocite{1990ApJ...348..371S}
\nocite{1997MNRAS.288..404H}
\nocite{metcalfe01}
\nocite{1986seg..work..439S}
\nocite{1986MNRAS.221..233P}
\nocite{1995MNRAS.274..769M}
\nocite{1991ApJ...369...79L}
\nocite{1997A&A...317...43B}
\nocite{1999A&A...341..641A}
\nocite{1998MNRAS.294..147M}
\nocite{1994MNRAS.268..393D}
\nocite{1993MNRAS.260..241C}
\nocite{1993AJ....105.2017S}
\nocite{1991AJ....102..445P}
\nocite{1995ApJ...449L.105S}
\nocite{1995ApJ...449L..23D}
\nocite{1995MNRAS.275L..19G}
\nocite{1995ApJ...455...60L}
\nocite{1998ApJ...506...33P}
\nocite{2000A&A...364..349C}
\nocite{1997ApJ...490...11W}
\nocite{1996MNRAS.282L...1G}
\nocite{1988ApJ...332L..29C}

\bibliographystyle{mnras} 
\bibliography{odtsurveyI} 

\begin{thebibliography}{{Fern{\' a}ndez-Soto}, {Lanzetta} \& {Yahil}<1999>}
\bibitem[{Anderson}<2001>]{anderson}{Anderson} S.~F. e.~a., 2001.\newblock {\rm
  AJ}, {\rm 122}, 503.
\bibitem[{Arnouts} {\rm et~al.}<1999>]{1999A&A...341..641A}{Arnouts} S.,
  {D'Odorico} S., {Cristiani} S., {Zaggia} S., {Fontana} A., {Giallongo} E.,
  1999.\newblock {\rm A\&A}, {\rm 341}, 641.
\bibitem[{Arnouts}<2001>]{eisdeep}{Arnouts} S. e.~a., 2001.\newblock {\rm A\&
  A}, {\rm 379}, 740.
\bibitem[{Baugh} \& {Efstathiou}<1994>]{BaughEfst}{Baugh} C.~M., {Efstathiou}
  G., 1994.\newblock {\rm MNRAS}, {\rm 267}, 323.
\bibitem[{Bertin} \& {Arnouts}<1996>]{sex}{Bertin} E., {Arnouts} S.,
  1996.\newblock {\rm A\&AS}, {\rm 117}, 393.
\bibitem[{Bertin} \& {Dennefeld}<1997>]{1997A&A...317...43B}{Bertin} E.,
  {Dennefeld} M., 1997.\newblock {\rm A\&A}, {\rm 317}, 43.
\bibitem[{Booth}<2001>]{booth}{Booth} J., 2001.\newblock {\rm PhD thesis}, {\rm
  Univ. Oxford}.
\bibitem[Brown {\rm et~al.}<2003>]{MLB02}Brown M., Taylor A., Bacon D., Gray
  M., Dye S., Meisenheimer K., Wolf C., 2003.\newblock {\rm MNRAS}, {\rm 341},
  100.
\bibitem[Budavari {\rm et~al.}<2001>]{sdssqso}Budavari T., Csabai I., Szalay A.
  {\rm et~al.}, 2001.\newblock {\rm AJ}, {\rm 122}, 1163.
\bibitem[{Cabanac}, {de Lapparent} \&
  {Hickson}<2000>]{2000A&A...364..349C}{Cabanac} R.~A., {de Lapparent} V.,
  {Hickson} P., 2000.\newblock {\rm A\&A}, {\rm 364}, 349.
\bibitem[{Cohen} {\rm et~al.}<2000>]{Cohen}{Cohen} J.~G., {Hogg} D.~W.,
  {Blandford} R., {Cowie} L.~L., {Hu} E., {Songaila} A., {Shopbell} P.,
  {Richberg} K., 2000.\newblock {\rm ApJ}, {\rm 538}, 29.
\bibitem[{Colless}<2001>]{Colless}{Colless} M. e.~a., 2001.\newblock {\rm
  MNRAS}, {\rm 328}, 1039.
\bibitem[{Couch}, {Jurcevic} \& {Boyle}<1993>]{1993MNRAS.260..241C}{Couch}
  W.~J., {Jurcevic} J.~S., {Boyle} B.~J., 1993.\newblock {\rm MNRAS}, {\rm
  260}, 241.
\bibitem[{Cowie} {\rm et~al.}<1988>]{1988ApJ...332L..29C}{Cowie} L.~L., {Lilly}
  S.~J., {Gardner} J., {McLean} I.~S., 1988.\newblock {\rm ApJ}, {\rm 332},
  L29.
\bibitem[{Cowie} {\rm et~al.}<2004>]{goods}{Cowie} L., {Barger} A., {Hu} E.,
  {Capak} P., {Songaila} A., 2004.\newblock {\rm AJ}, {\rm submitted}.
\bibitem[{Driver} {\rm et~al.}<1994>]{1994MNRAS.268..393D}{Driver} S.~P.,
  {Phillipps} S., {Davies} J.~I., {Morgan} I., {Disney} M.~J., 1994.\newblock
  {\rm MNRAS}, {\rm 268}, 393.
\bibitem[{Driver} {\rm et~al.}<1995>]{1995ApJ...449L..23D}{Driver} S.~P.,
  {Windhorst} R.~A., {Ostrander} E.~J., {Keel} W.~C., {Griffiths} R.~E.,
  {Ratnatunga} K.~U., 1995.\newblock {\rm ApJ}, {\rm 449}, L23+.
\bibitem[{Fern{\' a}ndez-Soto}, {Lanzetta} \& {Yahil}<1999>]{Fsoto}{Fern{\'
  a}ndez-Soto} A., {Lanzetta} K.~M., {Yahil} A., 1999.\newblock {\rm ApJ}, {\rm
  513}, 34.
\bibitem[Fioc \& Rocca-Volmerange<1997>]{pegase}Fioc M., Rocca-Volmerange B.,
  1997.\newblock {\rm A\&A}, {\rm 326}, 950.
\bibitem[{Gardner} {\rm et~al.}<1996>]{1996MNRAS.282L...1G}{Gardner} J.~P.,
  {Sharples} R.~M., {Carrasco} B.~E., {Frenk} C.~S., 1996.\newblock {\rm
  MNRAS}, {\rm 282}, L1.
\bibitem[{Geller}, {de Lapparent} \& {Kurtz}<1984>]{geller}{Geller} M.~J., {de
  Lapparent} V., {Kurtz} M.~J., 1984.\newblock {\rm ApJ}, {\rm 287}, L55.
\bibitem[{Gladders} \& {Yee}<2000>]{gladders}{Gladders} M.~D., {Yee} H.~K.~C.,
  2000.\newblock {\rm AJ}, {\rm 120}, 2148.
\bibitem[{Glazebrook} {\rm et~al.}<1994>]{glazebrook}{Glazebrook} K., {Peacock}
  J.~A., {Collins} C.~A., {Miller} L., 1994.\newblock {\rm MNRAS}, {\rm 266},
  65.
\bibitem[{Glazebrook} {\rm et~al.}<1995>]{1995MNRAS.275L..19G}{Glazebrook} K.,
  {Ellis} R., {Santiago} B., {Griffiths} R., 1995.\newblock {\rm MNRAS}, {\rm
  275}, L19.
\bibitem[{Guhathakurta}, {Tyson} \&
  {Majewski}<1990>]{1990ApJ...357L...9G}{Guhathakurta} P., {Tyson} J.~A.,
  {Majewski} S.~R., 1990.\newblock {\rm ApJ}, {\rm 357}, L9.
\bibitem[{Hill} \& {Rawlings}<2003>]{Hill}{Hill} G.~J., {Rawlings} S.,
  2003.\newblock {\rm New Astronomy Review}, {\rm 47}, 373.
\bibitem[{Hogg} {\rm et~al.}<1997>]{1997MNRAS.288..404H}{Hogg} D.~W., {Pahre}
  M.~A., {McCarthy} J.~K., {Cohen} J.~G., {Blandford} R., {Smail} I., {Soifer}
  B.~T., 1997.\newblock {\rm MNRAS}, {\rm 288}, 404.
\bibitem[{Jannuzi} {\rm et~al.}<2002>]{ndwfs}{Jannuzi} B.~T., {Dey} A., {Brown}
  M.~J.~I., {Tiede} G.~P., {NDWFS Team}, 2002.\newblock {\rm American
  Astronomical Society Meeting}, {\rm 201}, 0.
\bibitem[{Jones} {\rm et~al.}<1991>]{1991MNRAS.249..481J}{Jones} L.~R., {Fong}
  R., {Shanks} T., {Ellis} R.~S., {Peterson} B.~A., 1991.\newblock {\rm MNRAS},
  {\rm 249}, 481.
\bibitem[Kinney, Calzetti {\rm et~al.}<1996>]{kinney}Kinney A., Calzetti D.
  {\rm et~al.}, 1996.\newblock {\rm ApJ}, {\rm 467}, 38.
\bibitem[{Koo}<1986>]{1986ApJ...311..651K}{Koo} D.~C., 1986.\newblock {\rm
  ApJ}, {\rm 311}, 651.
\bibitem[{Landolt}<1992>]{landolt}{Landolt} A.~U., 1992.\newblock {\rm AJ},
  {\rm 104}, 340.
\bibitem[{Lasker} \& Team<1998>]{Lasker}{Lasker} B.~M., Team S. S.-S.,
  1998.\newblock {\rm Bulletin of the American Astronomical Society}, {\rm 30},
  912.
\bibitem[{Le Fevre} {\rm et~al.}<1995>]{1995ApJ...455...60L}{Le Fevre} O.,
  {Crampton} D., {Lilly} S.~J., {Hammer} F., {Tresse} L., 1995.\newblock {\rm
  ApJ}, {\rm 455}, 60.
\bibitem[{Lilly}, {Cowie} \& {Gardner}<1991>]{1991ApJ...369...79L}{Lilly}
  S.~J., {Cowie} L.~L., {Gardner} J.~P., 1991.\newblock {\rm ApJ}, {\rm 369},
  79.
\bibitem[{McCracken} {\rm et~al.}<2001>]{McCracken}{McCracken} H.~J., {Le F{\`
  e}vre} O., {Brodwin} M., {Foucaud} S., {Lilly} S.~J., {Crampton} D.,
  {Mellier} Y., 2001.\newblock {\rm ApJ}, {\rm 376}, 756.
\bibitem[{Metcalfe} {\rm et~al.}<1998>]{1998MNRAS.294..147M}{Metcalfe} N.,
  {Ratcliffe} A., {Shanks} T., {Fong} R., 1998.\newblock {\rm MNRAS}, {\rm
  294}, 147.
\bibitem[{Metcalfe} {\rm et~al.}<2001>]{metcalfe01}{Metcalfe} N., {Shanks} T.,
  {Campos} A., {McCracken} H.~J., {Fong} R., 2001.\newblock {\rm MNRAS}, {\rm
  323}, 795.
\bibitem[{Metcalfe}, {Fong} \& {Shanks}<1995>]{1995MNRAS.274..769M}{Metcalfe}
  N., {Fong} R., {Shanks} T., 1995.\newblock {\rm MNRAS}, {\rm 274}, 769.
\bibitem[{Monet}<2003>]{usno}{Monet} D.~G. e.~a., 2003.\newblock {\rm AJ}, {\rm
  125}, 984.
\bibitem[{Nonino}<1999>]{nonino}{Nonino} M. e.~a., 1999.\newblock {\rm A\&AS},
  {\rm 137}, 51.
\bibitem[{Olding}<2002>]{olding}{Olding} E.~J., 2002.\newblock {\rm PhD
  thesis}, {\rm Univ. Oxford}.
\bibitem[{Peterson} {\rm et~al.}<1986>]{1986MNRAS.221..233P}{Peterson} B.~A.,
  {Ellis} R.~S., {Efstathiou} G., {Shanks} T., {Bean} A.~J., {Fong} R.,
  {Zen-Long} Z., 1986.\newblock {\rm MNRAS}, {\rm 221}, 233.
\bibitem[{Picard}<1991>]{1991AJ....102..445P}{Picard} A., 1991.\newblock {\rm
  AJ}, {\rm 102}, 445.
\bibitem[Pickles<1998>]{pickles}Pickles A., 1998.\newblock {\rm Publications of
  the Astronomical Society of the Pacific}, {\rm 110}, 863.
\bibitem[{Postman} {\rm et~al.}<1998>]{1998ApJ...506...33P}{Postman} M.,
  {Lauer} T.~R., {Szapudi} I., {Oegerle} W., 1998.\newblock {\rm ApJ}, {\rm
  506}, 33.
\bibitem[{Schlegel}, {Finkbeiner} \& {Davis}<1998>]{schlegel}{Schlegel} D.~J.,
  {Finkbeiner} D.~P., {Davis} M., 1998.\newblock {\rm ApJ}, {\rm 500}, 525.
\bibitem[{Smail} {\rm et~al.}<1995>]{1995ApJ...449L.105S}{Smail} I., {Hogg}
  D.~W., {Yan} L., {Cohen} J.~G., 1995.\newblock {\rm ApJ}, {\rm 449}, L105+.
\bibitem[{Songaila}, {Cowie} \& {Lilly}<1990>]{1990ApJ...348..371S}{Songaila}
  A., {Cowie} L.~L., {Lilly} S.~J., 1990.\newblock {\rm ApJ}, {\rm 348}, 371.
\bibitem[{Steidel} \& {Hamilton}<1993>]{1993AJ....105.2017S}{Steidel} C.~C.,
  {Hamilton} D., 1993.\newblock {\rm AJ}, {\rm 105}, 2017.
\bibitem[{Stevenson}, {Shanks} \& {Fong}<1986>]{1986seg..work..439S}{Stevenson}
  P.~R.~F., {Shanks} T., {Fong} R., 1986.\newblock In: {\it ASSL Vol. 122:
  Spectral Evolution of Galaxies}, p.~439.
\bibitem[{Stoughton}, {Lupton} \& {Bernardi}<2002>]{stoughton}{Stoughton} C.,
  {Lupton} R., {Bernardi} M. e.~a., 2002.\newblock {\rm ApJ}, {\rm 123}, 485.
\bibitem[{Wirth}<2004>]{goods2}{Wirth} G. e.~a., 2004.\newblock {\rm preprint},
  {\rm astro-ph/0401353}.
\bibitem[{Wolf} {\rm et~al.}<2003>]{combo}{Wolf} C., {Meisenheimer} K., {Rix}
  H.-W., {Borch} A., {Dye} S., {Kleinheinrich} M., 2003.\newblock {\rm A\&A},
  {\rm 401}, 73.
\bibitem[{Woods} \& {Fahlman}<1997>]{1997ApJ...490...11W}{Woods} D., {Fahlman}
  G.~G., 1997.\newblock {\rm ApJ}, {\rm 490}, 11.

\end{thebibliography}

\end{document}